\documentclass[3p,number]{elsarticle}

\usepackage{amsmath}
\usepackage{amsfonts,dsfont}
\usepackage{amssymb}
\usepackage{graphicx}

\newlength{\graphwid}
\def\showgraph#1#2{
\settowidth{\graphwid}{\includegraphics[#1,clip=true]{#2}}
\parbox[c]{\graphwid}{\includegraphics[#1,clip=true]{#2}}}

\newcommand{\act}{\mathcal A} 
\newcommand{\Int}{\mathrm{int}} 

\newcommand{\bra}[1]{\left\langle{#1}\right\rvert} 
\newcommand{\ket}[1]{\left\lvert{#1}\right\rangle} 
\newcommand{\bracket}[2]{\left\langle{#1}\vert{#2}\right\rangle}

\newcommand{\dd}{\mathrm d} 
\newcommand{\DD}{\mathcal D} 
\newcommand{\diag}{\mathrm{diag}\,}

\newcommand{\Tr}{\mathrm{Tr}\,} 
\newcommand{\Ln}{\mathrm{Ln}\,} 
\newcommand{\Det}{\mathrm{Det}\,} 

\newcommand{\keldC}{{\mathcal C}} 

\newcommand{\xUnit}{\mathds{1}} 
\newcommand{\Texp}{\mathrm{Texp}\,} 

\newcommand{\In}{\mathrm{in}} 
\newcommand{\Out}{\mathrm{out}} 
\newcommand{\up}{\wedge} 
\newcommand{\down}{\vee} 
\newcommand{\updown}{{\wedge\!\vee}} 

\newcommand{\wire}{{\mathcal M}} 

\newcommand{\redS}{\regSM} 

\newcommand{\aT}{{\tilde T}} 
\newcommand{\T}{T} 

\newcommand{\stuff}[1]{}

\newcommand{\vF}{{v_F}} 

\newcommand{\tm}{{\mathcal M}}
\newcommand{\impS}{\Sigma} 
\newcommand{\scattPot}{\mathcal V} 
\newcommand{\sign}{\mathrm{sign}\,}

\newcommand{\ec}{E_C} 
\newcommand{\wc}{\omega_C} 
\newcommand{\nuEff}{\nu^\ast} 

\renewcommand{\Im}{\mathrm{Im}\,} 
\renewcommand{\Re}{\mathrm{Re}\,} 

\newcommand{\eqn}[1]{(\ref{eqn:#1})} 

\newcommand{\regSM}{{\tilde S}} 

\newcommand{\sr}{S} 
\newcommand{\dr}{D} 

\newcommand{\tun}{\mathrm{tun}}

\newcommand{\lorg}{\mathcal L} 

\newcommand{\leftexp}[2]{{\vphantom{#2}}^{#1}{#2}}
\def\clap#1{\hbox to 0pt{\hss#1\hss}}

\begin{document}

\begin{frontmatter}


\title{Nonequilibrium functional bosonization of quantum wire networks}

\author[TKM,DFG]{St\'ephane~Ngo~Dinh}
\address[TKM]{Institut f\"ur Theorie der \!Kondensierten \!Materie, Karlsruhe Institute of Technology, 76128 Karlsruhe, Germany}

\author[TH]{Dmitry A. Bagrets}
\address[INT]{Institut f\"ur Nanotechnologie, Karlsruhe Institute of Technology, 76021 Karlsruhe, Germany} 
\address[TH]{Institut f\"ur Theoretische Physik, Universit\"at zu K\"oln, Z\"ulpicher Str.~77, 50937 K\"oln, Germany} 

\author[TKM,INT,DFG,NPI]{Alexander D. Mirlin}
\address[DFG]{DFG Center for Functional Nanostructures, Karlsruhe Institute of
Technology, 76128 Karlsruhe, Germany}
\address[NPI]{Petersburg Nuclear Physics Institute, 188300 St.~Petersburg, Russia}

\begin{abstract}
We develop a general approach to nonequilibrium nanostructures formed
by one-dimensional channels coupled by tunnel junctions and/or by
impurity scattering. The formalism is based on nonequilibrium
version of functional bosonization. A central role in this approach is played
by the Keldysh action that has a form reminiscent of the theory of full
counting statistics. To proceed with evaluation of physical observables, we
assume the weak-tunneling regime and develop a real-time instanton method. 
A detailed exposition of the formalism
is supplemented by two important applications: (i) tunneling into a biased Luttinger liquid
with an impurity, and (ii) quantum-Hall Fabry-P\'erot interferometry.
\end{abstract}


\end{frontmatter}



\section{Introduction}
\label{sect:introduction}

Non-equilibrium electronic phenomena in nanostructures represent one of central
directions of the modern condensed matter physics
\cite{Levchenko09,Kamenev11}. Advances of nanofabrication have allowed
researchers
to explore experimentally transport properties of a great variety of
nanodevices. Many remarkable phenomena have been observed in
far-from-equilibrium regimes, i.e. for sufficiently large applied bias
voltages. 

A particularly important class of nanostructures is represented by
coupled one-dimensional (1D) channels (quantum wires). The coupling
may be due to tunneling between the wires or due to impurity-induced
backscattering. Realizations of 1D elements that may serve as
building blocks of such networks include, in particular, semiconducting and metallic
quantum wires, carbon nanotubes, and quantum Hall edge states.

The standard analytical approach to interacting 1D systems (Luttinger
liquids) is the bosonization \cite{giamarchi-book}. Recently, a
nonequilibrium generalization of the bosonization framework
\cite{gutman10} was developed for
setups where a nonequilibrium fermionic distribution is created outside of
the interacting region and ``injected'' into the Luttinger liquid. We
will focus here on a more complicated situation when the tunneling or the impurity
backscattering takes place inside the interacting part of the
system. Such coupling terms represent in general a very serious
complication for the full bosonization approach, and 
we are not aware of any way to solve the problem exactly. We choose
instead an alternative route based on the functional bosonization
formalism \cite{Grishin04} that  retains both fermionic and bosonic
degrees of freedom. Combining the functional bosonization idea with
the Keldysh nonequilibrium framework, we derive Keldysh action for
the considered class of problems. This action has a structure
reminiscent of that of the generating function of the full counting
statistics \cite{Levitov93,Klich}. Our action generalizes that of
Ref.~\cite{Nazarov08} where a local scatterer  under
nonequilibrium conditions was explored. 

To proceed with evaluation of the functional integral, we assume a weak
tunneling and develop a
real-time instanton (saddle-point) method. This allows us to determine
Keldysh Green functions characterizing physical observables under
interest (tunneling density of states, distribution functions,
current-voltage characteristics, etc.).

The goal of this article is twofold. First, we give a detailed
exposition of the theoretical framework. 
Second, to illustrate the method, we present
application of this approach to two important problems:   (i)
tunneling into a biased Luttinger liquid 
with an impurity, 
and (ii) quantum-Hall Fabry-P\'erot interferometry.

The structure of the paper is as follows. 
Our general formalism is presented in Sec.~\ref{sect:framework}. 
In Sec.~\ref{sect:TDOS} we use the approach to calculate the
tunneling density of states of a biased Luttinger liquid 
with an impurity. 
In Sec.~\ref{sect:FPI} we apply the method to explore a nonequilibrium
quantum-Hall Fabry-P\'erot interferometer. Section \ref{sect:conclusion}
includes a summary of our results and a discussion of prospective
research directions. Technical details are presented in five
Appendices. 

Some of the results of this article have been previously published in a
concise form in short communications \cite{Bagrets10,Ngo_Dinh12}. 

\section{General Framework} 
\label{sect:framework}

\subsection{Model and Functional Bosonization} 
\label{sect:model}

\begin{figure}[ht]
	\centering{\includegraphics[scale=0.5]{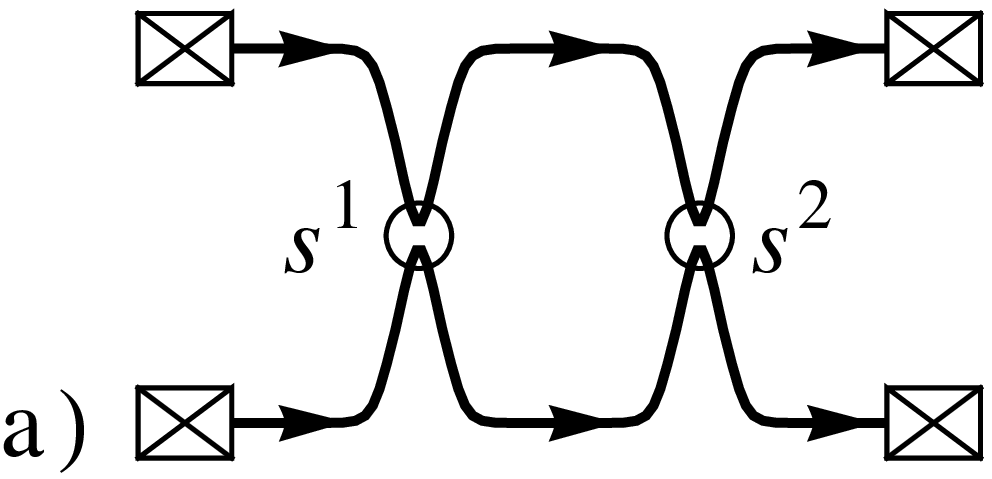}
          \quad\includegraphics[scale=0.5]{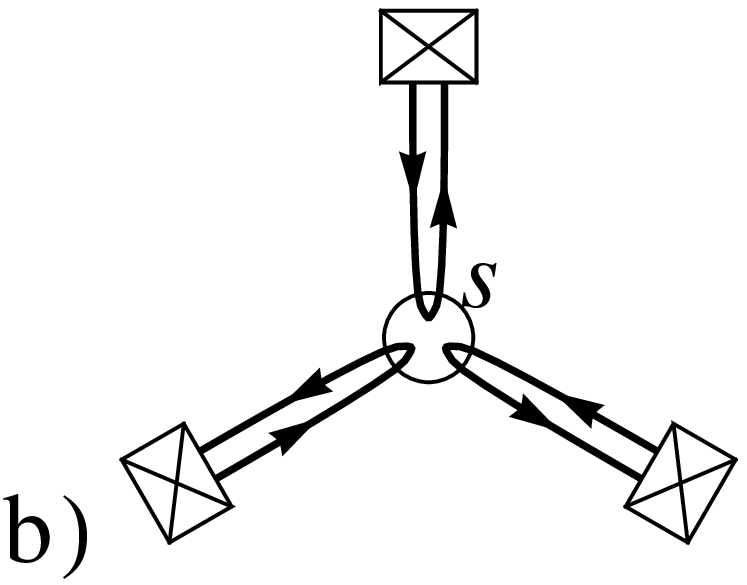}} 
	\caption{Two possible realizations of our model: \emph{(a)}
          Mach-Zehnder quantum Hall interferometer and \emph{(b)}
          junction of three quantum wires. Channels are represented by
          solid lines, point scatterers by white circles, reservoirs
          by boxes. Arrows indicate directions of
          motion.} 
\label{fig:realizations} 
\end{figure}

Let us consider a general model of the ballistic conductor, which can
be represented as a network of one-dimensional (1D) chiral channels
and point scatterers, as shown in Fig.~\ref{fig:realizations}. 
It is assumed that electrons propagate along the channels, denoted by
lower Greek index $\mu$,  
with constant velocity $v_\mu$ 
from source to drain reservoirs located at coordinates $x^\sr_{\mu}$
and $x^\dr_\mu$, respectively.  
In the physical world such channels are realized by quantum Hall (QH)
edge states or right/left-moving 1D states in quantum wires. At point
scatterers, denoted by Latin index $j$, instantaneous tunneling
between different channels occurs, which is described by the
scattering matrix $s^j$. Typical examples of scatterers are  
quantum point contacts (QPCs) or impurities in nanowires. Somewhat
less trivial type of scatterer  
is a multi-terminal junction that can be realized by a quantum dot under assumption that 
its Thouless energy is well above all typical energy
scales of the problem such as the temperature and the voltage.

Albeit quite simple, our quantum-wire network model covers a broad class of
mesoscopic ballistic devices, including QH interferometers and quantum
wire junctions (See Fig.~\ref{fig:realizations}). We note also
that the importance of
network models has been well appreciated in the context of the integer
QH effect, where the Chalker-Coddington network model~\cite{Chalker88}
serves as a highly useful starting point for numerical and analytical
investigation of the QH transition. 

To warm up, we briefly recall the construction of the bosonized Keldysh action
in the absence of tunneling, 
i.e. when all scattering matrices are trivial, $s^j=\xUnit$, and thus
different chiral channel are  
fully disconnected from each other.  We also require that each chiral
channel in the absence of tunneling is connected 
to one source and one drain reservoir (rather than forms a loop).  
Following the logic of the Keldysh formalism~\cite{Levchenko09, Kamenev11},
all field degrees of freedom are doubled: upper Greek indices
$\alpha=-/+$ denote fields $\psi^\alpha$ on the forward/backward
branch of Keldysh's time contour $\keldC$; integration along $\keldC$
is to be understood as $\int_\keldC \dd t' A(t')=\int\!\dd
t'\,(A^-(t')-A^+(t'))$.  

The fermionic action of the 1D fermions reads 
\begin{align*}
	\act_0[\psi,\bar \psi] = \sum_\mu \int_\keldC \dd t\,\dd x\,
        \bar \psi_\mu(i\partial_t+iv_\mu\partial_x)\psi_\mu + \frac 12
        \sum_{\mu\nu} \int_\keldC \dd t\, \dd x\, \dd x'\,
        U_{\mu\nu}(x,x')\varrho_\mu(x) \varrho_\nu(x'). 
\end{align*}
Electron-electron interaction is taken into account by potential
$U_{\mu\nu}$, charge density in channel $\mu$ is
$\varrho_\mu(x)=\bar\psi_\mu(x+0) \psi_\mu(x)$ with Grassmannian
fields $\psi_\mu(x)$, $\bar\psi_\mu(x)$. Spatial integration extends
along the corresponding channels: $x^\sr_\mu<x<x^\dr_\mu$. 
To proceed we apply the method of functional
bosonization~\cite{Grishin04} and decouple interaction via
Hubbard-Stratonovich (HS) transformation, introducing the bosonic
field $\varphi_\mu$: 
\begin{align*}
	\act_0[\psi,\bar\psi,\varphi] = \sum_\mu \int_\keldC \dd
        t\,\dd x\, \bar \psi_\mu(i\partial_t+iv_\mu
        \partial_x-\varphi_\mu)\psi_\mu + \frac 12
        \sum_{\mu\nu}\int_\keldC\dd t\,\dd x\,\dd x'\, \varphi_\mu(x)
        U^{-1}_{\mu\nu}(x,x') \varphi_\nu(x'). 
\end{align*}

The fact that we consider 1D chiral channels enables us to decouple
$\bar\psi_\mu\psi_\mu$ and $\varphi_\mu$ by a gauge transformation
$\psi^\mp_\mu \to e^{i\Theta^\mp_\mu}\psi^\mp_\mu$ requiring  
\begin{equation}
(\partial_t+v_\mu\partial_x)\Theta^\mp_\mu=-\varphi^\mp_\mu.
\label{eqn:Gauge_cond}
\end{equation} 
While resolving this gauge condition, one should properly take the Keldysh
structure into account, which yields 
\begin{align} \label{eqn:phaseHSField}
	\begin{pmatrix} \Theta^-_\mu \\ \Theta^+_\mu \end{pmatrix}_\xi
        = -\int\!\dd \xi'\,\begin{pmatrix} 
	           	D^T_{0\mu} & D^<_{0\mu}\\
			D^>_{0\mu} & D^\aT_{0\mu}
	           \end{pmatrix}_{\xi-\xi'}
		\begin{pmatrix}
			\varphi^-_\mu\\
			-\varphi^+_\mu
		\end{pmatrix}_{\xi'}
\end{align}
with $\xi=(x,t)$. Here the blocks of the
particle-hole propagator $D_{0\mu}$ satisfy the relations
\begin{eqnarray}
(\partial_t+v_\mu\partial_x) D^{T/\tilde T}(\xi,\xi') &=& \pm \delta(\xi-\xi'), \nonumber \\
(\partial_t+v_\mu\partial_x) D^{\gtrless}(\xi,\xi') &=& 0
\end{eqnarray}
and therefore Eq.~(\ref{eqn:phaseHSField}) indeed solves the gauge
condition~(\ref{eqn:Gauge_cond}). 
In the frequency-momentum representation the bare retarded/advanced
particle-hole propagator  
$D_{0\mu}$ in channel $\mu$ is given by
\begin{align} \label{eqn:ehpPropagator1}
	D_{0\mu}^{r/a}(\omega,q) = \frac i{\omega_\pm-v_\mu q},\quad \omega_\pm=\omega\pm i0.
\end{align}
The Keldysh propagator $D_{0\mu}^{K}$ depends on the nonequilibrium state of the system. 
In what follows, we consider the zero temperature limit. Under this
assumption, the electron distribution functions,
$f_\lambda^<=f_\lambda$, $f^>_\lambda=\xUnit-f_\lambda$, of source
reservoirs $\lambda$ are completely characterized by the applied bias
$\mu_\lambda=eV_\lambda$, 
\begin{align} \label{eqn:FermiDistribution}
	f^\gtrless_\lambda(t,t')=f^\gtrless_\lambda(t-t') = e^{i\mu_\lambda (t-t')} f^\gtrless_0(t-t')
	\quad \text{with}\quad f^\gtrless_0(t) = \mp \frac i{2\pi} \frac1{t\mp i a},
\end{align}
where $f_0(t)$ is the real-time representation of  the Fermi
distribution function and $a$ is a short-time cutoff. 
The components of the bare particle-hole propagators are then 
\begin{align} \label{eqn:ehpPropagator2}
	D^\gtrless_{0\mu}(x,t)=\lvert v_\mu\rvert^{-1}
        n_B^\gtrless(t-x/v_\mu),\quad
        D^{\T/\aT}_{0\mu}(x,t)=\theta(\pm t)D^>_{0\mu}(x,t)+\theta(\mp
        t)D^<_{0\mu}(x,t) \,,
\end{align}
with equilibrium Bose function $n_B^\gtrless(t)=-i/2\pi(t\mp i
a)$. Further, the Keldysh particle-hole propagator  
is given by the equilibrium relation
\begin{align}
 D_{0\mu}^k(\omega,q) = (D_{0\mu}^{r}(\omega,q)- D_{0\mu}^{a}(\omega,q))\,{\rm sgn}(\omega).
\end{align}
 
The gauge transformation has a non-trivial Jacobian contributing to
the action. According to the Dzyaloshinskii-Larkin 
theorem~\cite{Dzyaloshinskii73}, this contribution is Gaussian, and the
new action reads 
\begin{align} \label{eqn:cleanAction}
	\act_0[\varphi]=\frac 12 \sum_{\mu\nu} \int_\keldC \dd
        \xi\,\dd \xi'\, \varphi_\mu(\xi)
        V^{-1}_{\mu\nu}(\xi,\xi')\varphi_\nu(\xi') - \sum_\lambda
        \int_\keldC\dd \xi\,\varphi_\lambda(\xi) \varrho_{0\lambda}(\xi) \,,
\end{align}
with excess charge $\varrho_{0\lambda}=\mu_\lambda/(2\pi\lvert v_\lambda\rvert)$, effective interaction
\begin{align*}
	V^{-1}_{\mu\nu}(\xi,\xi')=U^{-1}_{\mu\nu}(x,x')\delta(t-t') -\delta_{\mu\nu} \Pi_\mu(\xi-\xi'),
\end{align*}
polarization operator $\Pi^{\alpha\beta}_\mu(\xi)=-i G^{\alpha\beta}_{0\mu}(\xi)G^{\beta\alpha}_{0\mu}(-\xi)$,
and free Green's function
\begin{align} \label{eqn:freeGF}
	G^\gtrless_{0\mu}(\xi)=-\frac 1{2\pi \lvert
          v_\mu\rvert}\frac{e^{-ieV_\mu(t-x/v_\mu)}}{t\mp i a
          -x/v_\mu},\quad G^{\T/\aT}_{0\mu}(\xi)=\theta(\pm
        t)G^>_{0\mu}(\xi)+\theta(\mp t)G^<_{0\mu}(\xi). 
\end{align}
After the standard rotation in the Keldysh space
\cite{Levchenko09,Kamenev11} and the transformation into
frequency-momentum representation, one obtains  
the retarded/advanced components of the 1D polarization operator 
\begin{align} 
\label{eqn:genericPolOp}
	\Pi_\mu^{r/a}(\omega,q) = \frac 1{2\pi \lvert v_\mu\rvert} \frac{v_\mu q}{\omega_\pm-v_\mu q}.
\end{align}

With the action $\act_0[\varphi]$ being Gaussian, the respective average value $\left\langle \varphi_\mu(\xi)\right\rangle_0$ and the correlator of the fluctuations $\delta\varphi_\mu(\xi)\equiv \varphi_\mu(\xi)-\left\langle \varphi(\xi)\right\rangle_0$ are simply given by $\left\langle \varphi(\xi)\right\rangle_0=\sum_\nu \int_\keldC\dd \xi'\, V_{\mu\nu}(\xi,\xi') \varrho_{0\nu}(\xi')$ and $\left\langle \delta\varphi_\mu(\xi) \delta\varphi_\nu(\xi')\right\rangle_0=iV_{\mu\nu}(\xi,\xi')$.

From (\ref{eqn:phaseHSField}), or symbolically $\Theta_\mu=-D_{0\mu}\varphi_\mu$, one obtains for the correlator $iD_{\Theta,\mu\nu}(\xi,\xi')=\left\langle \delta\Theta_\mu(\xi) \delta\Theta_\nu(\xi')\right\rangle$ the relation 
\begin{align} \label{eqn:DVD}
	D_{\Theta,\mu\nu}=-D_{0\mu}V_{\mu\nu} D_{0\nu}.
\end{align}

After having reviewed the results for the action in the absence of
tunneling, we are ready to consider the case of a connected quantum
network. This will be the subject of Sec.~\ref{sect:tunAct}.

\subsection{Keldysh Action} 
\label{sect:tunAct}

With all preliminaries we are now in a position to formulate the Keldysh
action of the connected quantum 
network when at least one node is characterized by a non-trivial
scattering matrix $s^j \neq \xUnit$. 
For the case of a single compact scatterer such an action has been
constructed in Ref.~\cite{Nazarov08} 
with the use of nonequilibrium Green's function method. The result
bears connection with the solution of 
the problem of full counting statistics \cite{Levitov93,Klich}. In our
paper we generalize this approach 
to the situation with many scatterers. It turns out that the Keldysh
action in this case can be written 
in terms of a full time-dependent single-particle scattering matrix
(S-matrix) of the system in a given configuration of field
$\varphi^\alpha$, which we denote $S^\alpha=S[\varphi^\alpha](t,t')$, where
$\alpha$ is the Keldysh index. Let us emphasize that 
the $S$-matrix is non-local in time and takes different values on
the forward and backward branches of  
the Keldysh contour. Our result reads:
\begin{multline} 
\label{eqn:fullTunAction}
	\act[\varphi]= \frac 12 \sum_{\mu\nu}\int_\keldC \dd \xi\,\dd
        \xi'\,\varphi_\mu U^{-1}_{\mu\nu}
        \varphi_\nu-\sum_\mu\int\!\!\dd
        \xi\,\dd\xi'\,(\varphi^c_\mu,\varphi^q_\mu)_\xi 
				\begin{pmatrix}
					0 & \Pi^a_\mu(\xi-\xi')\\
					\Pi^r_\mu(\xi-\xi') & 0
				\end{pmatrix}
				\begin{pmatrix}
					\varphi^c_\mu\\
					\varphi^q_\mu
				\end{pmatrix}_{\xi'}\\
		-i \ln \Det\left[\xUnit-f+S[\varphi^+]^\dagger S[\varphi^-]f\right].
\end{multline}
The last term in Eq.~(\ref{eqn:fullTunAction}) 
is a functional determinant with respect to (real) time
and channel indices. It is understood that $f$ in the expression for
the corresponding operator has the structure
 $f_{\mu\nu}(t,t')=\delta_{\mu\nu} f_\mu(t-t')$, i.e., it is
diagonal in channel representation, with $f_\mu(t)$ being the Fourier
transform of the source distribution function connected to channel
$\mu$.  The  second term in (\ref{eqn:fullTunAction}) represent anomalous contributions (related
to the Schwinger anomaly) that have been already encountered before,
see Eq.~(\ref{eqn:cleanAction}). They are
most transparently written in the rotated Keldysh representation:
$\varphi^{c(q)}= (\varphi^-\pm\varphi^+)/2$,
$\Pi^{r/a}_\mu=(\Pi^T_\mu-\Pi^\aT_\mu\pm(\Pi^>_\mu-\Pi^<_\mu))/2$ and
integration is performed along the real time axis.

 A detailed derivation of the result (\ref{eqn:fullTunAction}),
which employs ideas of Ref.~\cite{Klich},
is presented in Appendix A. In view of the importance of this result,
we give also its alternative proof (Appendix B), which follows
closely the method of Ref.~\cite{Nazarov08}.

\subsubsection*{Construction of Scattering Matrix}

Let us now discuss how the S-matrix for the systems under
consideration is constructed. The elements of
$S^\alpha_{\nu\mu}(t,t')$ give the amplitude that a wave packet
incident from source $\mu$ at time $t'$ leaves  
the system at time $t$ through the drain $\nu$. They are sums 
$S^\alpha_{\nu\mu}(t,t')=\sum_\mu A^{(p)\alpha}_\mu(t,t')$
over all corresponding paths $p$ formed by elements of the network.

\begin{figure}[ht]
	\centering{\includegraphics[scale=0.3]{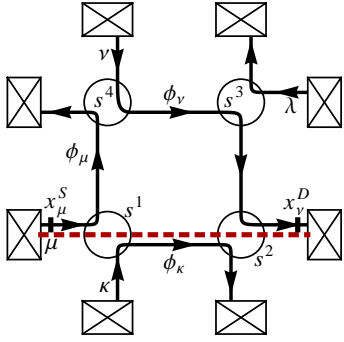}}
	\caption{Network built out of four channels $\mu,\kappa,\nu,\lambda$ and
          point scatterers $1,2,3,4$. Counting fields measure outgoing
          currents. Along interior parts of channels classical phases
          $\phi_{\mu/\kappa/\nu}$, e.g.\ due to magnetic field, are
          accumulated. An exemplary path $p$ is indicated by a dashed
          line.}\label{fig:examplePath} 
\end{figure}

Figure \ref{fig:examplePath} shows an exemplary path $p$ through a
network of channels and point scatterers.  
It consists of an alternating sequence of two types of processes:

\paragraph{Electron propagation in the potential $\varphi_\mu^\alpha$
  between $x^i$ and $x^j$,} leading to the accumulated phase
	\begin{equation}
	\label{eqn:kin_ph}
		\vartheta^\alpha_{\mu}(x^j,x^i;t)(t)\equiv -v^{-1}_\mu
                \int_{x^i}^{x^j}\!\dd
                x\,\varphi^\alpha_\mu(x,t-(x^j-x)/v_\mu). 
	\end{equation}
In addition to the Hubbard-Stratonovich field $\varphi^\alpha_\mu$,
there may be other time-depending phases $\phi^\alpha_\mu$ (e.g., induced by a  magnetic field)
contributing to Eq.~(\ref{eqn:kin_ph}). 
Note that $\vartheta^\alpha_\mu(x,x^i)$ satisfies the same
differential equation $(\partial_t+v_\mu\partial_
x)\vartheta^\alpha_\mu(x,x^i;t)=-\varphi^\alpha_\mu(x,t)$ as
$\Theta^\alpha_\mu(x,t)$, but has a simpler (``incomplete'') Keldysh
structure which involves only the retarded/advanced
components of the bare particle-hole propagator $D_{0\mu}$. We refer
to $\vartheta^\alpha_\mu(x,x^i)$ as a {\it kinematic} phase.  
To take a finite flight time of electron between $x^i$ and $x^j$ into
account, we introduce a ``delay operator''  
\begin{equation}
\Delta_\mu(x^j,x^i;t',t)=\delta(t'-t-(x^j-x^i)/v_\mu)
\end{equation}
Then the amplitude of this process reads
	\begin{equation}
	\label{eqn:M_matrix}
		\wire^\alpha_\mu(x^j,x^i;t,t')\equiv
                e^{i\vartheta^\alpha_\mu(x^j,x^i;t)}\Delta_\mu(x^j,x^i;t,t'). 
	\end{equation}
Indeed, consider the 1D version of the Schr\"odinger equation on a directed link $x^i \to  x^j$,
\begin{equation}
i\partial_t \psi_\mu = ( -i v_\mu \partial_x + \phi_\mu^\alpha) \psi_\mu .
\end{equation}
Using the
definition of the kinematic phase~(\ref{eqn:kin_ph}), this equation can be solved independently 
one each branch of the Keldysh contour yielding the relation
\begin{equation}
 \psi_\mu(x_j, t) = e^{i\vartheta_\mu^\alpha(x^j,x^i;\,t)} \psi_\mu(x_i, t- (x^j - x^i)/v_\mu),
\end{equation}
which implies that the scattering matrix is given Eq.~(\ref{eqn:M_matrix}).

\paragraph{Scattering/tunneling at point scatterer $j$:} The amplitude
of instantaneous scattering from channel $\mu$ to $\nu$ is 
\begin{align*}
	s^{j}_{\nu\mu}(t',t)=s^{j}_{\nu\mu} \delta(t'-t).
\end{align*}

\paragraph{Passing the charge detector at drain $\mu$, which is
  described by the counting field $\chi_\mu$.} As a special variant of
\emph{a.}, our formalism includes the theory of full counting
statistics. A counting field residing in the drain lead $\mu$ measures
the current flowing in that drain. The corresponding amplitude is 
\begin{align*}
	\mathcal X^\alpha_\mu(t',t)=e^{-i\frac\alpha 2\chi_\mu} \delta(t'-t).
\end{align*}
Then the action~(\ref{eqn:fullTunAction}) enables us to express the
cumulant generating function of the network  
as a functional integral over $\varphi$,
\begin{equation}
{\cal Z}(\vec\chi) = \int \prod_\mu{\cal D}\varphi^{\pm}_\mu(x,t)\,
\exp\left\{ i{\cal A}(\,\varphi,\vec\chi) \right\},
\end{equation}
where vector $\vec\chi$ combines counting fields $\chi_\mu$ in all drains.

Finally, the amplitude $A^{(p)\alpha}_{\nu\mu}$ of a path $p$ is the
path-ordered (real) time convolution (``latest to the left'') of the
amplitudes of its
constituent processes. As an example, the amplitude
indicated by the dashed line in Fig.~\ref{fig:examplePath} reads 
\begin{multline*}
	A^{(p)\alpha}_{\nu\mu}(t,t') =\left[\mathcal
          X^\alpha_\nu\ \wire^\alpha_\nu(x^\dr_\nu,x^2_\nu)\ s^{2}_{\nu\kappa}\ 
\wire^\alpha_\kappa(x^2_\kappa,x^1_\kappa)\  s^{1}_{\kappa\mu}\ 
\wire^\alpha_\mu(x^1_\mu,x^\sr_\mu)\right](t,t')\\     
	= \delta(t-t'-\tau)\ e^{-i\frac\alpha 2
          \chi_\nu+i\phi_\kappa}\ s^{2}_{\nu\kappa}s^{1}_{\kappa\mu}\ 
\exp\left\{i\vartheta^\alpha_\nu(x_\nu^\dr,x^2_\nu;t)
+i\vartheta^\alpha_\kappa(x^2_\kappa,x^1_\kappa;t-\tau_3)
+i\vartheta^\alpha_\mu(x^1_\mu,x_\mu^\sr;t-\tau_3-\tau_2)\right\} \,,   
\end{multline*}
where $\tau_{1/2/3}$ denote the flight times of the subpaths $x_\mu^\sr\to
x^1_\mu$, $x^1_\kappa\to x_\kappa^2$, $x^2_\nu\to x_\nu^\dr$, and
$\tau=\tau_1+\tau_2+\tau_3$ is the total flight time.

\subsection{Weak Tunneling Expansion} 
\label{sect:weakTun}

Due to the complex temporal behavior of the scattering matrix
analytical evaluation of the functional determinant  
(\ref{eqn:fullTunAction}) is not feasible in general. 
An approximate treatment is possible if a weak tunneling at the point
scatterers is assumed
(i.e. the scattering matrix $s^j_{\nu\mu}$ close to
$\delta_{\nu\mu}$), and the ultimate goal of this section is the
expansion of the action in the tunneling strength.  
Since in the absence of tunneling the network is described by the
Gaussian action~(\ref{eqn:cleanAction}), 
one can introduce the tunneling action $\act_t[\varphi]$, so that
$\act[\varphi]=\act_0[\varphi]+\act_t[\varphi]$, 
where the expansion of $\act_t[\varphi]$ starts from second-order
terms with respect to the tunneling amplitudes 
at the point scatterers. In Appendix \ref{app:regularization} we show
that an exact representation of 
$\act_t[\varphi]$ is given in terms of a modified (``regularized'') functional determinant
\begin{align} \label{eqn:regTunAct}
	\act_t[\varphi]=-i \ln \Det\left[\xUnit-f+\regSM^{+\dagger}\regSM^-f\right].
\end{align}
The new, ``regularized'' scattering matrix $\regSM$ here is
constructed similarly to $S$. Each of its elements
$\regSM^\alpha_{\nu\mu}(t,t') = \sum_p \tilde
A^{(p)\alpha}_{\nu\mu}(t,t')$ is a sum over the same paths $p$ which
contribute to $S^\alpha_{\nu\mu}(t,t')$ and connect the source $\mu$ with
the drain $\nu$. Full and regularized amplitudes, $A^{(p)}_{\nu\mu}$ and
$\tilde A^{(p)}_{\nu\mu}$ respectively, differ in the partial
amplitudes assigned to the elementary processes \emph{a.} and
\emph{b.} which constitute a path $p$: 
\setcounter{paragraph}{0}

\paragraph{Propagation between $x^i$ and $x^j$.} Only the time delay is taken
into account:
\begin{align*}
	\tilde \wire^\alpha_\mu(x^j,x^i) =\Delta_\mu(x^j,x^i),
\end{align*}
while phase accumulation is shifted to
\paragraph{Tunneling at point scatterers $j$.} The off-diagonal
tunneling amplitudes become ``dressed'' by tunneling phases
$\Phi^\alpha_{\nu\mu}(x^j,t)\equiv\Theta^\alpha_\mu(x^j_\mu,t) -
\Theta^\alpha_\nu(x^j_\nu,t)$: 
\begin{align*}
	\tilde s^{j\alpha}_{\nu\mu}(t,t')=e^{i\Phi^\alpha_{\nu\mu}(x^j, t)} s^j_{\nu\mu} \delta(t-t').
\end{align*}
The phases $\Theta_\mu=-D_{0\mu} \varphi_\mu$ are defined as in
Sect.~\ref{sect:model} and can be modified by additional
time-independent phase contributions due to e.g.\ magnetic or counting
fields as follows. If the additional phase accumulated by an electron which
propagates along a channel $\mu$ from a position $x$ to the drain lead $\mu$
is denoted as $\phi^\alpha_{\mu.\Out}(x)$,  then the phase
$\Theta^\alpha_\mu$ is modified according to 
\begin{align*}
	\Theta^\alpha_\mu(x,t)\mapsto \Theta^\alpha_\mu(x,t)-\phi^\alpha_{\mu.\Out}(x).
\end{align*}
In our previous example, Fig.~\ref{fig:examplePath}, the regularized scattering amplitudes read
\begin{align*}
	\tilde s_{\kappa\mu}^{1\alpha}(t,t') &=
        e^{i\Phi^\alpha_{\kappa\mu}(x^1,t)-i\left(\phi_\mu-\phi_\kappa\right)+i\frac\alpha2\left(\chi_\mu-\chi_\kappa\right)}
        s^2_{\kappa\mu}\delta(t-t'),\\ 
	\tilde s_{\nu\kappa}^{1\alpha}(t,t') &=
        e^{i\Phi^\alpha_{\nu\kappa}(x^2,t)+i\frac\alpha2\left(\chi_\kappa-\chi_\nu\right)}
        s^2_{\nu\kappa}\delta(t-t'). 
\end{align*}
The regularized scattering matrix becomes trivial in the ``clean'' limit, $\regSM=\xUnit$,
since all effects of interaction are now contained in the phases of the off-diagonal elements of the
regularized scattering matrices $\tilde s_{\nu\mu}^j$ of connectors. Thus
Eq.~(\ref{eqn:regTunAct}) can be expanded in (even) powers of the tunneling amplitudes:
\begin{align*}
	\act_t[\varphi]=i \sum_{n=1}^\infty \frac 1 n
        \Tr\left[\left(\xUnit- \redS^{+\dagger} \redS^-
          \right)f\right]^n. 
\end{align*}
We are now going to elaborate on the second-order terms in this series.

\subsubsection*{Second Order Expansion}

We introduce a notation $\mathcal A\equiv \tilde S^{+\dagger} \tilde
S^-$.  Up to third
order corrections in the tunneling amplitudes [that we denote
as $\mathcal O(\tun^3)$] the tunneling action is 
\begin{align} 
	\act_t[\varphi]=&i \Tr\left[(\xUnit-\mathcal A)f+\frac 12
          (\xUnit-\mathcal A)f(\xUnit-\mathcal A) f\right]+
        O(\tun^3) \nonumber\\
		=&i \Tr\left[(\xUnit-\mathcal A)_{\mu\mu}f_\mu+\frac
          12 \sum_{\mu\neq\nu}\mathcal A_{\mu\nu}f_\nu\mathcal
          A_{\nu\mu} f_\mu\right]+ O(\tun^3). \label{eqn:TunAct2Exp}
\end{align}
In the last expression, the trace is only taken with respect to
time. To reduce the tunneling action to this form,
we used $(\xUnit-\mathcal A)_{\mu\mu}=O(\tun^2)$ and $\mathcal
A_{\nu\mu}= O(\tun)$. 

It can be shown (see Appendix \ref{app:secondOrder} for
a detailed derivation) that $\act_t$
acquires  contributions from paths
which start in a certain source reservoir, evolve forward and
backward in time, undergoing in total exactly 2 tunneling events, and
eventually returning to the original source. Such paths involve
exactly 2 different channels, $\mu$ and $\nu$. Thus we can classify all
paths according to the pair $(\mu,\nu)$ of channels $\mu\neq\nu$ and
the pair $(i,j)$ of scatterers (possibly equal) at which the tunneling
takes place: $\mu\to\nu$ at $i$ and $\nu\to\mu$ at $j$. Of course, the
class $(ij;\mu\nu)$ coincides with the class $(ji;\nu\mu)$. 
The second order expansion of the tunneling action then is a sum over these classes:
\begin{align} \label{eqn:TunActPolOp}
	\act_t[\varphi]=-i\sum_{(ij;\mu\nu)}\int\!\!\dd t_1\dd
        t_2\, \begin{pmatrix}e^{-i\Phi^-_{\mu\nu}(x^i,t_1)} &
          e^{-i\Phi^+_{\mu\nu}(x^i,t_1)}\end{pmatrix} \begin{pmatrix}\Pi_{ij;\mu\nu}^\T(t_{12})
          & -\Pi_{ij;\mu\nu}^<(t_{12})\\ -\Pi_{ij;\mu\nu}^>(t_{12}) &
          \Pi_{ij;\mu\nu}^\aT(t_{12})\end{pmatrix} \begin{pmatrix}
          e^{i\Phi^-_{\mu\nu}(x^j,t_2)}\\ e^{i\Phi^+_{\mu\nu}(x^j,t_2)}\end{pmatrix}, 
\end{align}
$t_{12}\equiv t_1-t_2$, where the tunneling polarization operators are given by
\begin{align}
	\Pi^\gtrless_{ij;\mu\nu}(t) &=-s^i_{\nu\mu}\bar
        s^j_{\nu\mu}\ e^{i\Delta\phi^{ij}_{\mu\nu}} e^{\pm
          i(\chi_\mu-\chi_\nu)}
        f_\mu^\gtrless(t+\tau^j_{\mu.\In}-\tau^i_{\mu.\In})  
f_\nu^\lessgtr(\tau^i_{\nu.\In}-\tau^j_{\nu.\In}-t),\label{eqn:tunPolOp1}\\
	\Pi^{\T/\aT}_{ij;\mu\nu}(t) &=\left[\theta(\pm t)
          \Pi^>_{ij;\mu\nu}(t)+\theta(\mp t)
          \Pi^<_{ij;\mu\nu}(t)\right]_{\chi\equiv 0} \,,
\label{eqn:tunPolOp2}
\end{align}
where 
$\tau^k_{\lambda.\In}$ is the flight time from
the source to the scatterer $k$ along a channel $\lambda$,
$\tau^k_{\lambda.\In}=(x^k_\lambda-x_\lambda^\sr)/v_\lambda$. We have also taken
into account counting fields in the drain leads (which are not
contained in $\Pi^{\T/\aT}_{ij;\mu\nu}$) and classical phases, 
\begin{align*}
	\Delta\phi^{ij}_{\mu\nu}\equiv
        \phi_{\nu.\Out}(x^i)-\phi_{\mu.\Out}(x^i)-\phi_{\nu.\Out}(x^j)+\phi_{\mu.\Out}(x^j). 
\end{align*}
In the case $i=j$ we will also use the convention
\begin{align}\label{eqn:tunPolOp3}
 	\Pi^T_{ii;\mu\nu}(t)= \Pi^\aT_{ii;\mu\nu}(t)=\frac 12
        \left[\Pi^>_{ii;\mu\nu}(t)+\Pi^<_{ii;\mu\nu}(t)\right].
\end{align}
The comparison of this expression with the Eq.~(\ref{eqn:tunPolOp2}) shows that they
differ from each other by the singular term proportional to ${\rm sign}(t)h(t)\delta(t)=\pi \delta^2(t)$, where
we put $h(t) = f^>_0(t) - f^<_0(t)$. It gives some constant (albeit infinite) 
contribution to the tunneling action~(\ref{eqn:TunActPolOp})
and therefore both representation for $\Pi^{T/\tilde T}_{ii;\mu\nu}$ are equivalent.

\newcommand{\Phiphi}{{\mathcal D}}

\subsection{Real-Time Instanton Method}\label{sect:SPA}

On the level of the second order expansion, the action
$\act_t[\varphi]$ is expressed in terms 
of the tunneling phases $\Phi$,  which are linear functionals of $\varphi$:
\begin{equation}
\Phi_{\mu\nu}(\xi)=\sum_\lambda
\int_\keldC\dd\xi'\,\Phiphi_{\mu\nu;\lambda}(\xi,\xi')\varphi_\lambda(\xi'), 
\quad \Phiphi_{\mu\nu;\lambda} \equiv D_{0\mu}\delta_{\mu\lambda} - D_{0\nu}\delta_{\nu\lambda}.
\end{equation}
The action $\act_t[\varphi]$ is non-Gaussian in $\Phi$ and, in fact, is quite similar to
the Ambegaokar-Eckern-Sch\"on (AES) action~\cite{Eckern84}. 
The difference is that the kernel appearing in Eq.~(\ref{eqn:TunActPolOp}) is 
not only non-local in time (as in the case of AES) but in general is
non-local in space as well. In view of the non-Gaussian character of the action
an exact evaluation of physical quantities does not seem feasible in
general. For this reason, we will use a saddle-point approximation
that catches correctly the relevant interaction-induced physics,
including both the renormalization and the dephasing phenomena. 

To explain the idea of the method, let us consider some physical quantity 
$\mathcal
O[\varphi]=\mathcal O_0 e^{i\act_J[\varphi]}$, where
$\act_J[\varphi]=-\sum_\mu \int_\keldC \dd \xi\, J_\mu(\xi)
\varphi_\mu(\xi)$ is a linear functional of $\varphi$,
and the prefactor $\mathcal O_0$ is independent on $\varphi$. Important
examples, which are treated in the next sections, include the electronic
Green's function and the current. The quantum average value of ${\mathcal
  O}$ 
is given by the functional integral
\begin{align} \label{eqn:averageFunctInt}
		\left\langle \mathcal O[\varphi]\right\rangle =
                \int\DD \varphi\,
                e^{i\act_0[\varphi]+i\act_t[\varphi]+i\act_J[\varphi]}
                \mathcal O_0, 
\end{align}
which we estimate in the semiclassical
approximation~\cite{Levitov97}. In this case the path integral 
is contributed by the saddle-point  trajectory $\varphi_{\ast\ast}$ of
the full action $\act_0[\varphi]+\act_t[\varphi]+\act_J[\varphi]$ and
quantum fluctuations around it. Here free and tunneling actions
$\act_0[\varphi]$, $\act_t[\varphi]$ are given by  
Eqs. (\ref{eqn:cleanAction}) and (\ref{eqn:TunActPolOp}), respectively. 
In the limit of weak tunneling between chiral channels the
saddle-point trajectory (``instanton'' ) 
can be found approximately by requiring that it minimizes the Gaussian
contributions to the action $\act_0[\varphi]+\act_J[\varphi]$, which
gives  
\begin{align} \label{eqn:instantonPot}
	\varphi_{\ast\mu}(\xi)	=\sum_\nu \int_\keldC \dd\xi'\,
        V_{\mu\nu}(\xi,\xi') (\varrho_{0\nu}(\xi')+J_\nu(\xi')). 
\end{align}
As shown in Appendix \ref{app:SPA}, under such an approximation
Eq. (\ref{eqn:averageFunctInt}) simplifies to
\begin{align} \label{eqn:obsSPA}
	\left\langle \mathcal O[\varphi]\right\rangle = \left\langle
        e^{i\act_J[\varphi]}\right\rangle_0 e^{i\tilde
          \act_t[\varphi_\ast]}\mathcal O_0 \,, 
\end{align}
with
\begin{gather}
	 \left\langle
         e^{i\act_J[\varphi]}\right\rangle_0=\exp\left\lbrace
         i\left\langle \act_J[\varphi]\right\rangle_0-\frac 12
         \left[\left\langle\left(\act_J[\varphi]-\left\langle
           \act_J[\varphi]\right\rangle_0\right)^2\right\rangle_0\right]\right\rbrace
	, \label{eqn:SPA1stExp}\\
	\tilde\act_t[\varphi_\ast]=-i\!\!\!\!\sum_{\{\mu,\nu\},(i,j)}\int\!\!\dd
        t_1\dd t_2\, \begin{pmatrix}e^{-i\Phi^-_{\ast\mu\nu}(x^i,t_1)}
          &
          e^{-i\Phi^+_{\ast\mu\nu}(x^i,t_1)}\end{pmatrix} \begin{pmatrix}\tilde\Pi_{ij;\mu\nu}^\T(t_{12}) 
          &
          -\tilde\Pi_{ij;\mu\nu}^<(t_{12})\\ -\tilde\Pi_{ij;\mu\nu}^>(t_{12})
          &
          \tilde\Pi_{ij;\mu\nu}^\aT(t_{12})\end{pmatrix} \begin{pmatrix}
          e^{i\Phi^-_{\ast\mu\nu}(x^j,t_2)}\\ e^{i\Phi^+_{\ast\mu\nu}(x^j,t_2)}\end{pmatrix}, \label{eqn:SPA2ndExp} 
\end{gather}
where $t_{12}\equiv t_1-t_2$ and $\langle\ldots\rangle_0$ denotes
averaging with respect to $\act_0[\varphi]$. We have introduced the
instanton phases $\Phi_{\ast\mu\nu}=\Phiphi_{\mu\nu}\varphi_\ast$ and
the \emph{renormalized} tunneling polarization operators 
\begin{align} \label{eqn:renPolOp}
	\tilde \Pi^{\alpha\beta}_{ij;\mu\nu}(t_1-t_2) =e^{
          i\left(D^{\alpha\beta}_{\Phi\mu\nu}(x^i,x^j;t_1-t_2)-D^{\alpha\alpha}_{\Phi\mu\nu}(0,0)\right)}\  \Pi^{\alpha\beta}_{ij;\mu\nu}(t_1-t_2),   
\end{align}
obtained by dressing of the bare tunneling polarization operators by phase-phase correlators
\begin{align*}
	D_{\Phi\mu\nu}(\xi_1,\xi_2) =\sum_{\kappa\lambda}\int_C\dd
        \xi'\,\dd \xi''\,
        \Phiphi_{\mu\nu;\kappa}(\xi_1,\xi')V_{\kappa\lambda}(\xi',\xi'')
        \Phiphi_{\mu\nu;\lambda}(\xi'',\xi_2). 
\end{align*}
The meaning of Eq.~(\ref{eqn:renPolOp}) is that quantum fluctuations
of tunneling phases renormalize the temporal dependence of tunneling
polarization operators which lead to non-trivial (usually power-law)
energy-dependence of tunneling coefficients. 

In the next two sections we consider two important applications of our
general approach.

\newcommand{\R}{\lvert r_0\rvert^2}
\newcommand{\tip}{{\rm tip}}
\section{Tunneling Density of States of Luttinger Liquid with Single
  Impurity}
\label{sect:TDOS}

\begin{figure}[ht]
	\centering{\includegraphics[scale=0.3]{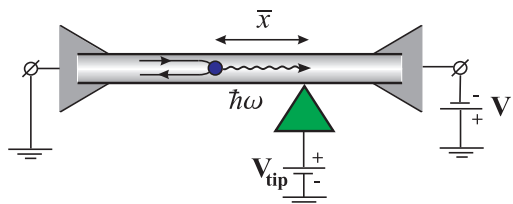}}
	\caption{Tunneling experiment with a voltage-biased quantum wire. }
\label{fig:LL_setup}
\end{figure}

In this section we show how our formalism can be applied to the evaluation of
the tunneling density of states (TDOS) of a nonequilibrium quantum wire
containing a single impurity \cite{Bagrets10}, as
depicted in Fig.~\ref{fig:LL_setup}. On the experimental side, the interest
to such a theoretical study is motivated by the recent development of  the
nonequilibrium tunneling spectroscopy of 1D nanostructures, including
carbon nanotubes~\cite{Chen09} and quantum Hall
edges~\cite{Altimiras10,Altimiras10a,Altimiras10b}.

\subsection{Model and Results}
\label{tdos-model-results}

 The wire is modeled as a network
of two channels: right- (left-) moving electrons, denoted with indices
$\mu=+(-)$.  Backscattering, i.e.\ tunneling between the two channels,
occurs at the impurity (at $x=0$) which is considered as a
point scatterer with scattering matrix $s$. Non-equilibrium conditions
are established by biasing the right reservoir with respect to the
left one: $eV_+=eV>0$, $eV_-=0$. 

The interplay of interaction, nonequilibrium, and impurity scattering
can be studied by placing a conducting tip  held at a voltage $V_\tip$
near a position $\bar x$ of the wire (we can assume without loss of
generality that $\bar x>0$) and measuring the
current $I_\tun$ between between tip and wire. Let us assume that the
tip can be described in terms of fermionic quasiparticles with density
of states $\nu_\tip(\epsilon)$ and distribution function
$f_\tip(\epsilon)$, as is the case, e.g., in the absence of
interaction within the tip. If the coupling $\lvert t\rvert^2$ between
the tip and the wire is weak, a simple perturbative expansion yields the tunneling current 
\begin{align*}
	I_\tun \propto \lvert t\rvert^2 \sum_{\mu=\pm} \int\!\!\dd
        \epsilon \left[\Gamma_\mu^>(\epsilon)
          f_\tip(\epsilon)-\Gamma^<_\mu(\epsilon)(1-f_\tip(\epsilon))\right]
        \nu_\tip(\epsilon). 
\end{align*}
The ``rates'' for tunneling into/out of the $\mu$-channel of the wire are defined as
\begin{align} \label{eqn:LLRatesDef}
	\Gamma^\gtrless_\mu(\epsilon)= \pm \frac i{2\pi} G_\mu^\gtrless(\bar x,\bar x,\epsilon).
\end{align}
Measurement of a differential tunneling conductance in the limit of a small temperature 
gives the access to the 
\emph{tunneling} density of states, since 
\begin{equation}
\frac{\partial I_\tun}{\partial V} \bigg|_{V_{\rm tip}=\epsilon} 
\propto \, \Gamma^>_\mu(\epsilon)+\Gamma^<_\mu(\epsilon) = \nu_\mu(\epsilon) \equiv
-\frac{1}{\pi} {\rm Im} \, G_\mu^r(\bar x,\bar x,\epsilon) \,.
\end{equation}
In the absence of interaction, the rates would simplify to
$\Gamma_\mu^\gtrless(\epsilon)=f^\gtrless_\mu(\epsilon)\nu_\mu(\epsilon)$
with the distribution functions $f_\mu^<(\epsilon)=f_\mu(\epsilon)$,
$f^>_\mu(\epsilon)\equiv 1-f_\mu(\epsilon)$ and the density of states
$\nu_\mu(\epsilon)$ in the channel $\mu$. 
Evaluation of tunneling rates and of the TDOS of the interacting
problem is the goal of this section.
 
We consider a spinless LL model with a point-like repulsive interaction,
$U_{\mu\nu}(x,x')= V_0 \delta(x-x')$,  that
does not discriminate between different channels.
The interaction strength in the LL model is conventionally
characterized by the constant $K=(1+V_0/\pi\vF)^{-1/2}$. The free electron
spectrum is linearized around the Fermi points, which requires the
introduction of a high-energy cutoff $\Lambda\sim E_F$ (which is of
the order of the bandwidth). In the absence of
backscattering, the tunneling rates exhibit the well-known zero bias
anomaly, i.e. a power-law suppression near the Fermi edges, 
\begin{equation}
	\Gamma_\mu^\gtrless(\epsilon) =\frac{\nu_0}\pi
        \Gamma(1+2\gamma)^{-1}\times \theta(\pm(\epsilon-eV_\mu))
        \left\lvert\frac{\epsilon-eV_\mu}\Lambda\right\rvert^{2\gamma}\quad
\end{equation}
  where $\nu_0=(2\pi\vF)^{-1}$ is the non-interacting density of
  states, and the exponent $\gamma$ is given by
\begin{equation}
 \gamma=\frac{(1-K)^2}{4K} \,.
\end{equation}

As is shown in the remainder of this section the tunneling rates
change considerably upon including the impurity. For $eV>0$ the rates
are given by 
\begin{equation} \label{eqn:LL_tunRates}
	\Gamma^\gtrless_\mu(\epsilon) = \pm \frac {\nu_0}\pi
        \left(\frac{eV}\Lambda\right)^{2\gamma}
        \Gamma(-2\gamma)\ \Im\left[(\mp z_\mu)^{2\gamma} +\mathcal
          C_\mu R_\ast(eV) (\pm 1)^{2\gamma} \Psi(-2\gamma,1-2\gamma +
          2\,\delta_\mu, -1-z_\mu)\right] \,,
\end{equation}
where we have introduced the following notations:
\begin{equation}
\mathcal C_\pm=\frac{\Gamma(2K)}{\Gamma\left( 1/2 \pm K/2 \right)^2},
\quad \delta_+= (1-K)/2,\quad \delta_-=1/2-K,\quad
z_\mu=(\epsilon-eV_\mu +\frac i2 \tau_\varphi^{-1})/eV.
\end{equation}
We have also introduced the renormalized reflection coefficient 
\begin{align} \label{eqn:LL_renRefl}
	R_\ast(eV) = \frac\R{\Gamma(2K)}  \left\lvert \frac{\Lambda}{eV}\right\rvert^{2(1-K)}
\end{align}
and the nonequilibrium dephasing rate 
\begin{align}
\label{eqn:LL_dephRate} 
	\tau_\varphi^{-1} \equiv R_\ast(eV) \frac{2 \sin^2\pi \delta_+}{\pi} \lvert eV\rvert.
\end{align}
\begin{figure}[ht]
	\centering{\includegraphics[scale=0.22]{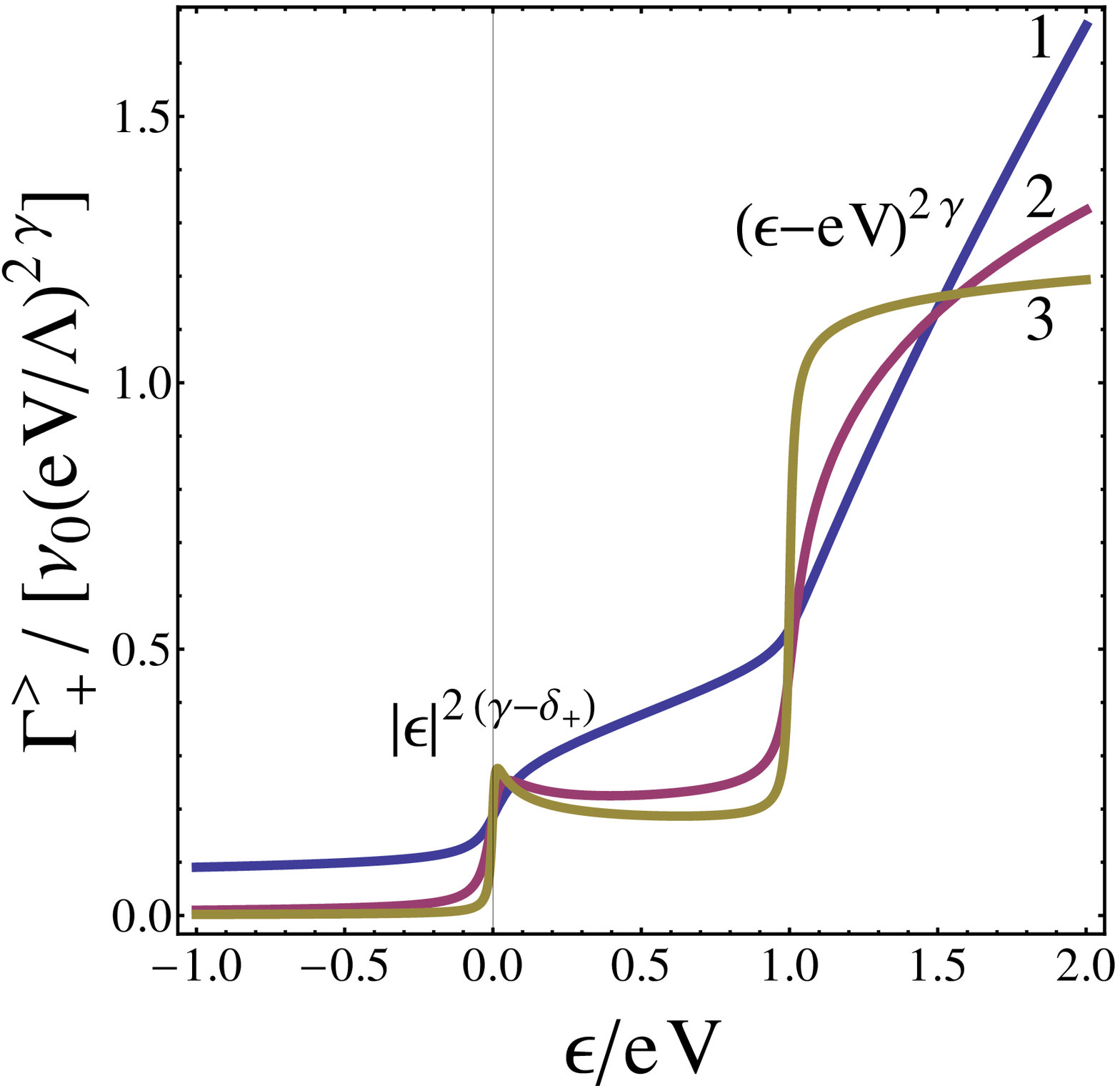}~\includegraphics[scale=0.22]{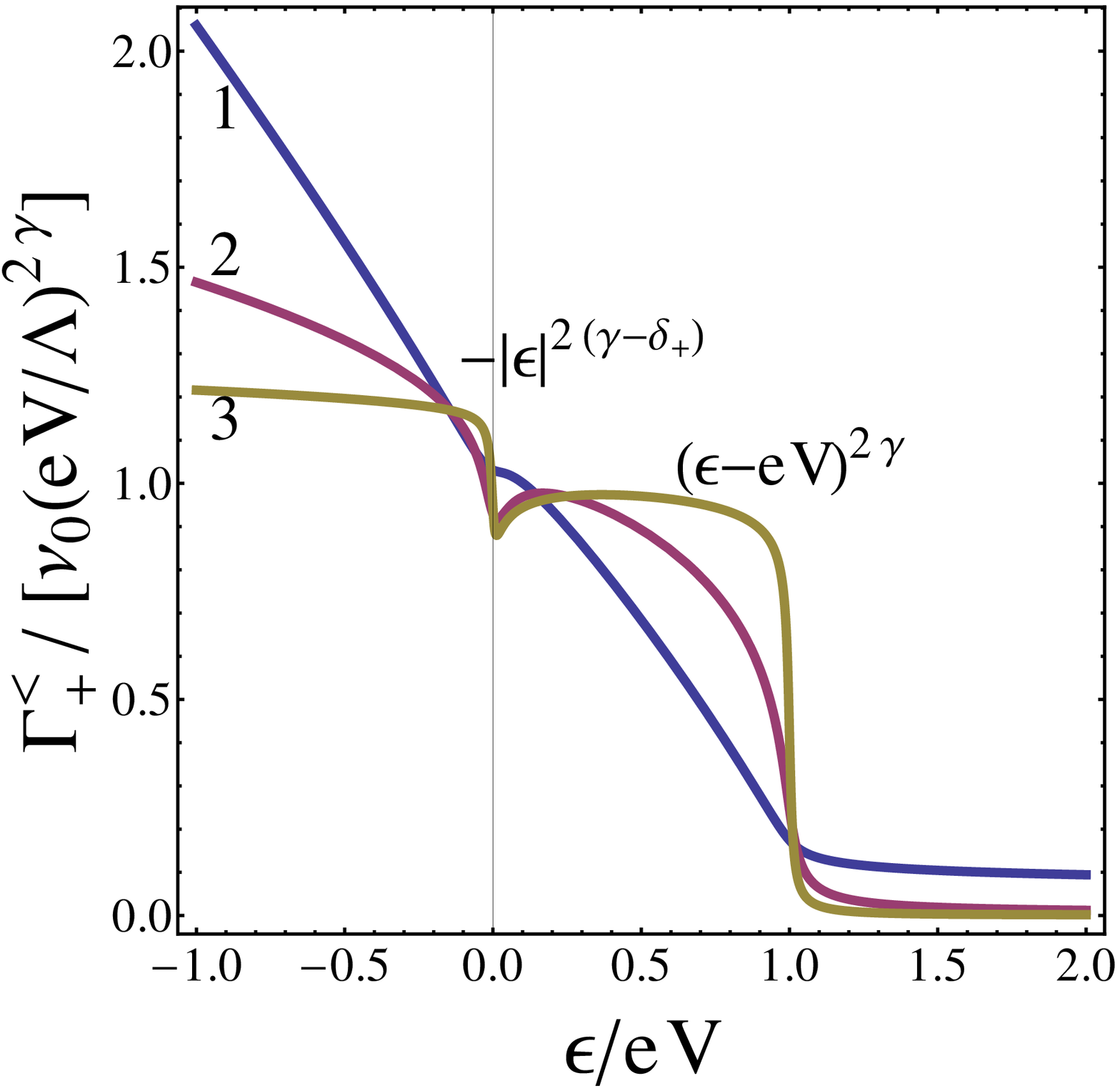}~\includegraphics[scale=0.22]{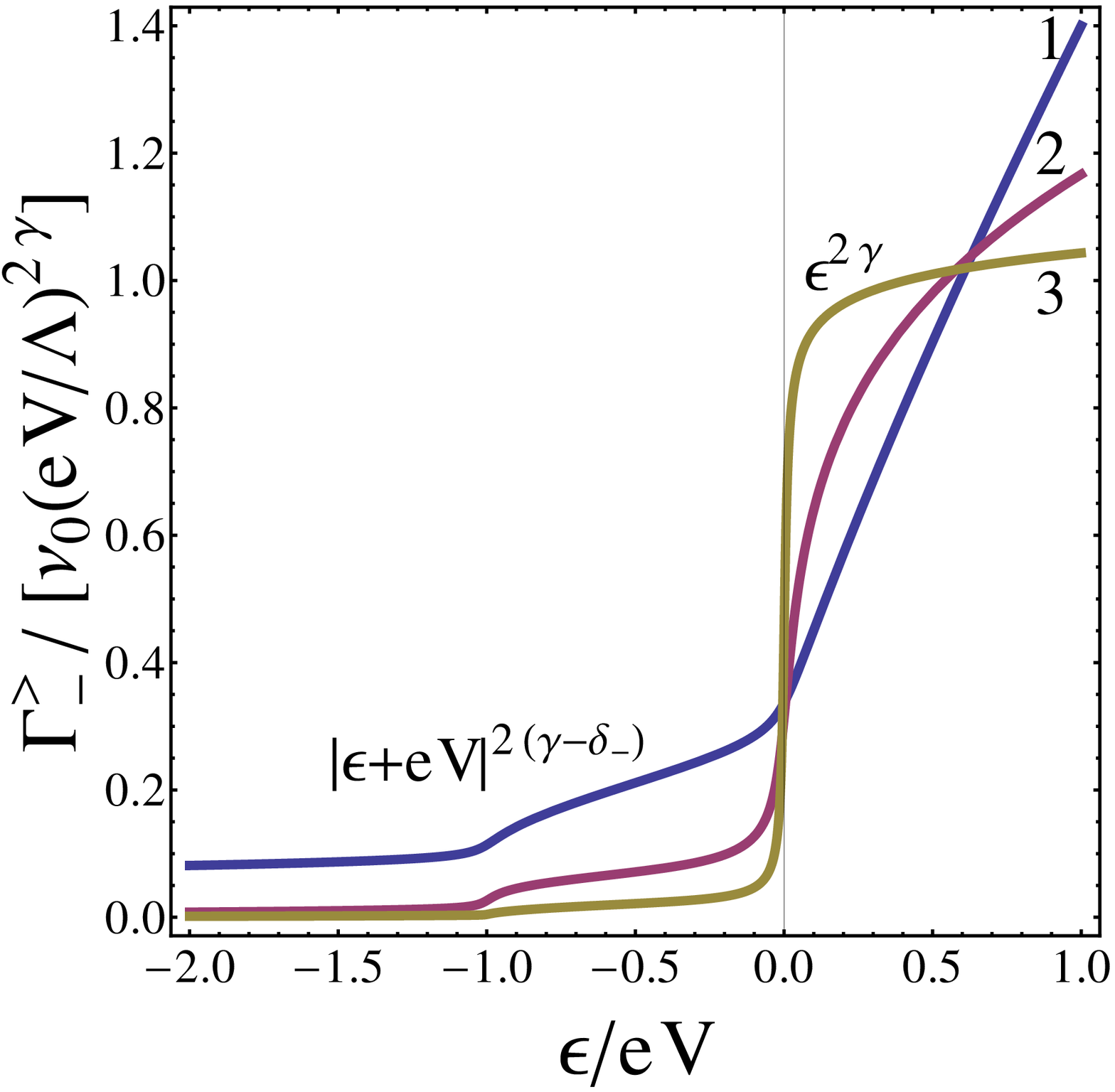}~ 
          \includegraphics[scale=0.22]{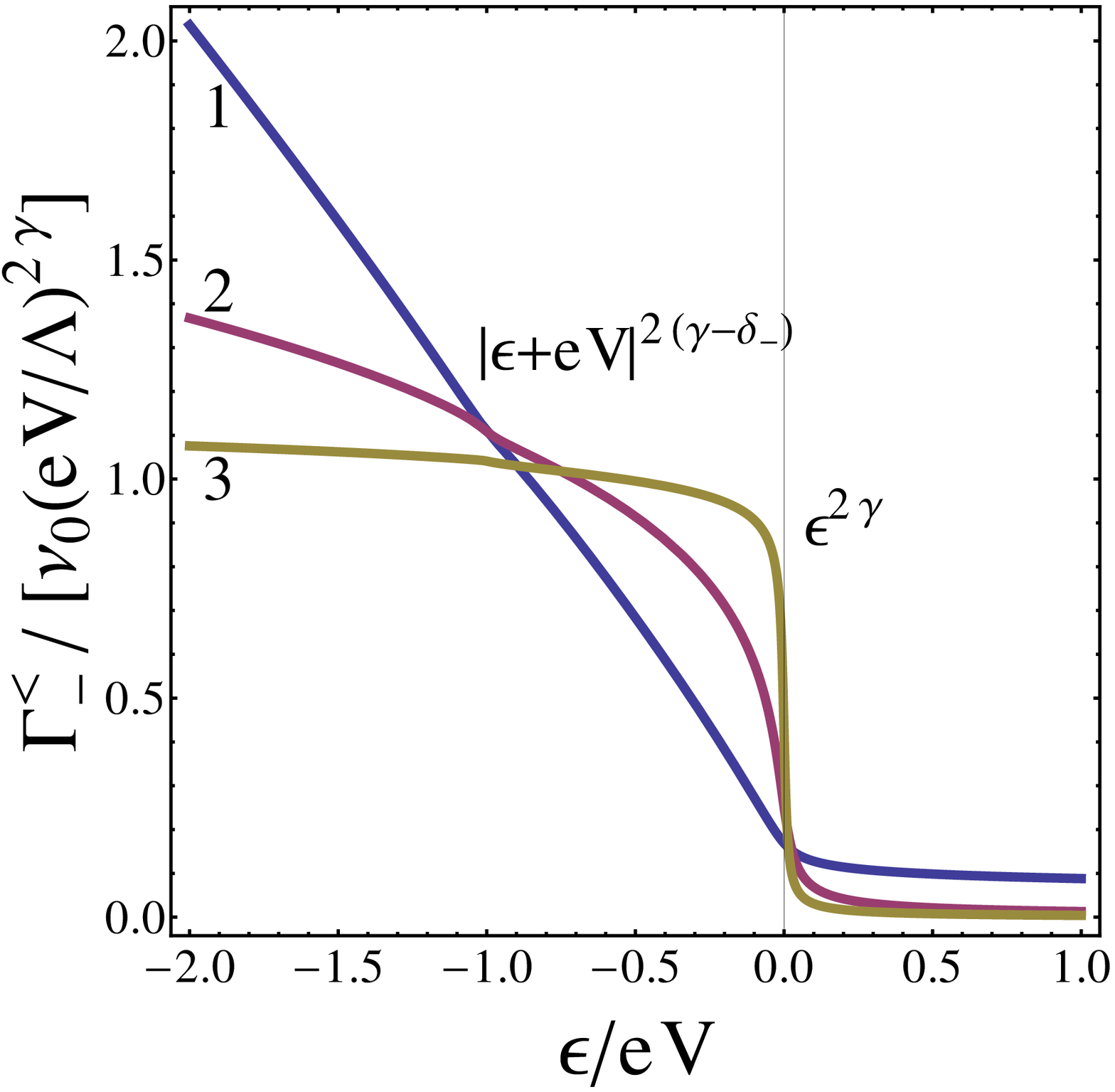}} 
	\caption{Tunneling rates for right- and left-movers with
          different interaction strengths: $K=0.3$ (1), $K=0.5$ (2),
          and $K=0.75$ (3). Fermi edge for right-(left-) movers is at
          $eV_+=eV$ ($eV_-=0$).}  
\label{fig:LL_tunRates}
\end{figure}
The energy dependence of rates $\Gamma^{\gtrless}_\pm(\epsilon)$ is
shown in Fig.~\ref{fig:LL_tunRates}.  
The main feature of these plots is that the tunneling rates have split
power-law singularities which  
are characterized by different exponents and are smeared by the
nonequilibrium dephasing rate $1/\tau_\varphi$.  
The main edges are located at the corresponding chemical potentials,
i.e., at $\epsilon=eV_\pm$  
in the case of right-/left-moving states, respectively, and are
characterized by the exponent equal (in the considered weak-back-scattering regime)
to its equilibrium value $\gamma$. 
The formation of the second (side) edge due to scattering off the
impurity occurs at $\epsilon = e(V_\pm - V)$. 
If the interaction is repulsive ($K<1$) then the corresponding exponent $2(\gamma-\delta_-)$
for left-moving electrons is always positive, 
hence the correction at the side edge $\epsilon=0$ is smooth. For
right-movers in the case of not too strong  
interaction, $K>1/3$, the nonequilibrium exponent
$2(\gamma-\delta_+)$ is negative, yielding a resonance  
in tunneling at the side edge $\epsilon=0$.

The presence of side edges in the tunneling rates can be understood in the following way.
Inelastic electron backscattering at the impurity at point $x=0$
induces the emission of {\it real}  
nonequilibrium plasmons with typical frequencies $\hbar \omega \le
eV$, which in the non-dissipative  
LL can propagate to the distant point of tunneling $\bar x$.  As the
result, inelastic tunneling with  
absorption or stimulated emission of these real plasmons become possible.    
For example, an electron tunneling into the  right/left moving state
of the LL with the energy $\epsilon<eV_\pm$ 
can accommodate itself above the corresponding Fermi energy ($eV_\pm$) by picking up the 
quantum $\hbar\omega>|\epsilon|$ from the nonequilibrium plasmon bath. 
Since the energy of out-of-equilibrium plasmons
is limited by the applied voltage, one has a threshold: $\epsilon>e(V_\pm-V)$, which is
developed into the power-law singularity typical for the LL. 
The singularity at the side edge of the tunneling rate-out describes
the inverse processes:
the inelastic tunneling from the LL accompanied by the stimulated emission
of nonequilibrium plasmons with typical energy $\hbar\omega \simeq
eV$. Such side edge is pronounced in the case of right-moving states
only,    
$\Gamma^<_+(\epsilon)$, and is not seen for the left-moving states,
$\Gamma^<_-(\epsilon)$, since 
the associated exponent $2(\gamma-\delta_-)$ is always positive in the latter case. 
 
Having announced the main results, we now turn to details of their derivation.

\subsection{Calculations}

\subsubsection{Action}

Since interaction does not discriminate between channels $\mu,
\nu=\pm$, it can be decoupled by a Hubbard-Stratonovich transformation
introducing a single field $\varphi$,
i.e.\ $\varphi_+=\varphi_-=\varphi$. 
For weak backscattering, i.e.\ weak tunneling between right- and
left-moving states at the impurity, the action
$\act[\varphi]=\act_0[\varphi]+\act_t[\varphi]$ is obtained according
to Sect.~\ref{sect:weakTun}. The free action (\ref{eqn:cleanAction})
is 
\begin{align*}
	\act_0[\varphi]= \frac 12 \int_\keldC \dd
        \xi\,\dd\xi'\,\varphi(\xi) V^{-1}(\xi-\xi') \varphi(\xi')  - \sum_\mu
        \int_\keldC\dd \xi\,\varphi(\xi) \left(\varrho_{0+}+\varrho_{0-}\right)
\end{align*}
with $\xi=(x,t)$, non-local effective
interaction $V^{-1}(\xi-\xi')=V_0^{-1} \delta(\xi-\xi')-\Pi(\xi-\xi')$
and the total polarization operator
$\Pi(\xi)=\Pi_+(\xi)+\Pi_-(\xi)$. Using (\ref{eqn:genericPolOp}) one
obtains for the retarded/advanced components of effective interaction 
\begin{align} \label{eqn:LL_effInt}
	V^{r/a}(\omega,p)=V_0 \frac{\omega^2-\vF^2p^2}{\omega_\pm^2-u^2 p^2}
\end{align}
with a plasmon velocity $u=\vF/K$. (For a repulsive interaction $K<1$,
so that $u>\vF$.)

Since we are dealing with a single scatterer, the tunneling (or,
equivalently, backscattering)
action $\act_t$, as given by (\ref{eqn:TunActPolOp}), consists of one
term [corresponding to class $(11;+-)$]: 
\begin{align} \label{eqn:LLTunAct}
	\act_t[\varphi]=-i\int\!\!\dd t_1\dd
        t_2\, \begin{pmatrix}e^{-i\Phi^-(t_1)} &
          e^{-i\Phi^+(t_1)}\end{pmatrix} \begin{pmatrix}\Pi_{+-}^\T &
          -\Pi_{+-}^<\\ -\Pi_{+-}^> &
          \Pi_{+-}^\aT\end{pmatrix}_{t_1-t_2} \begin{pmatrix}
            e^{i\Phi^-(t_2)}\\ e^{i\Phi^+(t_2)}\end{pmatrix}, 
\end{align}
where $\Phi^\mp(t)=\Theta^\mp_-(0,t)-\Theta^\mp_+(0,t)$ is the
tunneling phase evaluated at the impurity, $x=0$. The phases are
related to the Hubbard-Stratonovich field $\varphi$ according to
(\ref{eqn:phaseHSField}). The tunneling polarization operator
$\Pi_{+-}=\Pi_{11;+-}$ is given by Eqs.~(\ref{eqn:tunPolOp1}),
(\ref{eqn:tunPolOp2}).  The components of the polarization operator read 
\begin{align}
	\Pi^\gtrless_{+-}(t) &=-\R f_+^\gtrless(t) f^\lessgtr_-(-t) =
        -\R e^{-i eV
          t}\left[f^\gtrless_0(t)\right]^2, \label{eqn:LL_tunPolOp1}\\ 
	\Pi^{\T/\aT}_{+-}(t) &= \frac 12\left[\Pi^>_{+-}(t)+
          \Pi^<_{+-}(t)\right]. \label{eqn:LL_tunPolOp2} 
\end{align}
where $\R=\lvert s^1_{+-}\rvert^2=\lvert s^1_{-+}\rvert^2$ is the bare
reflection coefficient.
We note that in Ref.~\cite{Bagrets10} different
notations were used: $\act_b$ instead of
$\act_0[\varphi]$, $\act_{\rm imp}$ instead of $\act_t[\varphi]$, and
$\vartheta$ instead of $\Theta$. We also note that 
Eq.~(\ref{eqn:LL_tunPolOp2})  is slightly different from
Eq.~(\ref{eqn:tunPolOp2}); however, this difference leads only to
an additional constant contribution to the action that is physically irrelevant.  

\subsubsection{Green's Functions in Instanton Approximation}  

In order to find the tunneling rates, we represent the electron Green's function at the
point of tunneling $\bar x>0$ as a path integral over the field
$\varphi$, 
\begin{align*}
	G^\gtrless_\mu(\bar x,\bar x;\bar t)=\int \DD \varphi\,
        e^{i\act[\varphi]} e^{i\Theta^\pm_\mu(\bar x\bar
          t)}G^\gtrless_\mu(\bar x,\bar x;\bar t;[\varphi])
        e^{-i\Theta^\mp_\mu(\bar x,0)}. 
\end{align*}
Here, $G_\mu(\bar x,\bar x;\bar t,[\varphi])$ denotes the Green's
function for a given configuration of $\varphi$. It satisfies the
Dyson equation with the spatially local self-energy 
\begin{align*}
	\Sigma_\mu[\varphi](x,x';t,t')=i\delta(x)\delta(x') (\R \vF/2)
        e^{i\mu \Phi(t)} g_{-\mu}(t-t') e^{-i\mu \Phi(t')} \,, 
\end{align*}
where $g_\mu$ are the quasiclassical Green's functions of the source
reservoirs. 
Solving the Dyson equation to the first order in $\R$, we get
\begin{equation} 
\label{eqn:LLGFBornCorrection}
	G_\mu^{\alpha\beta}(\bar x,\bar x;\bar t) = \mathcal
        G^{\alpha\beta}_{0\mu} + \mathcal G^{\alpha\beta}_{1\mu} \,,
\end{equation}
where 
\begin{align}
	\mathcal G^{\alpha\beta}_{0\mu} =& \left\langle
        e^{i\Theta^\alpha_\mu(\bar x,\bar t)} G^{\alpha\beta}_{0\mu}
        (\bar x,\bar x;\bar t) e^{-i\Theta^\beta_\mu(\bar
          x,0)}\right\rangle,  \nonumber \\ 
	\mathcal G^{\alpha\beta}_{1\mu} =& i\frac{\R \vF}2
        \sum_{\gamma\delta=\mp} \gamma\delta \int \!\!\dd t_1\,\dd
        t_2\,\\ & \left\langle e^{i\Theta^\alpha_\mu(\bar x,\bar t)}
        G_{0\mu}^{\alpha\gamma}(\bar x,0;\bar t-t_1) e^{i\mu
          \Phi^\gamma(t_1)} g^{\gamma\delta}_{-\mu}(t_1-t_2) e^{-i\mu
          \Phi^\delta(t_2)} G_{0\mu}^{\delta\beta} (0,\bar x;t_2-\bar
        t) e^{-i\Theta^\beta_\mu(\bar x,0)}\right\rangle.
	\label{eqn:LLGFBornCorrection-2}
\end{align}
Here $G_{0\mu}^{\alpha\beta}$ are the Green's functions of free
electrons; in particular,
\begin{equation}
G^\gtrless_{0\mu}(x,t)=\pm f^\gtrless(t-\mu x/\vF)/i\vF.
\end{equation}

All averages $\langle\ldots\rangle$ in Eq.~(\ref{eqn:LLGFBornCorrection-2}) 
are taken with respect to the action
$\act_0[\varphi]+\act_t[\varphi]$. They are of the form
(\ref{eqn:averageFunctInt}) and can be evaluated with the real-time
instanton method described in Sect.~\ref{sect:SPA}. In this
approximation the first term in Eq.~(\ref{eqn:LLGFBornCorrection}) reads 
\begin{equation}
\label{eqn:CalG_0}
	\mathcal G_{0\mu}^{\alpha\beta} \approx e^{i\tilde
          \act_t[\varphi_{\ast}]} \times \left\langle
        e^{i\Theta^\pm_\mu(\bar x,\bar t)} G^\gtrless_{0\mu} (\bar
        x,\bar x;\bar t) e^{-i\Theta^\mp_\mu(\bar
          x,0)}\right\rangle_0. 
\end{equation}
The second factor here is the full Green's function of a \emph{clean} LL,
\begin{align} \label{eqn:LL_cleanGF}
	\tilde G_{0\mu}^\gtrless(\bar x,\bar x;\bar t)= e^{-\frac 12
          \left\langle\left[\Theta_\mu^\pm(\bar x,\bar
            t)-\Theta^\mp_\mu(\bar
            x,0)\right]^2\right\rangle_0}\ G_{0\mu}^\gtrless(\bar
        x,\bar x;\bar t)= \pm \frac{a^{2\gamma}}{2\pi i \vF}
        e^{-i\mu_\mu \bar t}\frac 1{(a\pm i \bar t)^{2\gamma+1}}. 
\end{align}
The first factor in Eq.~(\ref{eqn:CalG_0}) gives dephasing corrections
due to the interplay of tunneling and interaction. The instanton
action $\tilde \act_t[\varphi_{\ast}]$ is defined in
(\ref{eqn:SPA2ndExp}) and obtained by substituting the dressed
polarization operators into (\ref{eqn:LLTunAct}). The instanton phase
$\varphi_{\ast}$ is generated by the source
$i\act_{J}[\varphi]=i\Theta_\mu^\alpha(\bar x,\bar
t)-i\Theta^\beta_\mu(\bar x,0)$, 
\begin{align} \label{eqn:LL_smallInstanton}
	\Phi^\mp_{\ast}(t)=
	 \left\langle \Phi^\mp(t)\right\rangle_0-D_{\Phi\Theta\mu}^{\mp\alpha}(t-\bar t,-\bar x)+D^{\mp\beta}_{\Phi\Theta\mu}(t,-\bar x)\ \text{with}\ 
	D^{\gamma\delta}_{\Phi\Theta\mu}(t,x)=D^{\gamma\delta}_{\Theta,-\mu}(t,x)-D^{\gamma\delta}_{\Theta,+\mu}(t,x),
\end{align}
and depends on Keldysh indices $\alpha,\beta$, the direction $\mu$ of the
tunneling electron, as well as the time and the position of
tunneling $\bar t$, $\bar x$. Since the average value $\left\langle\Phi^\mp(t)\right\rangle_0$ does not depend on the Keldysh index $\mp$ and the time $t$ it will drop out when the instanton phase is substituted into (\ref{eqn:LLTunAct}). Therefore it will be omitted in what follows.

\subsubsection{Phase-phase Correlation Functions} 

To lay the groundwork for all further calculations we compute the
correlation functions of the phases $\Theta_\mu$ and
$\Phi=\Theta_--\Theta_+$. With the bare particle-hole propagator
(\ref{eqn:ehpPropagator1}), the effective interaction
(\ref{eqn:LL_effInt}), and the relation (\ref{eqn:DVD}), the retarded/advanced components of the correlators
$D_{\Theta,\mu\nu}$  
can be easily evaluated in the $(q,\omega)$ representation:
\begin{align}
	D^{r/a}_{\Theta,\mu\mu}(\omega,q) & = -\frac{2\pi}\omega
        \frac{\omega+\mu \vF q}\omega \mu \left[\frac{1+K}{4K}
          \frac1{q-\mu\omega_\pm /u}-\frac{1-K}{4K}\frac
          1{q+\mu\omega_\pm/u}-\frac 12 \frac
          1{q-\mu\omega_\pm/\vF}\right],\\ 
	D^{r/a}_{\Theta,-\mu\mu}(\omega,q) & = -\frac{2\pi}\omega
        \frac{1-K^2}{4K}\left[\frac 1{q-\omega_\pm/u} 
+\frac 1{q+\omega_\pm/u}\right].
\end{align}
Transforming the above relations into the mixed space-frequency representation, we obtain
\begin{equation}
D^{r/a}_{\Theta,\mu\nu}(\omega,x) = \mp \frac{2\pi i}\omega
\left\lbrace\theta(\pm \mu x)\left[c^+_{\mu\nu}\ e^{i\mu\omega
    x/u}-\delta_{\mu\nu} \ e^{i\mu\omega/\vF}\right]+\theta(\mp\mu
x)\ c^-_{\mu\nu}\ e^{-i\mu \omega x/u}\right\rbrace
\end{equation} 
with
\begin{equation}
c^\pm_{\mu\mu}=(1\pm K)^2/(4K), \quad c^\pm_{\mu,-\mu}=(1-K^2)/(4K).
\end{equation}
The Keldysh component of the phase correlator is given at zero
temperature by 
\begin{equation}
	D^{k}_{\Theta,\mu\nu}(\omega,x) = B(\omega)
        \left(D^r_{\Theta,\mu\nu}(\omega,x)-D^a_{\Theta,\mu\nu}(\omega,x)\right)
        =
        \sign\omega\  \left(D^r_{\Theta,\mu\nu}(\omega,x)-D^a_{\Theta,\mu\nu}(\omega,x)\right) \,.
\end{equation}
Performing the Keldysh rotation, we then arrive at
\begin{align}\label{eqn:D_mu_nu_wx} 
\begin{split}
	D^\gtrless_{\Theta,\mu\nu}(\omega,x)&=\frac12
        \left(D^k_{\Theta,\mu\nu}(\omega,x)\pm\left(D^r_{\Theta,\mu\nu}(\omega,x)-D^a_{\Theta,\mu\nu}(\omega,x)\right)\right)= 
        \pm
        \theta(\pm\omega)\left(D^r_{\Theta,\mu\nu}(\omega,x)-D^a_{\Theta,\mu\nu}(\omega,x)\right),\\ 
	D^{\T/\aT}_{\Theta,\mu\nu}(\omega,x)&=\frac12
        \left(D^k_{\Theta,\mu\nu}(\omega,x)\pm\left(D^r_{\Theta,\mu\nu}(\omega,x)+D^a_{\Theta,\mu\nu}(\omega,x)\right)\right)=
        \pm
        \theta(\pm\omega)D^r_{\Theta,\mu\nu}(\omega,x)\pm\theta(\mp
        \omega) D^a_{\Theta,\mu\nu}(\omega,x).
\end{split}
\end{align}
In the real-time representation $(x,t)$ these phase-phase correlation
functions can be decomposed  
into the plasmon (moving with velocity $u$) and particle-hole (having
velocity $v_F$) contributions 
\begin{align} 
\label{eqn:LL_DThetaLorg}
	iD_{\Theta,\mu\nu}^{\alpha\beta}(t,x)=
        c^+_{\mu\nu}\ \lorg^{\alpha\beta}_{\mu
          u}(t,x)-c^-_{\mu\nu}\ \lorg^{\alpha\beta}_{-\mu u}(t,x) -
        \delta_{\mu\nu}\ \lorg^{\alpha\beta}_{\mu\vF}(t,x) \,,
\end{align}
where for a given velocity $v$ the functions
$\lorg^{\alpha\beta}_v(t,x)$ read
\begin{align*}
	\lorg^\gtrless_v(t,x) = \ln\frac {\mp ia}{t\mp i a- x/v},\quad
        \lorg^{\T/\aT}_v(t,x) =\ln\frac {\mp i a\ \sign x/v}{t\mp i
          a\ \sign x/v -x/v} \,.
\end{align*}
This follows from the Eqs.~(\ref{eqn:D_mu_nu_wx}) after the
Fourier transformation from $\omega$ to $t$ 
with taking into account the high-energy cut-off. 
The functions $\lorg^{\alpha\beta}_v(t,x)$ satisfy
\begin{align*}
	\partial_t \lorg^\gtrless_v(t,x) = -(t\mp i a- x/v)^{-1},\quad
        \partial_t\lorg^{\T/\aT}_v(t,x) =-(t\mp i a\ \sign x/v
        -x/v)^{-1}. 
\end{align*}
It is worth
mentioning that the appearance of both 
plasmon and particle-hole ``light-cone'' singularities in the
phase-phase correlation function is a
special feature of the functional bosonization approach. 

For the $\Theta-\Phi$ phase-phase correlators we then obtain:
\begin{align}
\label{eqn:D_phi_theta}
	iD_{\Phi\Theta\mu}^{\alpha\beta}(t,x) &\equiv
        \left\langle\delta\Phi^\alpha(t,x)\delta\Theta^\beta_\mu(0,0)\right\rangle
        = i
        \left[D^{\alpha\beta}_{\Theta,-\mu}(t,x)-D^{\alpha\beta}_{\Theta,+\mu}(t,x)\right]\\ 
	&= -\mu\left[p\ \lorg^{\alpha\beta}_{\mu
            u}(t,x)-q\ \lorg^{\alpha\beta}_{-\mu
            u}(t,x)-\lorg_{\mu\vF}^{\alpha\beta}(t,x)\right]\,.
\end{align}
The correlation function of the tunneling phases $\Phi(t)=\Phi(t,0)$ at the
position of the impurity reads
\begin{equation}
	iD_\Phi^{\alpha\beta}(t) \equiv \left\langle \delta\Phi^\alpha(t) \delta \Phi^\beta(0)\right\rangle = i \lim_{x\to0}
        \left[D_{\Phi\Theta-}^{\alpha\beta}(t,x)-D_{\Phi\Theta+}^{\alpha\beta}(t,x)\right]
        = -2(1-K)\lorg_\Phi^{\alpha\beta}(t)\,, 
\end{equation}
where
\begin{equation}
\lorg^\gtrless_\Phi(t) = \ln\frac {\mp i a}{t\mp
          i a},\quad \lorg^{\T/\aT}_\Phi(t) = \frac 12 \left[\ln\frac
          {-ia}{t- i a} +\ln\frac {ia}{t+ia}\right] \,.
\end{equation}

\subsubsection{Instanton Action}
\label{sec:instanton-action}

The correlators obtained in the previous section reduce the ``dressed'' tunneling polarization 
operators~(\ref{eqn:renPolOp}) to the form
\begin{align} \label{eqn:LL_dressedPolOp}
	\tilde \Pi_{+-}^\gtrless(t)=-\R \frac 1{(2\pi a)^2} e^{-i eV t}\left(\frac a{a\pm i t}\right)^{2K},
\end{align}
or in the frequency representation
\begin{align*}
	\tilde \Pi^\gtrless_{+-}(\omega) =
        -\frac{R_{\ast}(eV)}{2\pi}\theta(\pm(\omega-eV)) \left\lvert
        \frac{\omega-eV}{eV}\right\rvert^{2K-1} \lvert eV\rvert \,,
\end{align*}
where we used the definition (\ref{eqn:LL_renRefl}) for the renormalized reflection coefficient $R_*$.
With the mixed phase-phase correlation function (\ref{eqn:D_phi_theta}) at hand we are also in
a position to write down the instanton trajectories (\ref{eqn:LL_smallInstanton}), 
\begin{align*}
	i\Phi_{\ast}^\mp(t) &=  \mu\left\lbrace p \ln\left[\frac{t\pm
            i a +\mu \bar x/u}{t\pm i a-\bar t+\mu\bar
            x/u}\right]-q\ln\left[\frac{a-i\beta(t-\mu\bar
            x/u)}{a-i\alpha(t-\bar t-\mu\bar
            x/u)}\right]-\ln\left[\frac{t\pm i a +\mu\bar x/\vF}{t\pm
            i a -\bar t +\mu \bar x/\vF}\right]\right\rbrace
        ,\quad \mu\bar x>0, \\ 
i\Phi_{\ast}^\mp(t) &= \mu\left\lbrace p \ln\left[\frac{a-i\beta(t
    +\mu \bar x/u)}{a-i\alpha(t-\bar t+\mu\bar
    x/u)}\right]\!-q\ln\left[\frac{t\pm i a-\mu\bar x/u}{t\pm i a-\bar
    t-\mu\bar x/u}\right]\!-\ln\left[\frac{a-i\beta(t+\mu\bar
    x/\vF)}{a-i\alpha(t -\bar t +\mu \bar x/\vF)}\right]\right\rbrace,
 \mu\bar x<0,
\end{align*}
where we have introduced $p=(1+K)/2$ and $q=(1-K)/2$. It is worth
emphasizing that these instantons represent non-classical solutions in the sense of the Keldysh nonequilibrium theory:
the phases $\Phi_*(t)$ are different on the upper and lower time contours, so that 
the quantum part is non-zero, $\Phi_*^q(t)\ne 0$. Because of this the corresponding tunneling action 
$\tilde \act_t[\varphi_*]$, which we are going to evaluate, is non-zero. 

\newcommand{\K}{\mathcal K}

\begin{figure}[ht]
	\centering{\includegraphics[scale=0.25]{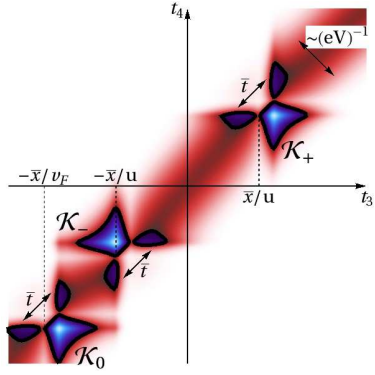}}
	\caption{Regions in the $t_3-t_4$-plane providing dominant
contributions to the integral for the instanton action,
Eq.~(\ref{instanton-action}) .} \label{fig:LL_instSingRegions}
\end{figure}

To exemplify the evaluation of the instanton
action $\tilde \act_t[\varphi_{\ast}]$,
let us consider the case of tunneling into/out of a right-moving state with the tip being placed on the 
right from the impurity, 
$\mu=+$ and $\bar x>0$. The phase factor is 
\begin{equation} 
\label{eqn:LL_smallInstantonRightFactors}
	e^{i\Phi_{\ast}^\mp(t)}=\kappa_+^\mp(t) \kappa_-^\mp(t)\kappa_0^\mp(t) 
\end{equation}
with
\begin{equation} 
	\kappa^\mp_+(t)= \left(\frac{t\pm i a +\bar x/u}{t\pm i a-\bar t+\bar
x/u}\right)^p,\ \kappa^\mp_-(t)=\left(\frac{a-i\alpha(t-\bar t-\bar
x/u)}{a-i\beta(t-\bar x/u)}\right)^q,\ \kappa^\mp_0(t)=\left(\frac{t\pm i a
-\bar t + \bar x/\vF}{t\pm i a +\bar x/\vF}\right). 
	\label{kappa-mp}
\end{equation}
The instanton action reads
\begin{gather}
	i\tilde\act_t[\varphi_{\ast}]=\sum_{\zeta,\eta=\mp}\, \zeta\eta \int\!\!\dd t_3 \dd t_4\, \tilde \Pi_{+-}^{\zeta\eta}(t_3-t_4)\  \K_+^{\zeta\eta}(t_3,t_4)\ \K_-^{\zeta\eta}(t_3,t_4)\ \K_0^{\zeta\eta}(t_3,t_4) \label{eqn:LL_instActK1}\\
	\text{with
}\K^{\zeta\eta}_\sigma(t_3,t_4)=\kappa^\zeta_\sigma(t_3)^{-1}
\kappa_\sigma^\eta(t_4)\quad \text{for $\sigma=+,-,0$.}\nonumber
\label{instanton-action}
\end{gather}
Since the polarization factor $\tilde \Pi_{+-}(t_3-t_4)$ comes with the factor
$e^{-ieV(t_3-t_4)}$ the integral is dominated by the region $\lvert
t_3-t_4\rvert \lesssim \lvert eV\rvert^{-1}$. Furthermore, important
contributions are expected to come from regions around the singularities of the
phase factors,
i.e.  $t_3,t_4\sim -\bar x/u$, $\bar t-\bar x/u$ for $\K_+$, $t_3,t_4\sim \bar
x/u$, $\bar t+\bar x/u$ for $\K_-$, and $t_3,t_4\sim -\bar x/\vF$, $\bar t-\bar
x/\vF$ for $\K_0$. These regions in the $t_3-t_4$-plane are sketched in
Fig.~\ref{fig:LL_instSingRegions}.

We will assume that the singularities are well separated, which imposes the
condition
\begin{equation}
\lvert \bar t\rvert, \lvert eV\rvert^{-1} \ll (1-K)\lvert \bar x\rvert/\vF, 
\label{singularities-separation-condition}
\end{equation}
For interaction strength of order unity, the dephasing time
$\tau_\varphi$ (which governs the relevant $\bar t$) is $\sim \lvert
eV\rvert^{-1}$, so that the conditions
(\ref{singularities-separation-condition}) reduce simply to $\lvert
eV\rvert^{-1} \ll \lvert \bar x\rvert/\vF$. This
condition, implying a sufficiently large voltage and/or tip-to-impurity
distance, can be easily satisfied.

Far from the singularities, the phase factors become trivial,
$\K_\sigma(t_3,t_4)\to1$, so that the integral (\ref{eqn:LL_instActK1})
approximately splits into
\newcommand{\I}{\mathcal I}
\begin{equation}
i\tilde \act_t[\varphi_{\ast}] \approx \I[\K_+]+\I[\K_-]+\I[\K_0],
\end{equation}
with
\begin{equation}
\I[\K_\sigma]= \sum_{\zeta,\eta=\mp}\zeta\eta\,\int\!\!\dd
t_3 \dd t_4\, \tilde \Pi_{+-}^{\zeta\eta}(t_3-t_4)
\left(\K_\sigma^{\zeta\eta}(t_3,t_4)-1\right).
\label{IKsigma}
\end{equation}
We have added $-1$ to the phase factor $\K_\sigma$ in Eq.~(\ref{IKsigma})
to make the convergence manifest. This does not change the value of the
integral in view of
$\sum_{\zeta\eta}\zeta\eta\, \tilde \Pi_{+-}^{\zeta\eta}(t)=0$. For the very same reason, the independence of $\K_-^{\zeta\eta}(t_3,t_4)$ of the Keldysh indices $\zeta$, $\eta$ implies $\I[\K_-]=0$.

According to Eqs.~(\ref{kappa-mp}), (\ref{instanton-action}), $\K_+$,
$\K_0$ have the form
$\K^{\zeta\eta}(t_3,t_4)\equiv\kappa^\zeta(t_3)^{-1}\kappa^\eta(t_4)$ with
$\kappa^\mp(t)=\left(\frac{t-t_a\pm i a}{t-t_b\pm ia}\right)^r$ and some
exponent $r>0$. Therefore, $\mathcal I[\K]$ is dominated by the regions
$\lvert t_3-t_4\rvert \lesssim \lvert eV\rvert^{-1}$ (singularity of $\tilde
\Pi_{+-}$) and $\lvert t_3-t_a\rvert, \lvert t_4-t_b\rvert\lesssim \lvert
eV\rvert^{-1}$ (singularity of $\K$) which determine the long-time asymptotics
$\lvert eV\bar t\rvert \gg 1$. 
This yields
\begin{align}
	\mathcal I[\K] \approx &\int\!\!\dd t\,\sum_{\zeta\eta}\zeta\eta\,\tilde \Pi_{+-}(t)\ \int\!\!\dd T\, \left(\kappa^\zeta(T)^{-1} \kappa^\eta(T)-1\right) \nonumber\\
	 & + \sum_{\zeta\eta}\zeta\eta\, \tilde \Pi_{+-}^{\zeta\eta}(t_a-t_b)\Big\rvert_{eV=0}\ \int\!\!\dd t_3\, e^{-i eVt_3} \kappa^\zeta(t_3)^{-1}\ \int\!\!\dd t_4\, e^{ieVt_4}\kappa^\eta(t_4) \nonumber \\
		\approx&-R_\ast(eV) \left\{\frac{\lvert eV(t_b-t_a)\rvert}{2\pi}\left(1-e^{-2\pi i r\ \sign[eV(t_b-t_a)]}\right)-\frac{\Gamma(2K)}{\Gamma(r)^2} e^{ieV(t_b-t_a)}\left[ieV (t_b-t_a)\right]^{2(r-K)}\right\}. \label{eqn:LL_IK}
\end{align}
Assuming for definiteness $\bar x>0$ and considering first 
right-movers, $\mu=+$, we thus obtain for the instanton action
$i\tilde\act_t[\varphi_{\ast}]=\mathcal I[\K_+]+\mathcal I[\K_0]$. The first
contribution here,
\begin{align} 
	\I[\K_+] &= -\frac{\lvert \bar t\rvert }{2\tau_\varphi} - iR_{\ast}(eV)\ \frac{\sin 2\pi p}{2\pi}\ eV\bar t\ +\ \mathcal C_+ R_\ast(eV) e^{ieV\bar t} [ieV \bar t]^{2q}, \quad \mathcal C_+ = \Gamma(2K)/\Gamma(p)^2,\label{eqn:LL_instActRightP}
\end{align}
encodes effects of real plasmons on tunneling which are generated because of backscattering off
the impurity. One of such effects is the shot-noise, which is represented by the
first term in Eq.~(\ref{eqn:LL_instActRightP}). This term is negative and linear
in time $\bar t$ and thus accounts for dephasing with the rate
(\ref{eqn:LL_dephRate}). The second term in
Eq.~(\ref{eqn:LL_instActRightP})represents a perturbatively small
renormalization of bias voltage and will be neglected in the following. The
third term in Eq.~(\ref{eqn:LL_instActRightP}) is subleading as compared to the
first one and we will treat it perturbatively in $R_{\ast}$. It shows
an oscillatory behavior
accounting for an energy transfer $\sim eV$ between the nonequilibrium bath of plasmons and a tunneling electron.
We will return to this point when discussing the tunneling rates.

We turn now to the contribution $\I[\K_0]$. In this case $r=1$, so that the
first term in (\ref{eqn:LL_IK}) vanishes. Thus, the the electron-hole pair
contribution reads
\begin{align}
	\I[\K_0] & = \sum_{\zeta\eta}\zeta\eta\, \tilde \Pi_{+-}^{\zeta\eta}(\bar t)\Big\rvert_{eV=0}\ \int\!\!\dd t_3\, e^{-i eVt_3} \kappa_0^\zeta(t_3)^{-1}\ \int\!\!\dd t_4\, e^{ieVt_4}\kappa_0^\eta(t_4) \label{eqn:LL_instActRightG1}\\
	&=R_\ast(eV) \Gamma(2K) e^{-ieV\bar t}\left[-ieV\bar t\right]^{2(1-K)}.
\label{eqn:LL_instActRightG2}
\end{align}
It is oscillatory and can be again treated perturbatively. In contrast to the
plasmon contribution, however, it seemingly corresponds to an energy transfer of
$\sim-eV$ and does not have a clear physical interpretation.
We will see below that this term is an artifact of functional bosonization which
will be canceled by the Born correction $\mathcal G_{1\mu}^{\alpha\beta}$. In
total, we have
\begin{align}\label{eqn:LL_instActRight}
	e^{i\tilde \act_t[\varphi_{\ast}]} \approx e^{-\lvert \bar t\rvert /2\tau_\varphi} \left(1+\mathcal C_+ R_\ast(eV)\ e^{ieV\bar t} \left[ieV\bar t\right]^{2q}+\I[\K_0]\right)\quad \text{for $\bar x>0$, $\mu=+$.}
\end{align}

The same considerations can be applied to left-movers, with the result
\begin{align} 
\label{eqn:LL_instActLeft}
	e^{i\tilde \act_t[\varphi_{\ast}]}\approx e^{-\lvert \bar t \rvert /2\tau_\varphi} \left(1\ +\  \mathcal C_-R_\ast(eV) e^{ieV\bar t} [ieV\bar t]^{1-2K}\right),\quad \mathcal C_-=\Gamma(2K)/\Gamma(q)^2, \quad \text{for $\bar x>0$, $\mu=-$.}
\end{align}

\subsubsection{Born Correction}

We evaluate the Born correction $\mathcal G_{1\mu}^{\alpha\beta}$ to the Green's function, (\ref{eqn:LLGFBornCorrection}), in leading order in $\R$, which amounts to taking averages with respect to the clean action $\mathcal A_0[\varphi]$ only. Then Wick's theorem yields
\begin{align} \label{eqn:LL_BornApprox}
	\mathcal G^{\alpha\beta}_{1\mu} \approx i\frac{\R \vF}2 \sum_{\gamma\delta=\mp} \gamma\delta \int \!\!\dd t_1\,\dd t_2\,  G_{0\mu}^{\alpha\gamma}(\bar x,0;\bar t-t_1) g^{\gamma\delta}_{-\mu}(t_1-t_2)G_{0\mu}^{\delta\beta} (0,\bar x;t_2-\bar t)\ \mathcal J_\mu^{\alpha\beta\gamma\delta}(\bar x,\bar t;t_1,t_2)
\end{align}
with
\begin{multline}
	\mathcal J_\mu^{\alpha\beta\gamma\delta}(\bar x,\bar t;t_1,t_2)=\left\langle e^{i\Theta^\alpha_\mu(\bar x,\bar t)} e^{i\mu \Phi^\gamma(t_1)} e^{-i\mu \Phi^\delta(t_2)} e^{-i\Theta^\beta_\mu(\bar x,0)}\right\rangle_0\\
	= e^{-\frac 12\left\langle\left[\Theta^\alpha_\mu(\bar x,\bar t)-\Theta^\beta_\mu(\bar x,0)\right]^2 \right\rangle_0-\frac12 \left\langle\left[\Phi^\gamma(t_1)-\Phi^\delta(t_2)\right]^2\right\rangle_0}\  e^{i\mu\left(\Phi^\gamma_{\ast}(t_1)-\Phi_{\ast}^\delta(t_2)\right)}
\end{multline}
and the instanton (\ref{eqn:LL_smallInstanton}). The appearance of the instanton makes the integral (\ref{eqn:LL_BornApprox}) quite similar to the instanton action and we will use an analogous approximations to deal with the time integrals.

We have already seen that the instanton phase factor factorizes into three contributions (\ref{eqn:LL_smallInstantonRightFactors})---two governed by plasmons and one by electron-hole pairs---and one might expect the integral (\ref{eqn:LL_BornApprox}) to split into three contributions in a way akin to the instanton action. However, it will turn out that the presence of the bare Green's functions $G^{\alpha\gamma}_{0\mu}(\bar x,0;\bar t-t_1)$ and $G^{\delta\beta}_{0\mu}(0,\bar x;t_2-\bar t)$ suppresses the plasmon contributions. Focusing again on $\mu=+$, $\bar x>0$, we show that the remaining electron-hole pair term cancels $\I[\K_0]$ in (\ref{eqn:LL_instActRight}).

The $t_1,t_2$-dependent contributions to (\ref{eqn:LL_BornApprox}) are
\begin{multline*}
	G_{0+}^{\alpha\gamma}(\bar x,0;\bar t-t_1) g_-^{\gamma\delta}(t_1-t_2) G_{0+}^{\delta\beta}(0,\bar x;t_2-\bar t) e^{-\frac 12\left\langle\left[\Phi^\gamma(t_1)-\Phi^\delta(t_2)\right]^2\right\rangle} e^{i\Phi_\ast^\gamma(t_1)-i\Phi_\ast^\delta(t_2)}\\
	=e^{-ieV\bar t} e^{ieV(t_1-t_2)}\times \tilde g_-^{\gamma\delta}(t_1-t_2) \times \kappa_0^\delta(t_2)^{-1} \kappa_+^\delta(t_2)^{-1} \kappa_-^\delta(t_2)^{-1}\times \tilde \kappa_0^\gamma(t_1)\kappa_+^\gamma(t_1)\kappa_-^\gamma(t_1).
\end{multline*}
We combined terms with similar pole structure, defining
\begin{align*}
	\tilde g^{\gamma\delta}_-(t_1-t_2) &\equiv 	e^{-\frac 12\left\langle\left[\Phi^\delta(t_2)-\Phi^\gamma(t_1)\right]^2\right\rangle_0} g^{\gamma\delta}_-(t_1-t_2)\Big\rvert_{eV=0}, \nonumber\\
	\tilde \kappa^\delta_0(t_2)^{-1} &\equiv G^{\delta\beta}_{0+} (-\bar x,t_2)\Big\rvert_{eV=0}\ \kappa_0^\delta(t_2)^{-1},\\
	\tilde \kappa_0^\gamma(t_1) &\equiv G^{\alpha\gamma}_{0+}(\bar x,\bar t-t_1)\Big\rvert_{eV=0}\ \kappa_0^\gamma(t_1).
\end{align*}
All voltage dependence has been singled out in the phase factors explicitly.

Let us examine the pole structure: $\tilde g^{\gamma\delta}_-(t_1-t_2)$ is divergent for $t_1\approx t_2$. This is reminiscent of $\tilde \Pi_{+-}(t_3-t_4)$ in (\ref{eqn:LL_instActK1}) which preferred $t_3\approx t_4$. The plasmon contributions $\kappa_+$ and $\kappa_-$ have been studied in the previous section where we already noted $\kappa_\pm(t)\to 1$ for $t$ far from their singularities. This is no longer true for $\tilde\kappa_0$. Indeed, leaving the Keldysh indices and the corresponding short-time regularizations aside for a moment, we have
\begin{align*}
	G_{0+}(x,t)\Big\rvert_{eV=0} = -\frac 1{2\pi\vF} \frac 1{t-x/\vF}\quad \text{and}\quad \kappa_0(t)=\frac{t-\bar t+\bar x/\vF}{t+\bar x/\vF},
\end{align*}
and hence,
\begin{align*}
	\tilde \kappa_0^\delta(t_2)^{-1}\sim -\frac 1{2\pi\vF}\frac 1{t_2-\bar t+\bar x/\vF},\ \tilde \kappa_0^\gamma(t_1) \sim \frac 1 {2\pi\vF} \frac 1{t_1+\bar x/\vF}.
\end{align*}
Similarly to the original phase factors $\kappa_0^\delta(t_2)^{-1}$, $\kappa_0^\delta(t_1)$, these new ones have the poles at $t_1\sim -\bar x/\vF$, $t_2\sim \bar t-\bar x/\vF$; however, at variance with 
$\kappa_0^\delta(t_2)^{-1}$, $\kappa_0^\delta(t_1)$, the factors $\tilde \kappa_0^\delta(t_2)^{-1}$, $\tilde \kappa_0^\delta(t_1)$ vanish far from the poles instead of converging to unity.
 Therefore, the poles of $\tilde \kappa_0^\delta(t_2)^{-1}$, $\tilde \kappa_0^\delta(t_1)$ are dominating the integral (\ref{eqn:LL_BornApprox}), while the plasmonic poles give subleading contributions (suppressed by the factor $\vF \bar t/(1-K)\bar x\ll 1$). With only leading terms taken into account the integrals simplify to
\begin{multline} \label{eqn:LL_BornDominant}
	\mathcal G_{1+}^{\alpha\beta} \approx  i \frac{\R \vF}2 e^{-\frac 12 \langle\left[\Theta^\alpha_\mu(\bar x,\bar t)-\Theta^\beta_\mu(\bar x,0)\right]^2\rangle}\\
	\sum_{\gamma,\delta=\mp} \gamma\delta\ e^{-\frac 12 \langle\left[\Phi^\delta(\bar t)-\Phi^\gamma(0)\right]^2\rangle}  \left[G_{0+}^{\alpha\gamma}(\bar x,\bar t+\bar x/\vF) g_-^{\gamma\delta}(-\bar t) G^{\delta\beta}_{0+}(-\bar x,\bar t-\bar x/\vF)\right]_{eV=0} e^{-ieV\bar t}\\
	\int\!\!\dd t_2\, e^{-ieVt_2}\ \kappa_0^\delta(t_2)^{-1} \int\!\!\dd t_1\, e^{ieVt_1}\ \kappa_0^\gamma(t_1)
\end{multline}
For large times $\lvert \bar t\rvert \gg \lvert eV\rvert^{-1}$, the short-time regularization and thus the distinction between different Keldysh components becomes immaterial, e.g.\ $\Pi_{+-}^{\gamma\delta}=\R e^{-ieV\bar t}/(2\pi\bar t)^2$, $G_{0+}^{\alpha\beta}(0,\bar t)=-1/(2\pi\vF \bar t)$, and one obtains
\begin{multline*}
	i\frac{\R\vF}2  \left[G_{0+}^{\alpha\gamma}(\bar x,\bar t+\bar x/\vF) g_-^{\gamma\delta}(-\bar t) G^{\delta\beta}_{0+}(-\bar x,\bar t-\bar x/\vF)\right]_{eV=0} e^{-ieV\bar t}\\
	=i\frac{\R \vF}2 \left(\frac 1{2\pi\vF} \frac 1{\bar t}\right)^2 \left(-\frac i\pi \frac 1{\bar t}\right) e^{-ieV\bar t} = -G_{0+}^{\alpha\beta}(\bar x,\bar x;\bar t) \left[\Pi^{\delta\gamma}_{+-}(\bar t)\right]_{eV=0}.
\end{multline*}
Substituting this into (\ref{eqn:LL_BornDominant}), taking into account the dressing of the Green functions (\ref{eqn:LL_cleanGF}) and polarization operators by phase factors and comparing with Eq.~(\ref{eqn:LL_instActRightG1}), we find 
\begin{align}
	\mathcal G_{1+}^{\alpha\beta} &\approx - \tilde G_{0+}^{\alpha\beta}(\bar x,\bar x;\bar t)\ \sum_{\delta\gamma}\delta\gamma\ \tilde \Pi^{\delta\gamma}_{+-}(\bar t)\Big\rvert_{eV=0} \ \int\!\!\dd t_2\, e^{-ieVt_2}\ \kappa_0^\delta(t_2)^{-1} \int\!\!\dd t_1\, e^{ieVt_3}\ \kappa_0^\gamma(t_1) \nonumber\\
	&\approx - \tilde G_{0+}^{\alpha\beta}(\bar x,\bar x;\bar t)\ \mathcal I[\K_0]. \label{eqn:LL_G1}
\end{align}

The  very same analysis can be performed for $\mu=-$. In this case, however, $\tilde\kappa^{\gamma}_0$ does not depend on the Keldysh index $\gamma$. Because of $\sum_{\gamma\delta} \gamma\delta\ g_+^{\gamma\delta}(t_1-t_2)=0$ the contribution $\mathcal G_{1-}^{\alpha\beta}$ is negligible.

Concluding, we obtain
\begin{align}\label{eqn:LL_instActSC}
	\mathcal G^{\alpha\beta}_{0+}&\approx e^{-\lvert \bar t\rvert /2\tau_\varphi}\ \tilde G_{0+}^{\alpha\beta}(\bar x,\bar x;\bar t)\ \left(1+\mathcal C_+ R_\ast(eV) e^{ieV\bar t} \left[ieV\bar t\right]^{2q}+ \I[\K_0]\right),\\
	\mathcal G^{\alpha\beta}_{0-} &\approx e^{-\lvert \bar t \rvert /2\tau_\varphi}\ \tilde G_{0-}^{\alpha\beta}(\bar x,\bar x;\bar t)\  \left(1\ +\  \mathcal C_-R_\ast(eV)\ e^{ieV\bar t} [ieV\bar t]^{1-2K}\right),\\
	\mathcal G_{1+}^{\alpha\beta} &\approx - \tilde G_{0+}^{\alpha\beta}(\bar x,\bar x;\bar t)\ \mathcal I[\K_0].
\end{align}
In leading order in $\R$, i.e. neglecting dephasing corrections, $G_{1+}^{\alpha\beta}$ cancels the $\I[\K_0]$ term of $\mathcal G_{0+}^{\alpha\beta}$, as was stated in the end of Sec.~\ref{sec:instanton-action}.

\subsubsection{Tunneling Rates}

We are now ready to evaluate the components of the Keldysh Green function and,
in particular, the tunneling density of states controlling the tunneling
current between the system and the tip. Let us note that because of the tunneling 
actions~(\ref{eqn:LL_instActRight}) and (\ref{eqn:LL_instActLeft}) 
the Green functions contain power-law terms $[ieV\bar t]^{2\delta}$,
$r>0$, which have apparent branchcut singularities near $\bar t=0$. However,
results~(\ref{eqn:LL_instActRight}) and (\ref{eqn:LL_instActLeft}) are only valid in the long-time limit
$\lvert eV\bar t\rvert\gg 1$. A regularization which takes this into account and
does not violate the symmetry
relation $\left[iG_\mu^\gtrless(\bar x, \bar x;\bar t)\right]^\ast = iG_\mu^\gtrless(\bar x, \bar x;-\bar t)$) is $[1+ieV\bar t]^{2\delta}$, which yields the Green's functions
\begin{gather}
	G^{\alpha\beta}_\mu (\bar x,\bar x;\bar t) = \tilde G^{\alpha\beta}_\mu (\bar x,\bar x;\bar t) e^{-\lvert \bar t \rvert /2\tau_\varphi}
	\left(1\ +\  \mathcal C_\mu R_\ast(eV)
	e^{ieV\bar t} [1+ieV\bar t]^{2\delta_\mu}\right)\\
	\text{with}\quad \mathcal C_+=\Gamma(2K)/\Gamma(p)^2,\quad \mathcal C_-=\Gamma(2K)/\Gamma(q)^2, \quad \delta_+=(1-K)/2,\quad \delta_-=1/2-K. 
\end{gather}
The tunneling rates are obtained by Fourier transformation to energy representation. 
Using the aforementioned symmetry property of the Green's function, we obtain
\begin{align}
	\Gamma_\mu^\gtrless(\epsilon) &= \pm \frac 1 {2\pi} \int_0^\infty\!\! \dd \bar t\, \left(e^{i\epsilon\bar t}\ iG_\mu^\gtrless(\bar x,\bar x;\bar t)+e^{-i\epsilon\bar t}\ iG_\mu^\gtrless(\bar x,\bar x;-\bar t)\right)
		=\mp\frac 1\pi \Im \int_0^{\infty}\!\!\dd \bar t\,e^{i\epsilon \bar t}G_\mu^\gtrless(\bar x,\bar x;\bar t)
\nonumber\\
		&= \pm\frac{\nu_0}\pi {\rm Im}\left[\mathcal J_0^\gtrless(\epsilon;eV_\mu)+\mathcal C_\mu R_\ast(eV)\ \mathcal J_{\delta\mu}^\gtrless(\epsilon;eV_\mu-eV)\right]
		\label{eq:Rates_Gamma}
\end{align}
	with
\begin{align*}
	\mathcal J_\delta^\gtrless(\epsilon;U)\equiv \pm i a^{2\gamma} \int_0^\infty\!\!\dd \bar t\, e^{i(\epsilon-U+i/2\tau_\varphi)\bar t}\ (1+ieV\bar t)^{2\delta}  (a\pm i \bar t)^{-(2\gamma+1)}, 
\end{align*}
	Since $\mathcal J_r^\gtrless(\epsilon;U)$ is an analytic function of all
parameters $\epsilon, \delta, \gamma,U, eV\in\mathbb C$ as long as $\Im
(\epsilon-U+i/2\tau_\varphi)>0$ and $\Im eV\le 0$,  we can consider here $\Re
\delta>-1$, $\Re (2\gamma+1)<1$, $\Im eV<0$ and
$\Re[(\epsilon-U+i/2\tau_\varphi)/eV]<0$ and deduce all relevant cases by
analytic continuation. Under these constraints, one can evaluate the integral by
rotating the integration contour, $\bar t=-is/eV$, $0<s<\infty$, into the
complex plane, and put $a\to0$. Writing $z=(\epsilon-U+i/2\tau_\varphi)/eV$ we
get
	\begin{align}
		\mathcal J^\gtrless_\delta(\epsilon;U) & = \pm\frac{a^{2\gamma}}{eV} \int_0^\infty\!\!\dd s\, e^{zs}\ 
(1+s)^{2\delta}(\pm s/eV)^{-(2\gamma+1)}\nonumber\\
			&= \left(\pm\frac{eV}\Lambda\right)^{2\gamma} \Gamma(-2\gamma) \Psi(-2\gamma,1-2\gamma+2\delta,-z)
			\label{eq:Psi_rates}
	\end{align}
with the confluent hypergeometric function $\Psi(a,b,z)$. 
In the case $\delta=0$ one has $\Psi(-2\gamma,1-2\gamma,z)=z^{2\gamma}$ and thus the term $\mathcal J^\gtrless_0$
in Eq.~(\ref{eq:Rates_Gamma}) yields the equilibrium zero-bias anomaly. 
Taking into account the 2nd (impurity) contribution we arrive at the results that have been presented in
Sec.~\ref{tdos-model-results} [ see Eq.~(\ref{eqn:LL_tunRates}) ]. We note here that the result for
the physical (real) voltage $eV$ is obtained from Eq.~(\ref{eq:Psi_rates}) by substituting there $eV-i0$.
One can also note that at $\delta\neq 0$
one gets $\Psi(-2\gamma,1-2\gamma+2\delta ,z) \sim z^{2(\gamma-\delta)}$ at $z\to 0$. Therefore 
Eq.~(\ref{eq:Psi_rates}) gives the impurity correction $\Delta \Gamma_\mu$ to tunneling rates which is singular at
$\epsilon = eV_\pm - eV$. Explicitly one has 
\begin{equation}
\Delta \Gamma_\mu^<(\epsilon) \propto - R_*(eV)\,
\bigl|\epsilon-eV_\pm + eV\bigr|^{2(\gamma-\delta_\mu)} \sin \left\{
\begin{array}{cc}
2\pi\delta_\mu, & \epsilon>eV_\pm - eV  \\
2\pi\gamma, & \epsilon< eV_\pm - eV 
\end{array}\right.
\label{OutR_Rate}
\end{equation}
in case of tunneling from the wire into the tip and
\begin{equation}
\Delta \Gamma_\mu^>(\epsilon) \propto R_*(eV)\,\,\theta(\epsilon-eV_\pm + eV) 
\bigl|\epsilon-eV_\pm + eV\bigr|^{2(\gamma-\delta_\mu)}
\label{InR_Rate}
\end{equation}
in case of tunneling from the tip into the wire.

\newcommand{\osc}{\mathrm{AB}}
\newcommand{\dir}{\mathrm{inc}}
\newcommand{\epsth}{\epsilon_{\rm th}} 
\newcommand{\regABNu}{{AB}} 
\newcommand{\regCBI}{{CD I}} 
\newcommand{\regCBNu}{{CD II}} 
\newcommand{\regABI}{{AB*}} 

\section{Quantum Hall Fabry-P\'erot Interferometer }
\label{sect:FPI}

In this section we study the role of Coulomb interaction 
in an electronic Fabry-P\'erot interferometer realized with chiral edge states in the integer 
QHE regime. Electronic Fabry-P\'erot (FPI)~\cite{MarcusWest09, ZhangMarcus09, Yamauchi09, Heiblum09} 
and Mach-Zehnder (MZI)~\cite{heiblum03, heiblum06, heiblum07,
heiblum07a, roche07, roche08, roche09, roche12, strunk07, strunk08, strunk10,
schoenenberger} interferometers are analogues of the optical 
interferometers, where the chiral edge states play the role of light beams while quantum point
contacts (QPCs) act as beam splitters. Electron interferometry provides a powerful tool for studying 
the quantum interference and dephasing in mesoscopic semiconductor devices.
Another motivation behind these experimental efforts stems from the recent interest in topological 
quantum computations, which propose to exploit the non-Abelian anyons in the fractional QHE regime~\cite{Nayak08}.   

The Coulomb interaction is of paramount importance in fractional QHE systems,
where it gives rise to quasi-particles with fractional charge obeying anyonic statistics. 
It came as a surprise that {\it e-e} interaction plays a prominent role in integer QHE interferometers as well,
even when their conductance is $\sim e^2/h$ so that the Coulomb blockade physics
seems to be inessential. 
For instance, visibility in the MZIs and FPIs strongly depends on the source-drain voltage 
showing decaying oscillations, which have been termed ``lobes''. 
The search for a resolution of this puzzle in the case of MZI has triggered a lot of 
attention~\cite{Sukhorukov07, Chalker07, Levkivskyi08, Neder08, Youn08, Kovrizhin09, Schneider11}. 
On the contrary, the extent of theoretical works on FPIs operating in the integer QHE regime is rather 
small~\cite{Chamon97,Halperin07, Halperin10}.

In this section we develop a capacitance model of the {\it e-e}
interaction in a FPI and apply it
to study the transport properties of the FPI in and out of
equilibrium in the limit of weak backscattering.
Our approach is inspired by the previous theoretical work~\cite{Halperin07}.
Its essential idea is that a compressible Coulomb island can be formed in the
center of the FPI between two constrictions (Fig.~1), which
strongly affects Aharonov-Bohm oscillations. Starting from this model, we
demonstrate that 
depending on the strength of the {\it e-e} interaction the FPI can fall into
``Aharonov-Bohm'' (AB) or ``Coulomb-dominated'' (CD)
regimes observed in the experiments~\cite{ZhangMarcus09,Heiblum09}. 
We also analyze the suppression of  nonequilibrium AB oscillations with
the increase of a source-drain 
voltage and find regions of both power-law and exponential decays, which
explains experiments of 
Refs.~\cite{MarcusWest09, Yamauchi09}. 

The brief account on results of this section has been reported by two of us previously~\cite{Ngo_Dinh12}.
Here we present technical details of our calculations and further elaborate
on the qualitative picture
of the interplay of interference and {\it e-e} interaction in the FPI
which explains well 
the plethora of experimental data on the flux and gate periodicity of Aharonov-Bohm oscillations.

\subsection{Model}

\begin{figure}[ht]
	\centering{\includegraphics[scale=0.25]{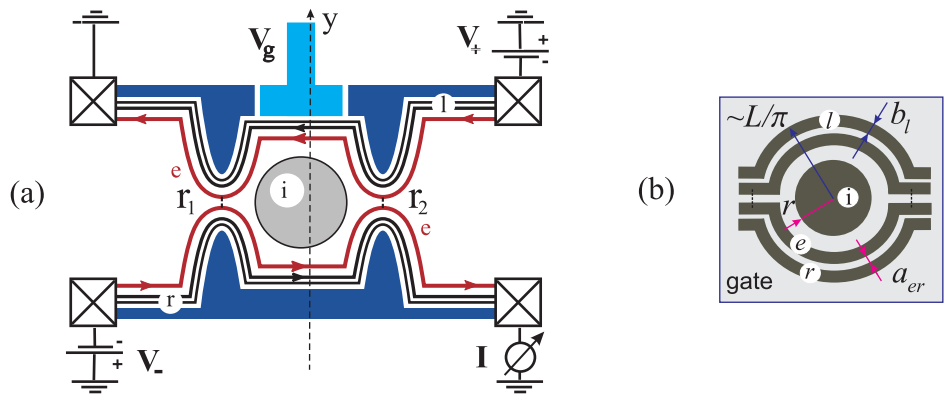}}
	\caption{(a) Fabry-P\'erot interferometer with compressible island $i$.
The innermost edge channels $e$ are subject to backscattering at the QPCs; the
remaining $f_T=\nu-1$ right($r$)- and left($l$)- moving channels are fully
transmitted; their role is in screening of the interaction between the
electrons of the channel $e$. Right- and left-moving channels are connected to
reservoirs with different chemical potentials $\mu_\eta = eV_\eta$, $\eta=\pm$.
Here, the gate $g$ is depicted as a ``plunger'' gate.
(b) Simple capacitance model, which uses geometry of the compressible regions.  
Right($r$)- and  left($l$)-channels are each joined into one conductor with the widths $b_r$ and $b_l$, respectively.}
	\label{fig:Setup}
\end{figure}

We consider an electronic FPI of size $L$ formed by a Hall bar 
with $\nu$ edge channels and two constrictions (QPCs) that allow for electron backscattering 
between the innermost right/left moving edge channels with amplitudes $r_{1(2)}$ as shown in Fig.~\ref{fig:Setup}~(a). 
Right- and left-moving channels are connected to leads with different chemical
potentials $\mu_+$ and $\mu_-$, respectively.
In what follows, we take into account the backscattering in the lowest order, 
thus accounting for interference of maximally two different paths. For
simplicity we assume the flight times along upper and lower arms (i.e. between
two QPCs) to be the same, $\tau_+=\tau_-=\tau=L/\vF$. We denote the magnetic
flux threading the interferometer cell by $\phi$, i.e.\ an electron which
encircles the cell once accumulates the phase $2\pi \phi/\phi_0$, where
$\phi_0=hc/\lvert e\rvert$ is the flux quantum.

\begin{figure}[ht]
 	\centering{\includegraphics[scale=0.4]{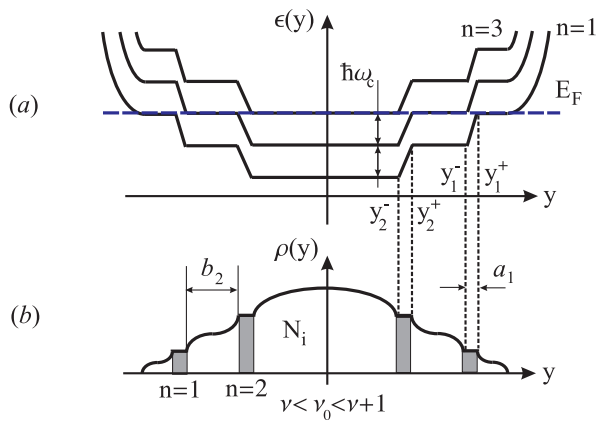}}
 	\caption{The structure of the edge states in the FPI. Cut through the center of an interferometer cell: 
(a) Landau level energies as the function of vertical coordinate $y$ and 
(b) electron density $\rho(y)$ (see also Fig.~\ref{fig:Setup}).
Besides $\nu$ completely filled LLs, the central region can sustain a partially filled Landau level 
whose occupation $N_i$ changes by tunneling. 
Here we denoted by $a_k = y_k^+ - y_k^-$ the width of the $k$-th incompressible strip (shown by grey), 
which has the integer filling factor $k$. 
Compressible regions (white) represent the regions with non-integer filling fraction and have the width
$b_k = y_{k-1}^- - y_{k}^+$. The picture above corresponds to the case $\nu=2$.
} 
\label{fig:cutDensity}
\end{figure}

The 2DEG in the QHE regime is divided into compressible and incompressible strips~\cite{Chklovskii92}.
The filling factor in the $n$-th incompressible strip is integer. These strips are separated by 
much wider regions of compressible Hall liquid with a non-integer filling factor (compressible strips).
The corresponding sketch of electron density profile $\rho(y)$ in the FPI along $y$-axis is shown in 
Fig.~\ref{fig:cutDensity}. Let us denote by $y_k^{\pm}$ the boundaries between compressible and incompressible 
regions. Then $a_k=y_k^+ - y_k^-$ is the width of the $k$-th incompressible strip while 
$b_k = y_{k-1}^- -y_k^+$ is the width of the $k$-th compressible one. As it was shown in 
Ref.~\cite{Chklovskii92},  in the situation of gate-induced
confinement of 2DEG in the QHE regime the widths $b_k \gg \lambda_B$, with $\lambda_B$ being the 
magnetic length. At the same time $a_k$ scales as  $a_k \sim \left(b_k\lambda_B\right)^{1/2}$, so that
in general the condition $b_k \gg a_k \gg \lambda_B$ is satisfied. 
In this picture  
compressible regions play the role of edge channels --- the self-consistent electrostatic potential
is constant through the compressible strips and can be controlled by connecting them to external leads.

We also assume that the filling fraction $\nu_0$ in the center of the FPI
exceeds $\nu$, giving rise to  
a compressible droplet (Coulomb island). The reason for that can be smooth (on a scale $\lambda_B$) disorder 
potential fluctuations~\cite{Cooper93}. Let us denote by $e N_i$ the excess charge on the island ($e<0$),
with $N_i$ being integer. On the scheme in Fig.~\ref{fig:cutDensity} the boundary
of the island is given by $y_\nu^-$. This value is quantized and changes abruptly when an electron
tunnels between innermost compressible strips and the island through the incompressible strip.
On the contrary, the boundaries of edge channels may change continuously 
because of a variation in external parameters, such as $\mu_\pm$ and $V_g$, or due to 
quantum fluctuations of electrostatic potentials on these compressible regions (see also discussion later).

\subsubsection*{Electrostatics of the FPI}

In the framework of the above model, we treat the {\it e-e} interaction in the
FPI by
using the constant interaction model with mutual capacitances $C_{\alpha\beta}$ between four 
compressible regions --- the interfering channel ($e$); 
right- and left-moving fully transmitted channels ($r$, $l$); the compressible island ($i$) ---
and the gate ($g$). These capacitances are denoted by $C_{eg}$, $C_{ei}$ etc.  
We assume a large capacitance between counter-propagating innermost channels ---
thus they share the same
electrostatic potential $\varphi_e$ --- and also consider $f_T=\nu-1$ right- and left-moving channels as 
joint conductors with potentials $\varphi_r$ ($\varphi_l$). 
Defining a capacitance matrix $\tilde C$ with elements $\tilde C_{\alpha\alpha}=\sum_\gamma C_{\alpha\gamma} + C_{\alpha g}$, and $\tilde C_{\alpha\beta}=-C_{\alpha\beta}$ ($\alpha\neq \beta$), 
where Greek indices span the set $\{e,r,l,i\}$ and $q_\alpha=-C_{\alpha g}V_g$ is an offset charge on the $\alpha$'s conductor, the electrostatic energy reads
\begin{align} \label{eqn:electrostaticE}
	E = \frac 12 \sum_{\alpha \beta} \left(Q_\alpha-q_\alpha\right) \left(\tilde C^{-1}\right)_{\alpha\beta} \left(Q_\beta-q_\beta\right).
\end{align}
Total charge $Q_i=e\left(N_i+\nu \phi/\phi_0\right)$ on the island is distributed on the highest partially filled Landau level (LL) and on $\nu$ fully occupied underlying LLs (cf.\ Fig.~\ref{fig:cutDensity}). 
Single electron tunneling is possible between interfering channels ($e$) and the island. We assume the rate of such 
tunneling process to be much smaller than all other energy scales in the problem, $\Gamma\ll \epsth$, hence $N_i$ is 
quantized and is fixed for given external parameters ($\phi$, $V_g$, $\mu_\pm$).

The mutual capacitances $C_{\alpha\beta}$ can be estimated from geometrical considerations~\cite{Evans93}.
We regard the island as a disc of radius $r$ and represent the compressible edge channels as concentric
rings of the width $b_\alpha$ and diameter $L$ (here $\alpha=e,r,l$) as depicted  
in Fig.~\ref{fig:Setup}~(b). The edge channels are assumed to be thin, $b_\alpha \ll L$.
Therefore for estimation of capacitances we can neglect the difference between the radii of the
island and those of edge channels, i.e. $L \simeq \pi r \simeq \pi y_k^{\pm}$.
A top gate, if present, is modeled by a plane situated at distance $d$ from the 2DEG. 
Since the size of FPI cell is much larger than $d$, we treat $C_{ig}$, $C_{eg}$ and $C_{r(l)g}$ 
as a parallel-plate capacitors and find an estimate 
\begin{equation}
C_{ig} \simeq \epsilon\frac{r^2}{4d}, \qquad C_{eg} \simeq \epsilon \frac{L b_{e}}{2\pi d},
\qquad C_{rg} \simeq \epsilon \frac{L b_{r}}{4\pi d},
\label{eqn:C_gate}
\end{equation}
where $\epsilon=12.6$ is the dielectric constant for GaAs. The estimate for edge-to-edge
($C_{er}$ and $C_{e\,l}$) and edge-to-island ($C_{ei}$) capacitances can be found as a mutual 
capacitance of two conducting rings. In the limit $b\ll a$ we obtain with logarithmic 
accuracy~\cite{Capacitances}
\begin{equation}
 C_{er} \simeq \frac{\epsilon L}{2\pi^2}\ln\left( \frac{b_e b_r}{a_{er}^2}\right), \quad
 C_{ei} \simeq \frac{\epsilon r}{\pi}\ln\left( \frac{r b_e}{a_{ei}^2}\right).
 \label{eqn:C_edge}
\end{equation}
Here $b_{r(l)} = \sum_k^{\nu-1} b_k$ are the total widths of fully transmitted edge channels.
Finding the mutual capacitance between a plunger gate and the island or the interfering edge channel 
is in general more difficult. Because of geometry, one can expect
that $C_{ig}$ and $C_{eg}$ in 
this case will be substantially smaller than the
above estimate~(\ref{eqn:C_gate}) for the case of a top gate.

Let us now comment on the flux dependence of the electrostatic energy,
Eq.~(\ref{eqn:electrostaticE}). When the magnetic flux through the island is
increased, $\delta\phi = \pi r^2 \delta B$ , the LLs are squeezed
and the charge on the island (for a fixed boundary $y_\nu^-$) varies as 
$\delta Q_i = e\nu \, \delta \phi/\phi_0$. A similar effect of magnetic field on
charges $Q_{e,r,l}$  distributed on compressible circular strips is negligibly small
because of the condition $b\ll r$.
Indeed, for a typical variation $\delta B$, such that $\delta \phi/ \phi_0 \sim 1$, the corresponding
modulation of these charges are
\begin{equation} 
\frac{\delta Q_{e,r,l}}{e} \sim  \frac{\delta\phi}{ \phi_0} \left(\frac{b}{\pi r}\right) \ll 1,
\end{equation}
and we do not include them into Eq.~(\ref{eqn:electrostaticE}).

\subsection{Results}
\label{sec:FPI-results}

In this subsection we summarize our results and give their physical
interpretation. The detailed derivation is presented in the next subsection. 
The qualitative behavior of the FPI crucially depends on the relative coupling
strength of the interfering edge ($e$) to the fully transmitted channels ($l$,
$r$), to the island, and to the gate. 
The essential parameters are the number of transmitted channels $\nuEff$ which screen the 
bare {\it e-e} interaction in the interfering channel --- as we demonstrate
in the section~\ref{sect:FPI_Calculations}, one has $\nuEff\simeq 1$ 
in the case of strong and $\nuEff\simeq\nu$ in the case of weak inter-edge interaction --- and the effective edge capacitance $\bar C_e$ as defined below by Eq.~(\ref{eqn:effCaps}). There are also two characteristic energy scales in our problem:
(i) charging energy $E_C=e^2/\bar C_e$, or charge relaxation frequency $\wc=({\nu^\ast}/\pi) \ec$; and 
(ii) the Thouless energy $\epsth=\tau^{-1}=\vF/L$. A relation between these two
parameters depends essentially on the geometry of the experiment (most
importantly, on the geometry of the gates). We will assume that the condition
$\epsth \ll \wc$ is always satisfied, which simplifies a lot our subsequent calculations
and enables us to get analytical results.
This appears to be a proper assumption
for most of available experiments.  In particular,
the value of the Thouless energy that can be deduced from the experiment of the
Harvard group is $\epsth \sim 50\:\mu{\rm V}$ \cite{MarcusWest09}, whereas the
charging energy is in the mV range \cite{ZhangMarcus09}.

\subsubsection{Visibility, dephasing and the ``lobe'' structure}

In the limit of weak backscattering, $r_j\ll 1$, the differential conductance of  the FPI,
$g=g_\dir+g_\osc$,  is the sum of incoherent and coherent contributions. The incoherent contribution is
\begin{align} \label{eqn:FPIcond}
	g_\dir(V) = \nu-R_{1\ast}(eV)-R_{2\ast}(eV),
\end{align}
where $R_{j\ast}(eV)$ are the renormalized reflection coefficients (see Eq.~(\ref{eqn:FPIrenRefl}) below).
The dependence of the AB conductance  on external parameters --- the gate voltage $V_g$, the variation
of the magnetic field $\Delta B$ and the bias $V$ --- factorizes into
\begin{equation}
g_\osc(\mu_+,\mu_-,\phi,V_g)= \tilde g(V) \cos[\varphi_{\rm AB}(V_g, \Delta B)]. 
\label{eqn:g_osc}
\end{equation}
The AB phase $\varphi_{\rm AB}$ will be discussed in the details shortly. The amplitude of the oscillations is
\begin{align} \label{eqn:condAmplitude}
	\tilde g(V)=e^{-\tau/\tau_\varphi} R_{12\ast}(eV)\ 2\left\lvert\cos\left(\lvert eV\tau\rvert +\frac\pi{4\nu^\ast}\right)\right\rvert
\end{align}
with the nonequilibrium dephasing rate given by
\begin{align} \label{eqn:FPIdephRate}
	\tau_\varphi^{-1} = \lvert eV\rvert \left(R_{1\ast}(eV)+R_{2\ast}(eV)\right) \frac 2\pi \sin^2\frac\pi{2\nu^\ast}.
\end{align}
In Eqs.~(\ref{eqn:FPIcond}) and (\ref{eqn:condAmplitude}) we have introduced the renormalized reflection
coefficients defined as
\begin{align}\label{eqn:FPIrenRefl}
\begin{split} 
	R_{j\ast}(eV)&=R_j\left\lvert\frac{\wc}{eV}\right\rvert^{1/\nu^\ast} \frac{e^{\gamma/\nu^\ast}}{\Gamma(2-1/\nu^\ast)} =R_{j\ast}(\epsth)\left\lvert\frac{\epsth}{eV}\right\rvert^{1/\nu^\ast},\\
	R_{12\ast}(eV)&=R_{12}\ \lvert\wc \tau\rvert^{1/2\nu^\ast} \left\lvert\frac{\wc}{eV}\right\rvert^{1/2\nu^\ast} \frac{2^{1/2\nu^\ast} e^{\gamma/\nu}}{\Gamma(1-1/2\nu^\ast)}= R_{12\ast}(\epsth)\left\lvert\frac\epsth{eV}\right\rvert^{1/2\nu^\ast}.
	\end{split}
\end{align}
Remarkably, in the last equation the amplitudes $r_1$, $r_2$ do not renormalize separately,
rather the renormalization operates non-locally. A similar result was found for FPIs in the fractional QHE 
regime in Ref.~\cite{Chamon97}. The relations~(\ref{eqn:FPIrenRefl}) are valid for bias
in the range $\epsth\ll eV\ll \wc$. The above renormalization comes from {\it virtual} electron-hole excitations  
(being a precursor of weak Coulomb blockade~\cite{Matveev95, Bagrets05}) and stops at $eV \simeq \epsth$.
On the contrary, the dephasing rate $\tau_{\varphi}^{-1}$ is caused by {\it real} e-h pairs excited by
backscattered electrons and is proportional to the shot noise of the QPCs.
There a simple linear dependence of the shot noise on voltage, which is valid in the absence of interaction, 
is modified because of the renormalization of reflection coefficients.

\begin{figure}[ht]
\centering{\includegraphics[scale=0.25]{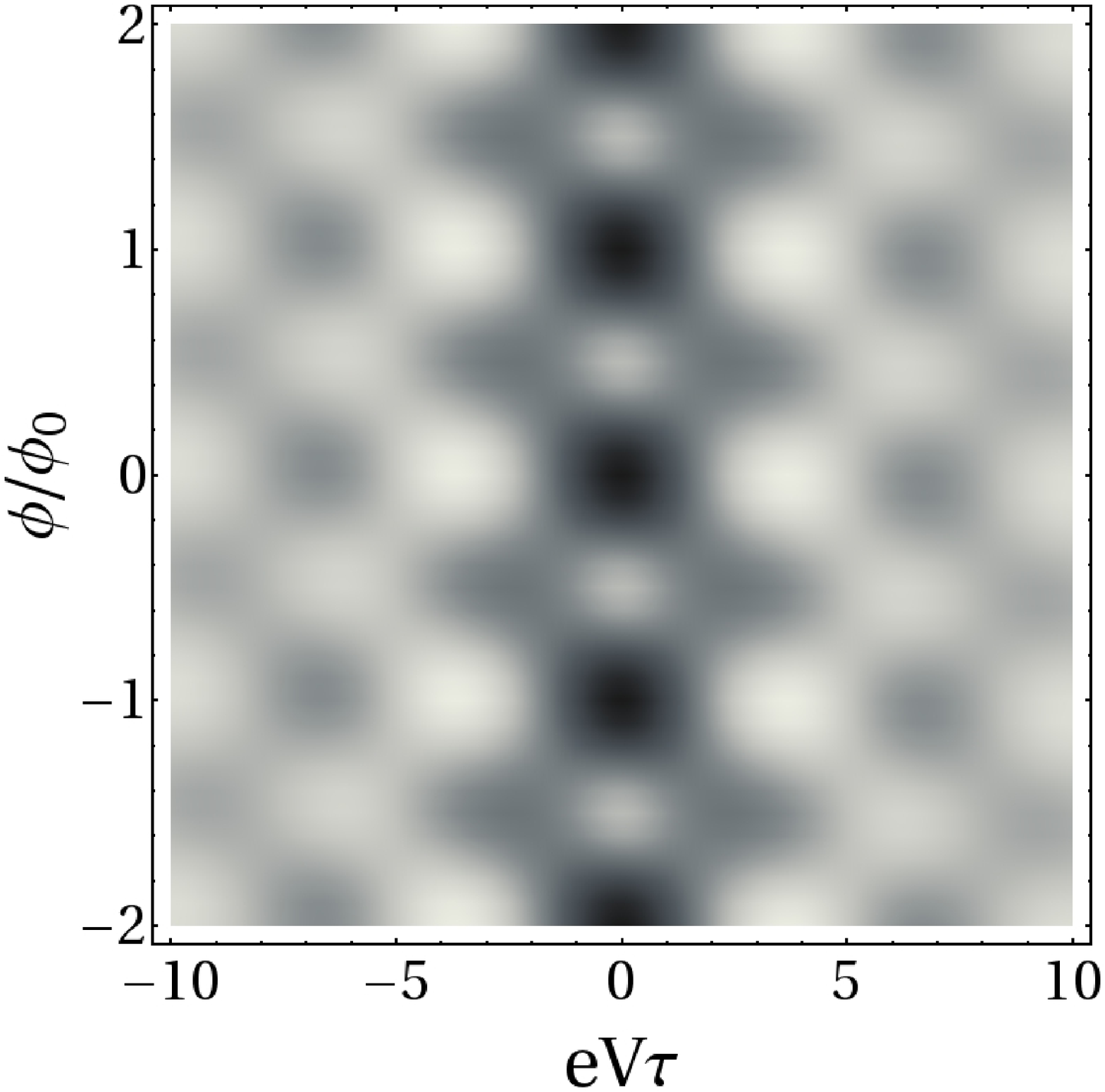}
\includegraphics[scale=0.35]{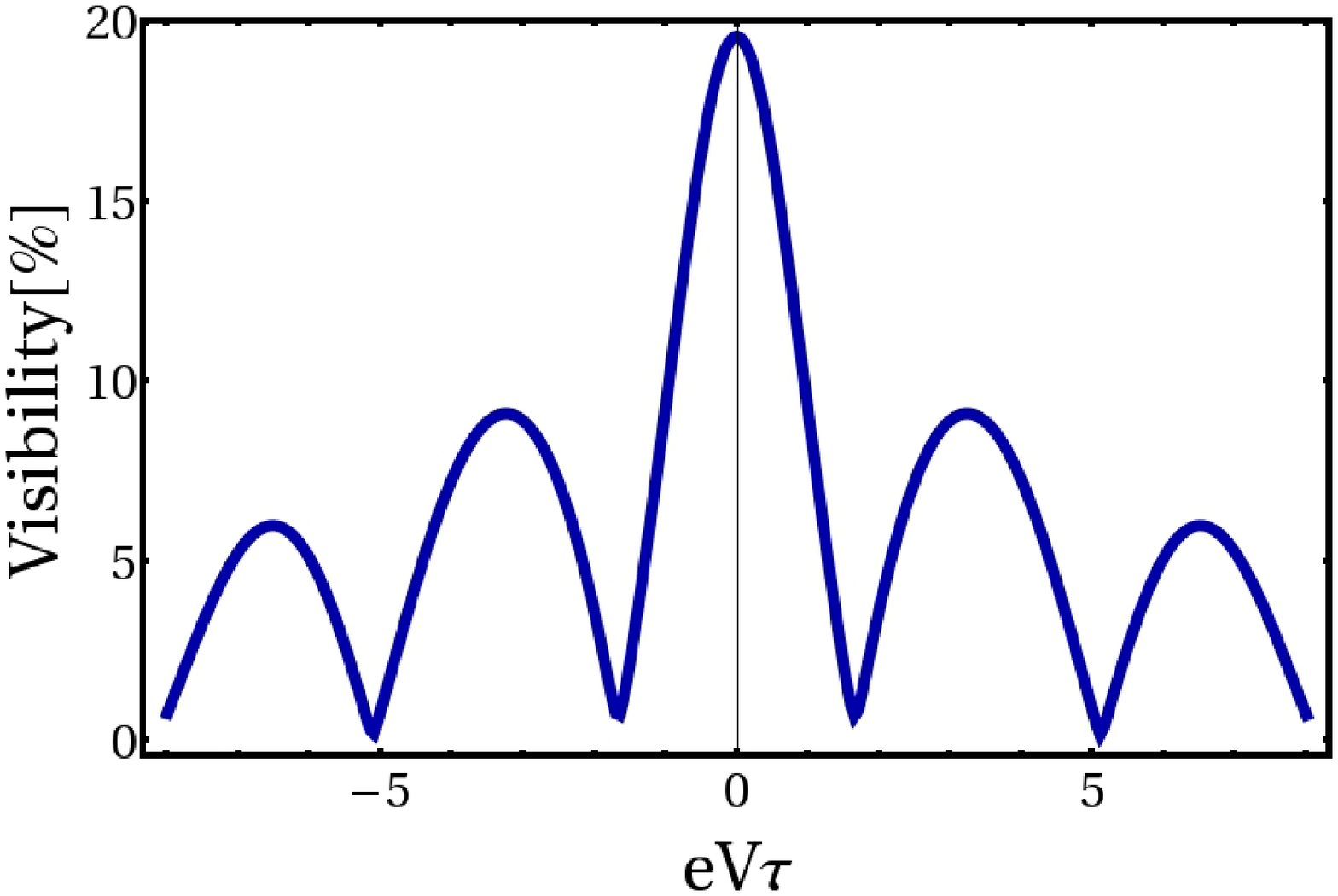}}
\caption{Total differential conductance $g$ as a function of bias and magnetic flux (left panel), 
and visibility (right panel). Parameters are: $\nuEff=2$, $\wc\tau=25$,
$R_{1\ast}(\epsth)=R_{2\ast}(\epsth)=0.2$.}
\label{fig:GGateBias}
\end{figure}

A functional dependence on bias in the conductance amplitude (\ref{eqn:condAmplitude}) stems from
an oscillatory prefactor which has a characteristic scale $\pi\epsth$. 
As a consequence, the amplitude or, equivalently, the visibility
$v(V)=\lvert\tilde g(V)\rvert/g_\dir(V)$ vanishes for certain equidistantly
distributed values of bias. The resulting characteristic ``lobe'' structure of
visibility is shown in Fig.~\ref{fig:GGateBias} and is in agreement with
experiments reported in Refs.~\cite{Yamauchi09,MarcusWest09}.

\subsubsection{Aharonov-Bohm oscillations} 
\label{sect:FPIABRes}

In experiment one usually characterizes the FPI in terms of a pattern of its
equilibrium conductance in the $(B,V_g)$~-~plane, which is governed by AB phase. We have identified four
different regimes where the behavior of AB oscillations is qualitatively
different (see Table~\ref{tab:table1}). In this table the parameter 
$\nu^\ast$ --- the effective number of transmitted channels which screen the
Coulomb interaction in the 
interfering channel  --- depends on the relative strength of the inter-edge {\it e-e} interaction.
To distinguish between the limits of weak and strong {\it e-e} interaction we compare an 
inter-channel interaction energy $\sim e^2/C_{e\alpha}$ (here $\alpha=r,l$) with a screened by the gate
charging energy of the interfering edge itself given by $\sim e^2/C^*_{\alpha g}$.
Here the edge-to-gate capacitance is effectively increased by the so-called ``quantum capacitance'': 
\begin{equation}
C^\ast_{\alpha g}= C_{\alpha g}+(\nu-1)\tau e^2/(\hbar \pi). 
\label{eq:Q_Capacitance}
\end{equation}
In the weak coupling limit
one has $C_{e\alpha}\gg C^\ast_{\alpha g}$. In this case the electrostatic potentials on all
edge channels ($r,l$, and $e$) are approximately equal to each other and we set $\nu^\ast\simeq\nu$.
In the opposite strong coupling limit we have $C_{e\alpha}\ll C^\ast_{\alpha g}$. The potential
$\varphi_e$ here fluctuates independently of potentials on other edge channels ($r$ and $l$), thus 
screening of {\it e-e} interaction by the latter channels is not effective and one gets $\nu^\ast\simeq 1$.
\begin{table}[h]
\caption{\label{tab:table1}
``Phase diagram'' of the FPI, which discriminate between two  Aharonov-Bohm (AB and AB*) 
and two Coulomb dominated (CD I and CD II) regimes.}
\begin{center}
\begin{tabular}{c||c|c}
	& \fbox{$C_{e\alpha}\gg C_{\alpha g}^\ast$} & \fbox{$C_{e\alpha}\ll C_{\alpha g}^\ast$} \\
	& $\nu^\ast=\nu$ & $\nu^\ast=1$ \\
	& $\bar C_{ei}=C_{ei}+C_{ri}+C_{li}$ & $\bar C_{ei}=C_{ei}$ \\
	& $\bar C_{eg}=C_{eg}+C_{rg}+C_{lg}$ & $\bar C_{eg}=C_{eg}$ \\
	\hline\hline
	$\bar C_{ei}\ll C_{ig}$ & \regABNu & AB* \\
	\hline
	$\bar C_{ei}\gg C_{ig}$ & \regCBNu &\regCBI
\end{tabular}
\end{center}
\end{table}
To make a distinction between the ``Aharonov-Bohm'' (AB) and ``Coulomb-dominated'' (CD) regimes
we now define an effective edge-to-island capacitance $\bar C_{ei}$ by the relation
$\bar C_{ei}=C_{ei}+C_{ri}+C_{li}$ in the weak coupling limit, i.e. at $\nu^\ast\simeq\nu$,
and set it to be $\bar C_{ei}=C_{ei}$ in the opposite case of strong coupling. Then the FPI
falls into AB or CD regimes depending on a ratio $\bar C_{ei}/C_{ig}$, as it is shown in the Table~\ref{tab:table1}.
As one can see from Eqs.~(\ref{eqn:C_gate}) and (\ref{eqn:C_edge}), for the device with a top gate
the capacitance $C_{ig}$ scales like $r^2$ when the FPI size grows, while $\bar C_{ei}$ increases
only as $r\ln r$. This suggests a simple rule of thumb: the AB regime occurs primary in large FPIs
with a top gate (in experiment ``large'' means a cell area $\sim$20$\mu{\rm m}^2$). 
In this situation, i.e. at $\bar C_{ei}/C_{ig} \ll 1$, fluctuations of charge on the island
are screened by the gate electrode and do not affect the AB conductance. In the case of opposite ratio
between the capacitances ($\bar C_{ei}/C_{ig} \gg 1$), as one will see shortly,
the AB conductance becomes linked to the Coulomb blockade on the compressible island.
This explains a terminology choice --- "Coulomb-dominated" --- for the above regime.

For a device without the top gate a bare edge-to-gate capacitance $C_{eg}$ is due to only
a plunger gate (see Fig.~\ref{fig:Setup}). Such gate is used to control the size of the interference
loop and because of geometry $C_{eg}$ typically is very small, so that one has $C^\ast_{eg} \simeq (\nu-1)\tau e^2/(\hbar \pi)$.
In this case our first condition of weak versus strong inter-edge {\it e-e} coupling can be simplified.
Defining the dimensionless coupling constant as 
\begin{align*}
	\alpha_\nu \equiv \frac{(\nu-1)e^2}{\epsilon \hbar \vF}
\end{align*}
and using the estimate~(\ref{eqn:C_edge}) for capacitances $C_{er}$ and $C_{el}$ one obtains the 
crossover value 
\begin{align*}
\alpha^\ast \sim \frac{1}{2\pi}\ln \left(\frac{b_e b_r}{a_{er}^2}\right),
\end{align*}
which sets the boundary between the weak and strong coupling regimes.

We name four regimes AB, AB*, CD I and CD II according to the Table~\ref{tab:table1} above which
for the benefit of the reader lists the values of parameters $\nu^\ast$, $\bar C_{ei}$ and $\bar C_{eg}$.
The capacitance $\bar C_{eg}$ here is defined in analogy to  $\bar C_{ei}$. 
In addition to these effective edge-to-island and edge-to-gate capacitances
we now define full island and edge capacitances as
\begin{align} \label{eqn:effCaps}
	\bar C_i=\bar C_{ei}+C_{ig}, \quad \bar C_e=\bar C_{eg}+\bar C_{ei}C_{ig}/\bar C_i.
\end{align}
Then the AB phase in Eq.~(\ref{eqn:g_osc}) reads
\begin{align} \label{eqn:FPIABPhase}
	\varphi_{\rm AB} = 2\pi \phi/\phi_0 - \frac{2\pi}{\nu^\ast} \frac{\bar C_{ei}}{\bar C_i}(N_{i\ast}+\nu \phi/\phi_0) 
 + \frac{|e|}{\omega_C} \left( 2 V_g - V_+ - V_-\right)
\end{align}
with integer $N_{i\ast}$ minimizing the charging energy of the Coulomb island
\begin{align} \label{eqn:islandChargingEnergy}
	E_i = \frac{e^2}{2 \bar C_i}(N_{i\ast}+\nu \phi/\phi_0 - C_{ig} V_g/\lvert e\rvert)^2,
\end{align}
and we have defined $\omega_C = \nu^* E_C/\pi$ and $E_C = e^2/\bar C_e$.

\begin{figure}[ht]
\centering{
\includegraphics[scale=0.23]{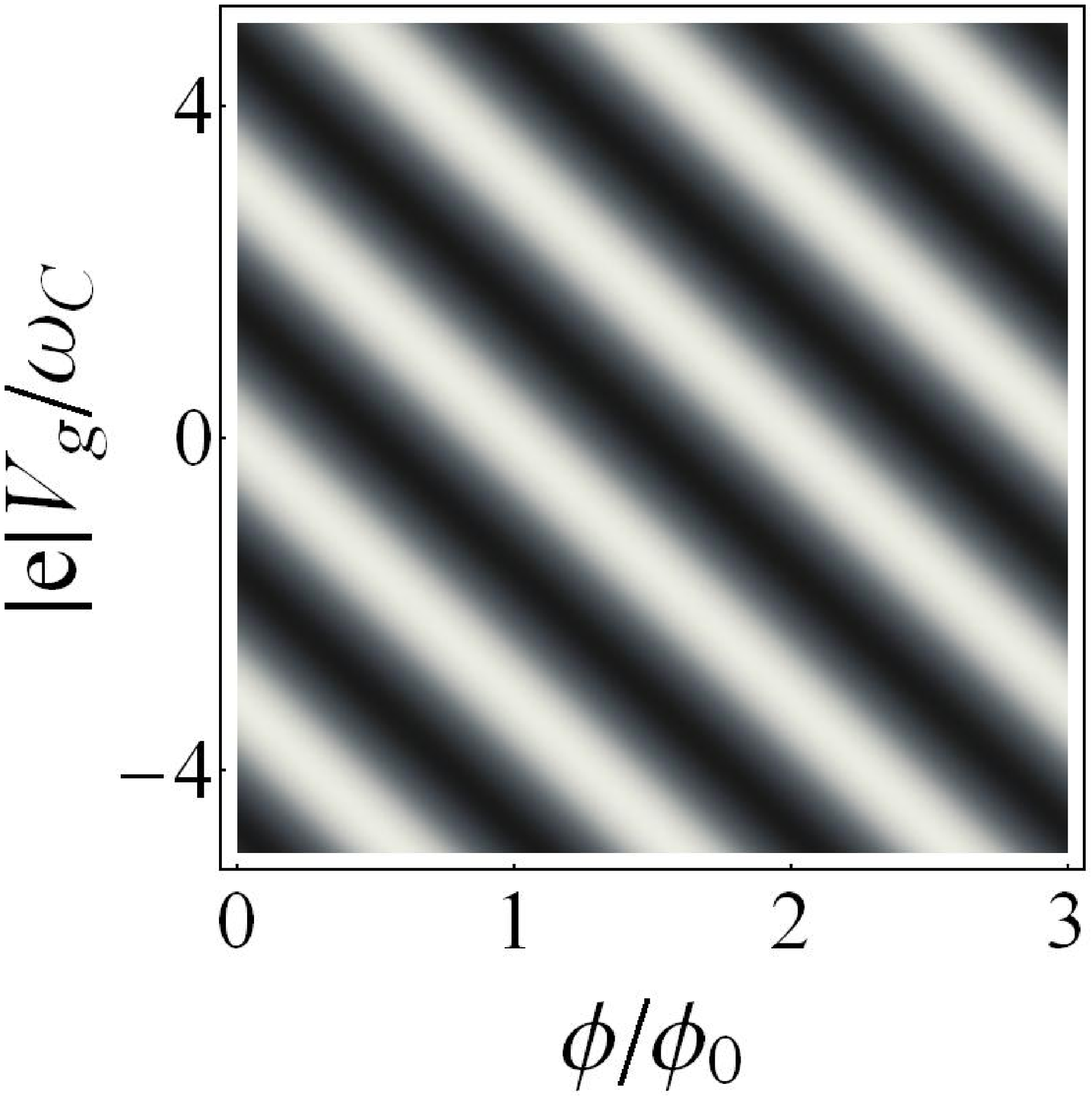}
\hspace{10mm}
\includegraphics[scale=0.23]{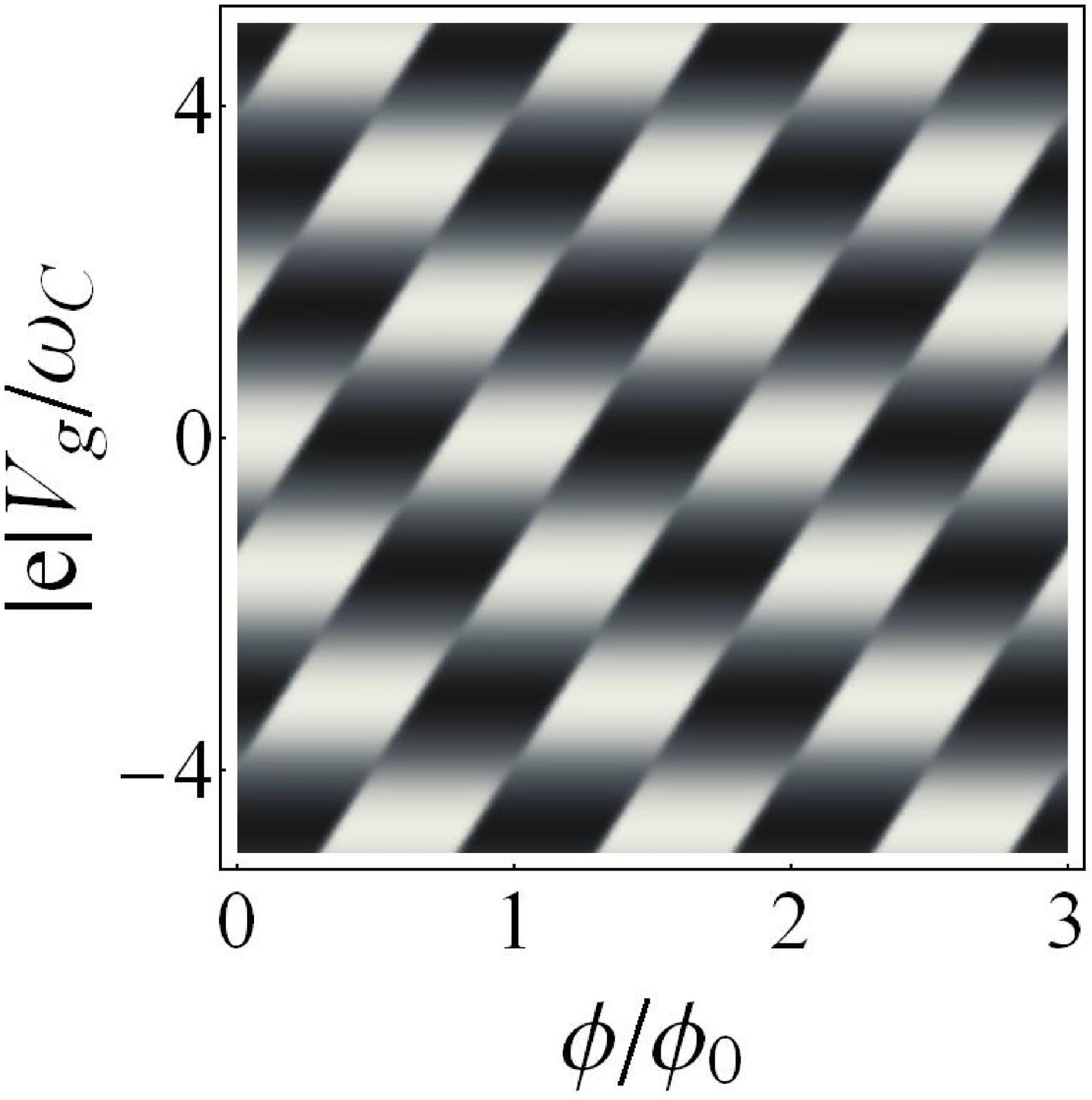}
\hspace{10mm}
\includegraphics[scale=0.23]{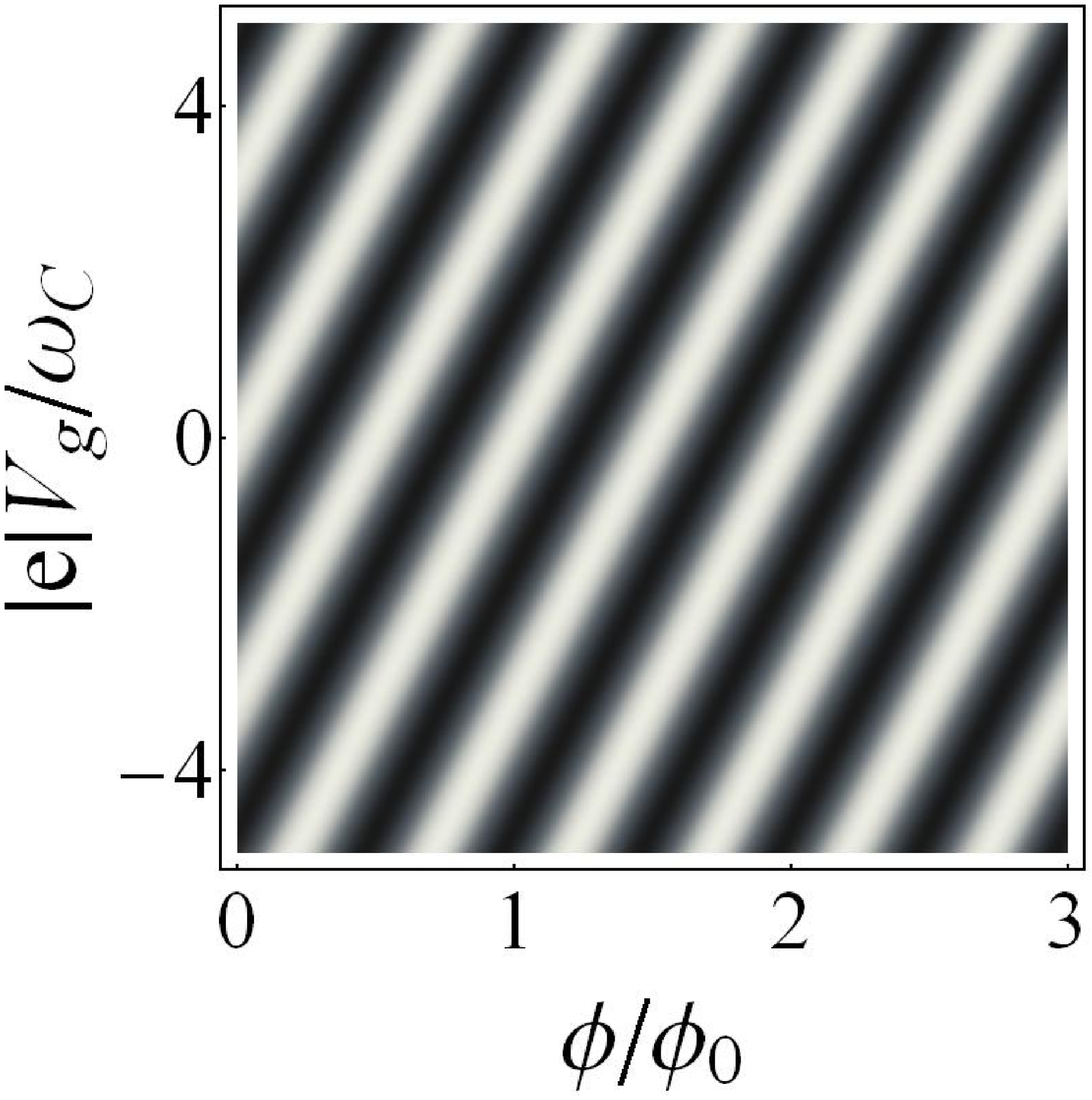}
}
\caption{Aharonov-Bohm conductance: 
AB regime (left), $\nuEff=\nu=2$, stripes of constant conductance have a negative slope, 
flux and gate voltage periods are $\Delta \phi=\phi_0$ and $\Delta V_g=\nu\ec/\lvert e\rvert$;
type-I CD regime (middle), $\nuEff=1$, $\nu=3$, stripes of constant conductance have now a positive slope, 
flux and gate voltage periods are $\Delta \phi=\phi_0/(1-\nu)$ and $\Delta  V_g= \ec/\lvert e\rvert$;
type-I CD regime (right), $\nuEff=\nu=2$, $C_{ig}/\bar C_e=0.6$, at fixed gate voltage $V_g$ 
conductance depends discontinuously 
on flux --- phase jumps occurs at every $\Delta \phi=\phi_0/\nu$.}
	\label{fig:AB_conductance}
\end{figure}

In Fig.~\ref{fig:AB_conductance} we show the conductance $g_\osc(\phi,V_g)$ in the ($B$,$V_g$)--plane for 
three regimes: AB, CD-I and CD-II. The plots display significant differences.
In particular, the lines of constant phase have a different slope in the AB and type-I CD regimes. 
The flux periodicity is also different in these two cases. The AB conductance in the case of type-II CD regime  
shows the ``rhomb-like'' pattern. The pattern of equilibrium conductance in the AB* case is the same as
in the AB regime, provided one sets $\nu^*=1$.
 
\subsubsection{Discussion and comparison with experiment}

Let us now discuss the physics which underlies the rich phenomenology of our rather simple model. To appreciate the role of interaction we consider first the non-interacting case, where the interfering channel couples neither to the fully transmitted ones nor to the island.  Consider electron contributing to the tunneling current which is, say,
incident from the left source and leaks into the left drain. It may tunnel
either at the left or right QPC. The latter path is longer than the first by
$2L=2\vF \tau$ and encircles a magnetic flux $\phi$. Along this path the
electron accumulates a dynamic (``Fabry-P\'erot'') phase $2\epsilon \tau$
($\epsilon$ is the energy of the electron) and a magnetic AB phase $2\pi
\phi/\phi_0$. According to quantum mechanics, the current results from
interference of both paths. Integration over all energies in the range
$\mu_-<\epsilon<\mu_+$ gives the back-scattering current
\begin{equation}
	I_b = -e \int_{\mu_-}^{\mu_+}\frac{\dd \epsilon}{2\pi}\, \left\lvert r_1+e^{2\pi i \phi/\phi_0+2i\epsilon\tau}r_2\right\rvert^2=I_{\dir}+I_\osc,
\end{equation}
that we have split into incoherent and coherent contributions,
\begin{equation}
	I_{\dir} = -\frac{e^2}{2\pi} V(R_1+R_2),\quad I_\osc =\frac{e}{2\pi} \frac{2R_{12}}\tau \cos[(\mu_++\mu_-)\tau-2\pi \phi/\phi_0] \sin [eV\tau].
	\label{eqn:I_AB}
\end{equation}
While the incoherent part of the current $I_{\dir}$ is expected already on the
classical level, $I_\osc$ stems from interference and is sensitive
to magnetic flux. The dynamic phase accumulated by an electron depends on its
``absolute'' energy $\epsilon$, and hence the current $I_\osc$ depends on both
chemical potentials $\mu_+$, $\mu_-$, but not just 
on their difference $eV=\mu_+-\mu_-$.  Clearly, the sum ($\mu_++\mu_-$)
enters only into the phase shift of the AB pattern, but not in the amplitude.
However, this independence of the amplitude of oscillations on the bias
does not in general hold for the differential conductance $g_\osc$. 
Specifically, when the differential conductance is calculated in the framework
of the model of non-interacting electrons, the amplitude of the corresponding
AB oscillations $\tilde g$ does depend on the manner in which bias is applied.

Experimentally, the bias is applied asymmetrically: $\mu_+=eV$, $\mu_-=0$. The
expected conductance then is
\begin{align*}
	g_\osc=G_Q^{-1} \partial_V I_\osc = -2 R_{12}\cos[2eV\tau+2\pi\phi/\phi_0],
\end{align*}
i.e. bias merely controls the phase shift of the AB oscillation pattern. The amplitude $\tilde g=-2 R_{12}$ is independent of bias. This clearly contradicts to our results presented above as well as to experimental observations.

The situation would change essentially if the bias were applied symmetrically:
$\mu_+=eV/2$, $\mu_-=-eV/2$. Then the conductance would be
\begin{equation}
	g_\osc= -2  R_{12}\cos[eV\tau] \cos[2\pi\phi/\phi_0].
	\label{g-osc}
\end{equation}
Now, the amplitude would oscillate with bias on the scale $\pi \epsth$, yielding a visibility with a 
``lobe'' structure. This result is apparently much more similar to our findings
(albeit without dephasing and renormalization of $R_{12}$) as well 
to the experimental observations. On the basis of the similarity between
Eq.~(\ref{g-osc}) and the experimental observations it was conjectured
in Ref.~\cite{ZhangMarcus09} that the electron-electron interaction
effectively symmetrizes the bias even if the latter is applied asymmetrically.

To see how this works, assume that a charge within the interferometer cell
produces a (for simplicity constant) self-consistent potential $\varphi_0$. An
electron which propagates in this potential during a time $2\tau$ accumulates
the ``electrostatic'' AB phase $-2\varphi_0\tau$. Hence, the dynamic phase would
be $2(\epsilon-\varphi_0)\tau$ and instead of the bare chemical potentials the
relative potentials $(\mu_\pm-\varphi_0)$  
enter the result~(\ref{eqn:I_AB}). Such a mean-field potential is indeed generated within our model. 
For instance, in the generic limit of large charging energy $\ec\gg\epsth$ our calculations in the 
subsection~\ref{sect:FPI_Calculations} yield $\varphi_0\simeq (\mu_+ + \mu_-)/2$ in the case of AB regime.
Therefore without a need of any fine tuning, the bias is effectively
symmetrized, which explains the appearance of the ``lobe'' structure.

The mean-field potential $\varphi_0$ on the compressible strip corresponding to the interfering edge
channel is in general the function of applied chemical potentials $\mu_\pm$, the gate voltage $V_g$
and the magnetic flux $\phi$. The most general expression found in the subsection~\ref{sect:FPI_Calculations}
reads
\begin{equation}
\varphi_0(\mu_\pm, V_g, \phi) = \frac 1{1+\wc \tau}\left[eV_g +\frac{\mu_++\mu_-}2 \wc\tau+\ec\frac{\bar C_{ei}}{\bar C_i}(N_{i\ast}+\nu\phi/\phi_0)\right].
\label{eqn:vaphi_0}
\end{equation}
Here $N_{i\ast}$, as before, provides the minimum for the Coulomb energy $E_i$ of the island given by 
Eq.~(\ref{eqn:islandChargingEnergy}). If we introduce the electrochemical potential
\begin{equation}
\tilde\varphi_0 = \varphi_0 - (\mu_+ + \mu_-)/2, 
\label{eqn:tvarphi0}
\end{equation}
then the AB phase, given by Eq.~(\ref{eqn:FPIABPhase}), is equivalently represented by relation
\begin{equation}
 \varphi_{AB}(\mu_\pm, V_g, \phi)= 2\pi \phi/\phi_0 - 2\tau \tilde\varphi_0(\mu_\pm, V_g, \phi).
 \label{eqn:AB_phase1}
\end{equation}
The first contribution here is the magnetic AB phase accumulated along a fixed reference loop with
area $A_0$, i.e. $\phi = A_0 \delta B$, with $\delta B \ll B$ being a weak modulation of magnetic field
on top of the high field $B$ which drives the 2DEG into the QHE regime. Because of the condition
$b\ll L$ (see Fig.~\ref{fig:Setup}~b) imposed in our model, and since a typical variation  $\delta B$
is such that $\phi$ changes on a scale of a few flux quanta only, one can use any boundary $y_k^\pm$
to define $A_0$. For example, one can set $A_0 = \pi (y_\nu^+)^2$. Let us further show, that
the second ``electrostatic'' contribution $(-2\tau\tilde\phi_0)$ to the phase $\varphi_{AB}$ can be
interpreted in terms of a motion of edge states which leads to the variation of a relevant 
FPI area when the magnetic field $B$ or gate voltage $V_g$ are varied.

\begin{figure}[b]
	\centering{\includegraphics[scale=0.5]{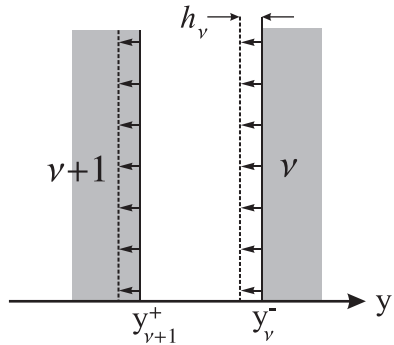}}
	\caption{When the boundaries of the compressible strip (white) separating incompressible regions (gray) with filling factors $\nu$ and $\nu+1$ move towards the $(\nu+1)$-liquid (indicated by arrows), the latter shrinks while the $\nu$-liquid's area grows: as the result the charge on the incompressible strip, defined by the boundaries
$y_{\nu+1}^-$ and $y_\nu^+$,  decreases (in this figure $h_\nu<0$). } \label{fig:movingEdge}
\end{figure}

First, we note that in a stationary limit an imbalance of electron density per unit length on the
interfering edge channel $\delta\rho_e$ is related with the corresponding electrochemical
potential~(\ref{eqn:tvarphi0}) by a simple relation 
$\delta \rho_e = -\tilde\varphi_0/(2\pi v_F)$,
since $(2\pi v_F)^{-1}$ is just but the 1D thermodynamic density of states in our model. 
As it is always the case in QHE systems,
this charge density can be translated into the variation $h_\nu$ of the boundary between
the compressible and incompressible strips (Fig.~\ref{fig:movingEdge}),
$\delta \rho_e = n_L h_\nu$, 
where $n_L = B/\phi_0 = (2\pi\lambda_B^2)^{-1}$ is the electron concentration on one completely filled LL,
and we have assumed the fluctuations of inner and outer boundaries of the compressible strip to be
the same, $\delta y_{\nu+1}^+ =  \delta y_{\nu}^- = h_\nu$. Therefore ``electrostatic''
part of the AB phase reads
\begin{equation}
 -2\tau\tilde\varphi_0 = 4\pi L \delta \rho_e = 4\pi n_L (L h_\nu) = 2\pi (B/{\phi_0})\delta A.
\end{equation}
Here $\delta A = 2 L  h_\nu$ is the change in area enclosed by interfering edge state. 
To recapitulate the logic, we have thus related the self-consistent electrostatic potential
$\varphi_0$ to a variation of the FPI area. With these arguments at hand one can now rewrite 
Eqs.~(\ref{eqn:vaphi_0}) and (\ref{eqn:AB_phase1}) in the equivalent form
\begin{equation}
 \phi_{AB} = 2\pi \bigl( {A_0 \delta B} +  {B\delta A}\bigr)/{\phi_0} + {\rm const}\,,
\end{equation}
where the variation of area reads
\begin{equation}
\delta A = -\frac{1}{\nu^*}\frac{\bar C_{ei}}{\bar C_i}
\left( \nu \frac{A_0}{B}\, \delta B + \frac{\phi_0}{B}\, \Delta N_i\right) + 
\frac{|e|}{\omega_C}\frac{\phi_0}{\pi B}\, \delta V_g.
\label{eq:dA}
\end{equation}
In the last equation we have introduced an integer $\Delta N_i$ which is a deviation in excess
number of electrons on the Coulomb island with respect to the excess charge corresponding to some 
initially chosen reference gate voltage $V_g$. We have also took into account the 
source-drain bias is small on a scale of the charging energy, $|e|V\ll E_C$. 

Let us now discuss our theory of the FPI in relation to the recent experiments by considering separately
each of the four regimes in the Table~\ref{tab:table1}. 
In the \emph{AB regime} one has $\bar C_{ei}/C_i \ll 1$, thus the coupling of the interfering edge
to the Coulomb island is negligible and the area $A$ does not change with $B$. 
The AB phase then simplifies to
\begin{equation}
	\varphi_{\rm AB}= 2\pi \phi/\phi_0 +\frac{2\pi}\nu \frac{\lvert e\rvert V_g}{\ec},
	\label{eqn:phase_AB_regime}
\end{equation}
yielding the lines of constant phase with a negative slope~(Fig.~\ref{fig:AB_conductance}, left)
and a magnetic field period $\Delta B = \phi_0/A_0$ which is independent of $\nu$.
The second term in the above equation describes a modulation in space of electron trajectory under variation 
of the gate voltage. If $\delta V_g > 0$ then the interfering edge state moves outwards and thus
encloses a larger flux as it is seen from Eq.~(\ref{eq:dA}).  The AB regime was observed in large devices 
(cell area $\sim$18$\mu{\rm m}^2$) with a top gate~\cite{MarcusWest09, ZhangMarcus09, Heiblum09}, 
where the condition $\bar C_{ei} \ll C_{gi}$ is satisfied. In Ref.~\cite{ZhangMarcus09} it
has been also found that magnetic field period $\Delta B$ is independent of $B$ but gate
voltage periodicities (both top- and plunger- ones) scale as $\Delta V_g \propto 1/B \propto \nu $. 
These observation are consistent with our Eq.~(\ref{eqn:phase_AB_regime}) if the
charging energy $E_C = e^2/{\bar C_e}$ is $\nu$~-~independent. It is so, provided the
full edge capacitance $\bar C_e \simeq C_{eg} + \bar C_{ei}$ stays approximately the same at
each Hall plateau. 

In the \emph{CD regime} one has $\bar C_{ei}/C_i \simeq 1$ and the area of the interfering loop
shrinks with the magnetic field. This is because interfering edge is now electrostatically coupled to
the charge $Q_i = e(N_i + \nu \phi/\phi_0)$ on the Coulomb island, which has explicit dependence on flux.
The AB phase in this regime reads
\begin{align*}
	\varphi_{\rm AB} = 2\pi (1-\nu/\nu^*)\phi/\phi_0 -\frac{2\pi}{\nu^*} N_i + 
\frac{2\pi}{\nu^*} \frac{\lvert e\rvert V_g}{\ec},
\end{align*} 
In the type-II CD regime $\nu^* \simeq \nu$ and at fixed $V_g$ the AB phase stays piecewise constant 
when the magnetic field is varied. The dependence of phase on $B$ exclusively enters via $N_i$. 
Indeed, as it follows from Eq.~(\ref{eq:dA}) the shrinkage of the
interfering loop in this case, $\delta A = - (A_0/B)\delta B $, exactly compensates 
a change in magnetic phase $\varphi = A_0\delta B$. When the FPI is brought close to a charge degeneracy 
point of the island by varying $V_g$ or $B$, electron tunneling becomes possible between the droplet and
interfering channels (i.e. $\Delta N_i = \pm 1$) resulting in abrupt change of $A$. This creates a phase lapse
(or jump) $\Delta\phi_{AB} = \pm 2\pi/\nu$ giving rise to the ``rhomb-like'' pattern shown
in Fig.~\ref{fig:AB_conductance} (middle) at $\nu \ge 2$. 

In the type-I CD regime $\nu^* \simeq 1$ and a change in AB phase caused by area shrinkage when rising $B$
overcompensates the magnetic AB phase, since now $\delta A = - \nu(A_0/B)\delta B$.
Counterintuitively, the phase decreases for increasing the magnetic field.
At the same time, whenever an electron tunnels into the island from the interfering edge channel ($e$),
the boundary of this edge state contracts so as to expel exactly one flux quantum from the AB loop.
The phase lapse, being equal to $\Delta\phi_{AB} = (- 2\pi)$ in this tunneling process, is therefore 
invisible in the interference conductance. 
As the result, one has the diagonal stripe pattern with lines of constant phase having positive slope 
(Fig.~\ref{fig:AB_conductance}, right).
The periods are $\Delta B = \phi_0/(f_T A)$ and $\Delta V_g = e/(C_{eg} + C_{gi})$, with $f_T=\nu-1$
being the number of fully transmitted edge channels (note, that at $\nu=1$ the lines of constant phase 
are vertical).

In the limit of weak backscattering at QPCs, the Coulomb-dominated regime has been observed in 
Ref.~\cite{Heiblum09}. In this work measurements were performed with the set of edge state configurations (including fractional fillings), classified by bulk filling $f_b$ and number $f_T$ of fully transmitted edges. 
We focus on the results obtained with a 4.4\,$\mu{\rm m}^2$-device without a top (but with plunger-) gate. They indicate that interaction plays a major role (i.e. CD-I and CD-II regimes). 
For integer $f_b$ and $f_T=f_b-1$ the results coincide with our Fig.~\ref{fig:AB_conductance}~(right),
including the period in magnetic field which scales as $\Delta B \propto 1/f_T$. 
The gate period in Ref.~\cite{Heiblum09} was found to be weakly increasing with $f_T$. 
We can explain this dependence if one assumes that the full edge capacitance, which equals to 
$\bar C_e = C_{gi} + C_{eg}$ in the CD-I regime, 
decreases with $f_T$, since a mutual coupling of the plunger-gate to the inner
interfering edge channel ($C_{eg}$) becomes less efficient at high $\nu$.  
Quite ``exotic'' behavior was observed for more than one number of channels trapped in the interferometer cell. 
In case of $f_b=4$ and $f_T=1$ experimental findings resemble very much Fig.~\ref{fig:AB_conductance}~(middle) 
corresponding to our CD-II regime. In such setting one fully transmitted and one partially reflected edge channel can be
described by our model with $\nu=2$ assuming that other two inner trapped channels play a role of
a compressible island, where the excess charge is quantized. 

Non-equilibrium transport measurements in the FPIs in the AB regime have been performed 
in Refs.~\cite{MarcusWest09, Yamauchi09}.
Their main findings can be summarized as follows:~(i)~the dependence of AB conductance versus $B$ and the bias $V$ 
factorizes into a product of two terms yielding a ``checkerboard" pattern in the $(\delta B,V)$-plane
(cf. Fig.~\ref{fig:GGateBias}, left); 
(ii) the scale of the ``lobe"-structure is set by $\epsilon_{\rm Th}\sim v_D/L$; (iii)~a visibility decay
with bias is stronger at higher magnetic fields. Results of our theory, 
Eqs.~(\ref{eqn:g_osc}) - (\ref{eqn:FPIrenRefl}), are in full accord 
with these observations. In particular, a suppression of visibility in our model at $|r_j|^2\ll 1$ is mainly due to a power-law
decay with the {\it negative} exponent $(-1/2\nu^*)$. 
Since in the AB regime $\nu^*=\nu\propto 1/B$, 
this decay is stronger in case of a small number of edge channels, i.e. at higher $B$, 
in agreement with Ref.~\cite{MarcusWest09}.

It is interesting to note, that our theory predicts a fourth regime (AB$^\ast$, see Table~\ref{tab:table1}).
It is characterized by the same equilibrium conductance pattern as the AB regime 
(Fig.~\ref{fig:AB_conductance}, left),  but in contrast to the latter, the power-law 
decay of the visibility oscillations corresponds to $\nu^\ast=1$, and thus is
independent of $B$. Such a behavior of the FPI has not yet been observed in the experiment.  

Closing this section we have to mention that 
a crossover from the AB to the type-I CD regime (in our) terminology has been recently discussed in
details in Ref.~\cite{Halperin10}. We note that our capacitance model is very similar in
spirit to the one used in that paper. However, the important difference is that our approach
takes explicitly into account quantum corrections to classical geometrical capacitances, 
given by Eq.~(\ref{eq:Q_Capacitance}).
As the result we obtain the extra type-II CD regime which may arise because of
screening of Coulomb interaction by the fully transmitted edge channels.

\subsection{Calculations}
\label{sect:FPI_Calculations}

\begin{figure}[ht]
	\centering{\includegraphics[scale=.4]{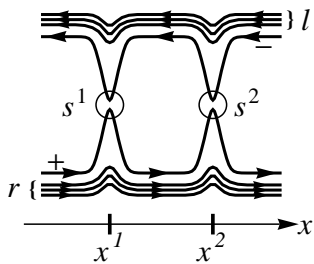} }
	\caption{Network model of the FPI.} \label{fig:FPISketch}
\end{figure}

We show here how to derive the above results using the formalism developed in
Sect.~\ref{sect:framework}. For simplicity we assume all edge channels to have
the same length $L$ and same velocity $\lvert v_\mu\rvert =\vF$
(with $v_\mu>0$ in right-moving channels and $v_\mu<0$ in left-moving ones).
Consequently, all flight times $\tau=L/\vF$ are the same. Further, the scatterer
1 has a coordinate $x^1$ and scatterer 2 has the coordinate $x^2=x^1+L$ for
each of the edges, see Fig.~\ref{fig:FPISketch}.

We remind that two characteristic energy scales play an important role in our analysis, 
namely, Thouless energy $\epsth=\tau^{-1}$ and charging energy $\ec=e^2/\bar
C_e$ (or charge relaxation frequency $\wc = \frac\nuEff\pi
\ec$), see Sec.~{sec:FPI-results}. As has been discussed there, we will assume
that the charging energy is much higher than the Thouless energy, and will
consider the range of voltages intermediate between these two scales,
$\wc\gg\lvert eV\rvert \gg \epsth$.

\subsubsection{Electrostatic Action}

Our network consists of $\nu$ right-moving and $\nu$ left-moving chiral channels
which we label with index $\mu$, see Fig.~\ref{fig:FPISketch}. The innermost
right-moving channel ($\mu=+$) is coupled to the innermost left-moving channel
($\mu=-$) by two scatterers $i=1,2$ with $2\times 2$-scattering matrices $s^i$.
The remaining chiral channels (right-moving ones labeled
$\mu=r1,\ldots,r(\nu-1)$, left-moving ones by $\mu=l1,\ldots,l(\nu-1)$) connect
sources to drains without any possibility of tunneling.

Interaction is taken into account by the electrostatic model (\ref{eqn:electrostaticE}) described in the beginning. For simplicity we assumed that electrostatic coupling between all fully transmitted right-moving channels ($\mu=r1,\ldots,r(\nu-1)$) is strong such that they share a common electrostatic potential $V_r$. This enables us 
to merge them into one conductor (labeled $\alpha=r$). We proceeded in the same
way with the fully transmitted left-moving ones ($\mu=l1,\ldots,l(\nu-1)$ are
now merged into $\alpha=l$) and the two innermost channels ($\mu=+,-$ are merged
into $\alpha=e$). This reduces the number of charge degrees of freedom
characterizing the edge channels down to three:
\begin{gather*}
	Q_e=\int\!\!\dd x\, \bar\psi_+\psi_++\int\!\!\dd x\, \bar\psi_-\psi_-,\quad Q_r=\sum_{\kappa=1}^{\nu-1} \int\!\!\dd x\, \bar\psi_{r\kappa}\psi_{r\kappa},\quad Q_l=\sum_{\kappa=1}^{\nu-1} \int\!\!\dd x\, \bar\psi_{l\kappa}\psi_{l\kappa}.
\end{gather*}
As a fourth conducting element, we introduce a central compressible island with
total charge $Q_i=e(N_i+\nu\phi/\phi_0)$. While the second contribution, the
charge in the $\nu$ fully occupied LLs of the central region, is fixed by
external parameters, the occupation $N_i$ of the partially filled LL of the
island is an (integer) degree of freedom to begin with. Since it is assumed to
fluctuate via very slow tunneling, $\Gamma\ll \epsth$, we will, however, treat
it in a mean-field approximation.

The fermionic action we start with reads as follows
\begin{align*}
	\act_0[\psi,\bar \psi,N_i]=\sum_\mu\,\int_\keldC\dd t \dd x\, \bar
\psi_\mu\left(i\partial_t+iv_\mu\partial_x\right)\psi_\mu - \frac 12
\sum_{\alpha\beta} \left(Q_\alpha-q_\alpha\right) \left(\tilde
C^{-1}\right)_{\alpha\beta} \left(Q_\beta-q_\beta\right)\,,
\end{align*}
where the first sum extends over the $2\nu$ chiral channels $\mu=r1,\ldots,r(\nu-1),l1,\ldots,l(\nu-1),+,-$ and the second one over the 4 conductors $\alpha,\beta=e,r,l,i$. The electrostatic part of $\act_0[\psi,\bar \psi,N_i]$ is, of course, a direct consequence of (\ref{eqn:electrostaticE}) and we refer to the corresponding section for definitions of $q_\alpha$ and $\tilde C$. Charges $Q_\alpha$ are, of course, dynamic, i.e.\ time-dependent, quantities, but for the sake of readability we leave time-dependence implicit (as we did with time integration in the above electrostatic action).

Our interest lies in interference effects which manifest themselves in tunneling corrections to current. Tunneling phases respond only to the Hubbard-Stratonovich field $\varphi=eV_e$ on the interfering edges (note that the electrostatic merging of channels $\mu=+,-$ allows us to use just one field $\varphi=\varphi_+=\varphi_-$). The short-term goal of the present section is to integrate out all other degrees of freedom.

\paragraph{Potentials $V_c=(V_e, V_r,V_l)^t$: } First, we decouple the quadratic charge terms via a (multidimensional) Hubbard-Stratonovich transformation, thereby (re)introducing the potentials $V_\alpha$ on the conductors. Since we do not need the potential $V_i$ on the island, we single out the island degrees of freedom beforehand, writing
\begin{align} \label{eqn:FPIES1}
	\tilde C^{-1} = \begin{pmatrix}
	                 	p_{cc} & p_{ci}\\
				p_{ic} & p_{ii}
	                \end{pmatrix},\ V_{ci} = p_{ci} (Q_i-q_i).
\end{align}
The index $c$ refers to the 3 indices $e$, $r$, $l$, that means $p_{cc}$, $p_{ci}=p_{ic}^t$, $p_{ii}$ are $3\times3$-, $3\times 1$-, and $1\times1$-matrices. With that the action reads
\begin{align*}
	\act_\Int[\psi,\bar\psi,N_i] &= -\frac 12 (Q_i-q_i)^2 p_{ii} -\frac 12 (Q_c-q_c)^t p_{cc} (Q_c-q_c) - V_{ci}^t (Q_c-q_c) \label{eqn:originalESAction}
\end{align*}
and becomes upon Hubbard-Stratonovich decoupling:
\begin{gather*}
	\act_\Int[\psi,\bar\psi, V_c, N_i]  -\frac 12 (Q_i-q_i)^2 p_{ii} + \frac 12 (V_c-V_{ci})^t p_{cc}^{-1} (V_c-V_{ci}) -V_c^t (Q_c-q_c)\\
	\text{with }V_c=(V_e, V_r,V_l)^t.
\end{gather*}

\paragraph{Integrating out $Q_e$, $Q_r$, $Q_l$: } Next, we integrate out the
charges $Q_e$, $Q_r$, $Q_l$. As explained in Sect.~\ref{sect:model} and
\ref{sect:tunAct} charges and potentials are decoupled by a gauge transformation
(\ref{eqn:phaseHSField}) which generates a tunneling term $\act_t$ (see below)
and, according to the Dzyaloshinskii-Larkin theorem, quadratic and linear (in
voltages) terms $V_\alpha^2 S_\alpha$ and $\bar Q_\alpha V_\alpha$. The former
is given by the polarization operators (\ref{eqn:genericPolOp}) and amounts for
screening, thus an ``enhancement'' of the capacitances (in fact, the
capacitances become complex, ``Keldysh''- and energy-dependent, but in the
static limit the corrections are indeed positive). Then the retarded/advanced
components of the ``screening capacitances'' read
\begin{align} \label{eqn:FPIScreenCap}
	\begin{split}
	S^{r/a}_\alpha(\omega) &=  -e^2\sum_{\kappa=1}^{\nu-1}\int\!\dd \xi_1\dd \xi_2\, \Pi_{\alpha\kappa}^{r/a}(\omega,\xi_1,\xi_2) = \pm i (\nu-1) e^2 \frac{1-e^{\pm i\omega \tau}}{2\pi \omega}, \quad \alpha=r,l,\\
	S^{r/a}_e(\omega) &= -e^2\sum_{\mu=\pm}\int\!\dd \xi_1\dd \xi_2\, \Pi_\mu^{r/a}(\omega,\xi_1,\xi_2) = \pm i  2e^2 \frac{1-e^{\pm i\omega \tau}}{2\pi \omega}.
	\end{split}
\end{align}
Charges injected from the reservoirs due to nonequilibrium boundary conditions (in excess of the equilibrium charge which is canceled by the positive background charge) are
\begin{align} \label{eqn:FPIExcessCharge}
	\bar Q_r =  (\nu-1)e \frac{\mu_+\tau}{2\pi},\quad \bar Q_l =  (\nu-1)e \frac{\mu_-\tau}{2\pi}, \quad \bar Q_e = e \frac{(\mu_++\mu_-)\tau}{2\pi}.
\end{align}

We collect them in the diagonal matrix $S=\diag(S_e,S_r, S_l)$ and the vector 
$\bar Q_c= (\bar Q_e,\bar Q_r, \bar Q_l)^t$. Subsequent elimination of charge degrees of freedom transforms $\act_\Int[\psi,\bar \psi, V_c,N_i]$ into
\begin{gather}
	\act_0[V_c,N_i] 
		 = -\frac 12 \bar p_{ii} (Q_i-q_i)^2+ \frac 12 V_c^t (p_{cc}^{-1}+S) V_c - V_c^t (\bar Q_c - q_c +p_{cc}^{-1} V_{ci}) \nonumber\\
		\text{with } \bar p_{ii}=p_{ii}-p_{ic} p_{cc}^{-1} p_{ci}.
\end{gather}

\paragraph{Integrating out $V_r$, $V_l$: } The final, and somewhat cumbersome
step is to integrate out 
the voltages $V_r$, $V_l$. In order to do that we again split the degrees of freedom, writing
\begin{align} \label{eqn:FPIES2}
	p_{cc}^{-1} = \begin{pmatrix}
	              	c_{ee} & c_{et}\\
			c_{te} & c_{tt}
	              \end{pmatrix}, \quad S=\begin{pmatrix}
	              													S_e & \\
	              													& S_t
	              												\end{pmatrix},
\end{align}
where the index $t$ refers to the 2 indices $r$, $l$, and $c_{ee}$, $c_{et}=c_{te}^t$, $c_{tt}$ are corresponding $1\times 1$-, $1\times 2$-, and $2\times 2$-matrices.
The action then reads
\begin{align*}
	\act_0[V_c,N_i]  =& -\frac 12 \bar p_{ii} (Q_i-q_i)^2 +\frac 12 (c_{ee}+S_e)V^2_e - V_e (\bar Q_e -q_e +c_{ee} V_{ei} +c_{et}V_{ti})\\
		& + \frac 12 V_t^t (c_{tt} +S_t) V_t - V_t^t (\bar Q_t-q_t +c_{te} V_{ei} + c_{tt} V_{ti}-c_{te} V_e).\\
\end{align*}

Performing the Gaussian integration over $V_r$, $V_l$ is a straightforward, albeit cumbersome calculation which in the end yields,
\begin{multline}
	\act_0[V_e,N_i] = -\frac 1{2\bar C_i} (Q_i - q_i)^2 - (\bar Q_t-q_t)^t \bar p_{ti} (Q_i-q_i)+\frac 12 \bar C^\ast_e V_e^2 -V_e (\bar Q_e-q_e+x_{et}(\bar Q_t-q_t)+x_{ei}(Q_i-q_i)) 
\end{multline}
where we have introduced the effective capacitances and coupling strengths
\begin{align*}
\bar C_i^{-1} &\equiv 
		p_{ii} -p_{ic} p_{cc}^{-1} p_{ci}+ (p_{ie}c_{et} + p_{it} c_{tt})(c_{tt}+S_t)^{-1} (c_{te} p_{ei} + c_{tt} p_{ti}),\\
\bar C^\ast_e &\equiv c_{ee}+S_e - c_{et} (c_{tt}+S_t)^{-1} c_{te},\quad \bar C_e\equiv\bar C_e^\ast\Big\rvert_{S_e=S_r=S_l=0}\\
x_{et} &\equiv -c_{et} (c_{tt}+S_t)^{-1},\\
x_{ei} &\equiv c_{ee} p_{ei} + c_{et} p_{ti} - c_{et} (c_{tt} +S_t)^{-1} (c_{te} p_{ei} + c_{tt} p_{ti}),\\
\bar p_{ti}&\equiv (c_{tt}+S_t)^{-1} (c_{te}p_{ei}+c_{tt}p_{ti}).
\end{align*}
\paragraph{Regimes ABI, CDI, CDII: } In the
limits of very strong, i.e.
	\begin{align*}
		C_{\alpha e} \ll C_{eg}+S_e,
	\end{align*}
	and very weak, i.e.
	\begin{align*}
		C_{\alpha e} &\gg \frac{(C_{\alpha g}+S_\alpha)^2}{C_{eg}+C_{rg}+C_{lg}+S_e+S_r+S_l},
	\end{align*}
	coupling  between fully transmitted edges $\alpha=r, l$ and interfering edge $e$ the expressions simplify to\\
	\begin{center}
	\begin{tabular}{c||c|c}
		& Strong coupling & Weak coupling\\\hline\hline
		$\bar C_i$ & $C_{ei}+C_{ig}$ & $C_{ei}+C_{ri}+C_{li}+C_{ig}$ \\
		$\bar C_e^\ast$ & $C_{eg}+S_e+C_{ig}-C_{ig}^2/\bar C_i$ &  $C_{eg}+C_{rg}+C_{lg}+S_e+S_l+S_r+C_{ig}-C_{ig}^2/\bar C_i $\\
		$x_{et}$ & $\begin{pmatrix} 0&0 \end{pmatrix}$ & $\begin{pmatrix} 1& 1 \end{pmatrix}$\\
		$x_{ei}$ & $C_{ei}/\bar C_i$ & $(C_{ei}+C_{ri}+C_{li})/\bar C_i$\\
		$\bar p_{ti}$ & 0 & 0,
	\end{tabular}
	\end{center}
	and using Eqs. (\ref{eqn:FPIES1})-(\ref{eqn:FPIES2}), the action becomes
\begin{gather}
	\act_0[\varphi,N_i]= \frac 12 \int_\keldC\dd t\,\dd t'\,\varphi(t) V^{-1}(t-t') \varphi(t') - \int_\keldC\dd t\, \varphi(t) N_0(t) -\frac 1{2 \bar C_i} (Q_i-q_i)^2\label{eqn:FPIfreeAct}\\
	\begin{split} 
		\text{with}\quad V^{r/a}(\omega) &= \ec \frac\omega{\omega\pm i \wc (1-e^{\pm i \omega\tau})},\label{eqn:FPIEffInteraction}\\
		N_0 &\equiv \frac\nuEff\pi \frac{\mu_++\mu_-}2\tau-\frac{\lvert e\rvert V_g}\ec-\frac{\bar C_{ei}}{\bar C_i} Q_i/\lvert e\rvert
	\end{split} 
\end{gather}
Here $\bar C_i$, $\bar C_e$, $\bar C_{eg}$, $\bar C_{ei}$, and $\nuEff$ are given in Sect.~\ref{sect:FPIABRes} (Eq.~(\ref{eqn:effCaps}) and table above).


\subsubsection{Tunneling Action}
To construct the tunneling action $\act_t[\varphi]$ in lowest order we use  Eq.~(\ref{eqn:TunAct2Exp}) which makes it necessary to identify the paths $(ij;\mu\nu)$. Only the innermost chiral channels $\mu,\nu=\pm$ allow for tunneling between each other at scatterers $i,j=1,2$ which gives 4 classes: $(11;+-)$, $(22;+-)$, $(12;+-)$, and $(21;+-)$. Classical phases are accumulated due to magnetic flux $\phi$:
\begin{align*}
	\Delta \phi^{11}_{+-}=\Delta \phi^{22}_{+-}=0,\quad \Delta \phi^{12}_{+-}=-\Delta\phi^{21}_{+-}=-2\pi \phi/\phi_0.
\end{align*}
At zero temperature the distribution functions read $f^\gtrless_\pm(t)=e^{-i\mu_\pm t} f_0^\gtrless(t)$ with Fermi distribution function $f_0^\gtrless(t)$. Writing for short $r_i\equiv s^i_{-+}$, $\chi\equiv\chi_+-\chi_-$, $\epsilon_{ij}=\epsilon_{ij3}$ (the right-hand side being the 3-dimensional Levi-Civita symbol), and $\Pi_{ij}\equiv\Pi_{ij;+-}$ we obtain the tunneling operators (\ref{eqn:tunPolOp1}, \ref{eqn:tunPolOp2})
\begin{align}
	\Pi_{ij}^\gtrless(t) &=-r_i \bar r_j\ e^{\pm i\chi}\ e^{-i\epsilon_{ij} \left[2\pi \phi/\phi_0+(\mu_++\mu_-)\tau\right]}\ e^{-ieVt} f_0^\gtrless(t+\epsilon_{ij}\tau)f_0^\lessgtr(t-\epsilon_{ij}\tau), \label{eqn:FPITunPolOp1}\\
	\Pi^{\T/\aT}_{ij}(t) &=\left[\theta(\pm t) \Pi^>_{ij}(t)+\theta(\mp t) \Pi^<_{ij}(t)\right]_{\chi\equiv 0}. \label{eqn:FPITunPolOp2}
\end{align}
Writing the tunneling phases $\Phi\equiv\Phi_{+-}$ the tunneling action reads
\begin{align} \label{eqn:FPITunAct}
	\act_t[\varphi]=-i\sum_{i,j=1,2}\int\!\!\dd t_1\dd t_2\, \begin{pmatrix}e^{-i\Phi^-(x^i,t_1)} & e^{-i\Phi^+(x^i,t_1)}\end{pmatrix} \begin{pmatrix}\Pi_{ij}^\T(t_{12}) & -\Pi_{ij}^<(t_{12})\\ -\Pi_{ij}^>(t_{12}) & \Pi_{ij}^\aT(t_{12})\end{pmatrix} \begin{pmatrix} e^{i\Phi^-(x^j,t_2)}\\ e^{i\Phi^+(x^j,t_2)}\end{pmatrix}
\end{align}
with $t_{12}\equiv t_1-t_2$. According to Sect.~\ref{sect:weakTun} and Eq.~(\ref{eqn:phaseHSField}) the tunneling phases are related to the potential $\varphi$ via
\begin{gather}
	\Phi(x^i,t)=\Theta_-(x^i,t)-\Theta_+(x^i,t) = \int_\keldC\!\!\dd t'\, \mathcal D_{+-}(x^i;t,t') \varphi(t') \label{eqn:FPIPhiphi}\\
	\text{with}\quad \mathcal D_{+-}(x;t,t') \equiv \int_{x_1}^{x_2}\dd x'\, D_{0-}(x-x';t,t')-\int_{x_1}^{x_2}\dd x'\, D_{0+}(x-x';t,t')\nonumber
\end{gather}
with bare particle-hole propagator $D_{0\mu}$ (Eq.~(\ref{eqn:ehpPropagator1})). Note that $x'$ is integrated over because potential $\varphi(x',t)=\varphi(t)$ in  our model does not vary in space.

Defining $\epsilon_i=\pm$ for $i=1,2$ retarded and advanced components of $\mathcal D_{+-}$ read in energy representation
\begin{align}
	\mathcal D^{r/a}_{+-}(\omega;x^i)= \pm i\epsilon_i \frac{e^{\pm i \omega\tau}-1}\omega. \label{eqn:FPIDPhiphi}
\end{align}

\newcommand{\strc}{\infty} 
\newcommand{\chF}{{\mathds 1}} 
\newcommand{\sgn}{{\mathrm{sign}\,}} 

\subsubsection{Current in Instanton Approximation}
Current is measured via the counting fields $\chi$ in the tunneling polarization operators (\ref{eqn:FPITunPolOp1}). We use the adiabatic approximation where measuring time $t_0$ is much larger than all intrinsic time scales of the system and transient effects due to switching of the counting fields are negligible. The tunneling correction to current is the derivative
\begin{gather}
	I_t = -i \frac e{t_0} \partial_\chi \ln\mathcal Z\Big\rvert_{\chi=0} = \sum_{i,j=1,2} \left(I_{ij}^<-I_{ij}^>\right)\\
	\text{with } I_{ij}^{\alpha\beta} = \frac e{t_0} \int\!\!\dd t_1\dd t_2\, \int\DD\varphi \,\sum_{\lbrace N_i\rbrace}\, e^{i\act_0[\varphi,N_i]+i\act_t[\varphi]}\ e^{-i\Phi^\alpha(x^i,t_1)} \Pi_{ij}^{\alpha\beta}(t_{12}) e^{i\Phi^\beta(x^j,t_2)} \Big\rvert_{\chi=0}.
\end{gather}

The average $\int\DD\varphi \sum_{\lbrace N_i\rbrace}$ is treated in real-time instanton approximation as outlined in Sect.~\ref{sect:SPA}:
\begin{align}
	I_{ij}^{\alpha\beta} &\approx \frac e{t_0} \int\!\!\dd t_1\dd t_2\, e^{i\tilde \act_t[\varphi_\ast]} \left< e^{-i\Phi^\alpha(x^i,t_1)} \Pi_{ij}^{\alpha\beta}(t_{12}) e^{i\Phi^\beta(x^j,t_2)}\right>_0\Big\rvert_{\chi=0,\ N_i=N_{i\ast}}\nonumber\\
	&= \frac e{t_0} \int\!\!\dd t_1\dd t_2\, e^{i\tilde \act_t[\varphi_\ast]} e^{-i\Phi_0(x^i)+i\Phi_0(x^j)}\ \tilde \Pi_{ij}^{\alpha\beta}(t_{12})\Big\rvert_{\chi=0,\ N_i=N_{i\ast}} \label{eqn:FPICurrent}
\end{align}
As always $\langle\ldots\rangle_0$ denotes averaging with respect to $\act_0[\varphi,N_{i\ast}]$ given in (\ref{eqn:FPIfreeAct}). Because of the linear-in-$\varphi$ contribution potential $\varphi$ and hence tunneling phase $\Phi$ have non-vanishing expectation values $\varphi_0=\varphi_0[N_i]$, $\Phi_0=\Phi_0[N_i]$ which minimize
$\act_0[\varphi,N_i]$ for given $N_i$. At the saddle-point $N_{i\ast}$ in turn minimizes $\act_0[\varphi_0[N_i],N_i]$. For strong coupling $\wc\tau\gg 1$ the mean-field reads
\begin{align} \label{eqn:FPImeanfield}
	\begin{split}
		\varphi_0 = \frac 1{1+\wc \tau}\left[eV_g +\frac{\mu_++\mu_-}2 \wc\tau+\ec\frac{\bar C_{ei}}{ \bar C_i}(N_{i\ast}+\nu\phi/\phi_0)\right], \\
		\Phi_0(x^{1/2})=\mp \tau  \varphi_0\quad \Rightarrow\quad -i\Phi_0(x^k)+i\Phi_0(x^l)=2i\epsilon_{kl} \tau\varphi_0
	\end{split}
\end{align}
with $N_{i\ast}\in \mathbb Z$ minimizing the electrostatic energy $E_i$, (\ref{eqn:islandChargingEnergy}).

Due to the presence of the source term $i\act_J[\varphi]=-i\Phi^\alpha(x^i,t_1)+i\Phi^\beta(x^j,t_2)$ the instanton phase $\Phi_\ast=\Phi_0+\delta \Phi_\ast$, (\ref{eqn:instantonPot}), deviates from the mean-field by
\begin{align} \label{eqn:FPIinstantonPhase}
	\delta \Phi_\ast^\gamma(x^k,t)= D_\Phi^{\gamma\alpha}(t-t_1,x^k,x^i)-D_\Phi^{\gamma\beta}(t-t_2,x^k,x^j).
\end{align}
The instanton action thus reads
\begin{align}
	i\tilde \act_t[\varphi_\ast]=\sum_{kl}\sum_{\gamma\delta}\gamma\delta \int\!\!\dd t_3 \dd t_4\, e^{-i\Phi_0(x^k)+i\Phi_0(x^l)} e^{-i\delta\Phi_\ast^\gamma(x^k,t_3)+i\delta\Phi_\ast^\delta(x^l,t_4)} \tilde \Pi^{\gamma\delta}_{kl}(t_3-t_4)\label{eqn:FPIinstantonAct}
\end{align}

We will evaluate the time integrals in (\ref{eqn:FPICurrent}) and (\ref{eqn:FPIinstantonAct}) approximately. They will be dominated by the singularities of the instanton and the polarization operators. To identify and characterize them more precisely it is indispensable to compute the phase correlator $D_\Phi\equiv-i\left\langle (\Phi-\Phi_0)(\Phi-\Phi_0)\right\rangle_0$. It will turn out that the singularities (branchcuts) of $\tilde \Pi_{kk}(t)$ around $t\sim 0$ and of $\tilde \Pi_{kl}(t)$, $k\neq l$, around $t\sim \pm\tau$ dominate all integrals.

\subsubsection{Correlation Functions}
In this section we calculate the correlation function of the tunneling phases $\Phi(x^i)$ which according to (\ref{eqn:FPIPhiphi}) is $D_\Phi=-\mathcal D_{+-} V \mathcal D_{+-}$. Details of the calculation are not important for the rest of the paper and may be safely skipped. The final results for zero temperature and the strong coupling limit, $\wc\tau\gg1$, are
	\begin{align} 
		D^{\gtrless}_\Phi(t=0,x^i,x^j)&=D^{T/\aT}_\Phi(0,x^i,x^j)\equiv D_\Phi(0;x^i,x^j)=A_{ij}\frac i{\nuEff}\left(\gamma+\ln[\wc\tau]\right)\label{eqn:FPIDPhiZero}\\
		\intertext{and} \nonumber\\
		D^{\gtrless}_{\Phi}(t,x^i,x^j) &= A_{ij}\frac i{2\nuEff} \left\{\ln\left[\frac{a\pm i (t-\tau)}{a\pm i t}\right]+\ln\left[\frac{a\pm i (t+\tau)}{a\pm i t}\right]\right\}.\label{eqn:FPIDPhi}
	\end{align}
for large times, $\lvert \wc t\rvert\gg1$.

We start the computation by combining (\ref{eqn:FPIEffInteraction}) and (\ref{eqn:FPIDPhiphi}),
\begin{align*}
	D_\Phi^{r} (\omega, x_i,x_j) = - \mathcal D_{+-}^{r}(\omega;x^i) V^{r}(\omega) \mathcal D_{+-}^{r} (\omega;x^j)
		 = - i A_{ij} \frac\pi {\nuEff} \frac{i \wc \left(1-e^{i\omega \tau}\right)}{\omega\left[\omega+i \wc \left(1-e^{i\omega \tau}\right)\right]} \left(e^{i\omega\tau}-1\right)	
\end{align*}
	with $A_{ij}=\epsilon_i \epsilon_j$.
	In time representation the relevant correlation functions are the $\gtrless$-components which, at zero temperature, read
	\begin{gather}
		\begin{split}
		D_\Phi^\gtrless(t) &=  \pm \int\!\frac{\dd \omega}{2\pi}\, e^{-i\omega t} \left[D^r_\Phi(\omega)-D^a_\Phi(\omega)\right] \theta(\pm\omega)\\
			&=  - A_{ij} \frac i{2\nuEff}\left\{ J^\gtrless(t-\tau)-J^\gtrless(t)+\left[J^\gtrless(-t^\ast-\tau)\right]^\ast-\left[J^\gtrless(-t^\ast)\right]^\ast\right\}
		\end{split}\label{eqn:FPIDPhiphiJ}\\
	\text{with} \quad J^\gtrless(t) \equiv  \int \!\! \dd \omega\, \left[\pm \theta(\pm \omega)\right] \frac{i\wc \left(1-e^{i\omega \tau}\right)}{\omega\left[\omega+i\wc \left(1-e^{i\omega \tau}\right)\right]} \left(e^{-i\omega t}-1\right). \label{eqn:FPIJDef}
	\end{gather}
	The integral defining $J^>$ ($J^<$) is perfectly convergent for all times with non-positive (non-negative) imaginary part, $\Im t\le0$ ($\Im t\ge0$), thus ensuring the analyticity of $J^\gtrless$ in this region. Apparently, we have $J^\gtrless(t)^\ast=J^\lessgtr(t^\ast)$. 

	\begin{figure}[ht]
		\centering{\includegraphics[scale=0.6]{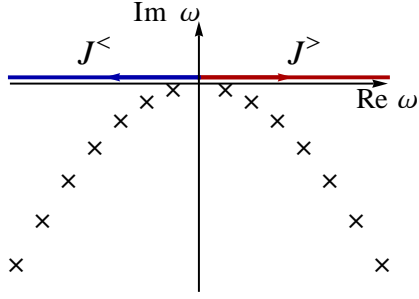}}
		\caption{Analytic structure of the integrand in (\ref{eqn:FPIJDef}): it is analytic in the upper half of the $\omega$-plane and possesses poles in the lower half. Contours of integration for $J^>$ and $J^<$ are indicated by arrows.} \label{fig:FPIJPoles}
	\end{figure}

	First, we perform the integration for $\Re t<0$. Under this assumption the contour of integration can be rotated into the upper half of the complex $\omega$-plane where the integrand is analytic (see Fig.~\ref{fig:FPIJPoles}). Defining dimensionless time and charging frequency, $z\equiv -t/\tau$, and $y\equiv \wc \tau$, respectively, and integrating along the imaginary axis, one obtains for $y\gg1$

	\begin{multline}
		J^\gtrless (t) = \int_0^\infty\!\!\dd s\, \left(e^{-z s}-1\right) \frac{y\left(1-e^{-s}\right)}{s\left[s+y\left(1-e^{-s}\right)\right]}
		\approx   \int_0^\infty\!\!\dd s\, \left(e^{-z s}-1\right) \frac{y}{s\left(s+y\right)}\\
		=  -e^{y z}\Gamma(0,yz)-\gamma-\ln yz \equiv g(y z) \label{eqn:FPIgDef}
	\end{multline}
	with the incomplete Gamma function $\Gamma(\alpha,x) = \int_{x}^\infty\!\dd s\, e^{-s} s^{\alpha-1}$, $x\ \in \mathbb R$, and the 
	Euler-Mascheroni constant $\gamma$.

	The asymptotic behavior of $g$ is
	\begin{align}
		g(yz)\equiv \to & (-1+\gamma+\ln yz) yz, & yz\to 0+,\nonumber\\
				\to & -\gamma -\ln yz - \frac 1 {yz} , & yz\to \infty. \label{eqn:FPIgLarge}
	\end{align}


	We now proceed with the case $\Re t>0$ where the contour of integration can be rotated into the lower half of the complex plane $\omega$-plane. In contrast to the previous case the integrand does possess poles in this region (see Fig.~\ref{fig:FPIJPoles}), around which, therefore, the integral has to be taken additionally. Since both pole and imaginary axis contribution, $J_0^\gtrless$ and $J_1^\gtrless$ respectively, separately diverge for large $\omega$ we have to introduce an auxiliary ultraviolet cutoff, $e^{\mp a \omega}$, $a=\tilde a \tau$. Then, defining $z\equiv t/\tau$,  the imaginary axis contribution reads for $y=\wc \tau\gg 1$
	\begin{align*}
		J^\gtrless_1(t) = & - y \int_0^\infty\!\!\dd s\, \frac{1-e^s}{s\left[s-y\left(1-e^s\right)\right]} \left(e^{-s z}-1\right) e^{\pm i s\tilde a} \approx \int_0^\infty \!\!\dd s\, \frac{e^{-s z} -1} s e^{\pm i s \tilde a} =  - \ln \frac{\tilde a\pm i z}{\tilde a}.
	\end{align*}

	The poles are defined as roots of equation 
$\omega+i\wc \left(1-e^{i\omega \tau}\right)=0$, $\omega\neq 0$, and writing $x=i\omega\tau$, they are given by $x_n = y-W_{-n}(y e^y)$, $n\in \mathbb Z\!\!\setminus\!\!\{0\}$, where the product log function is defined by $W_n(x) e^{W_n(x)}=x$. We choose the numbering such that $\Im x_{n+1} > \Im x_n$, $\Im x_1>0 > \Im x_{-1}$. As can be deduced already from the defining equation the roots satisfy $\Re x_n \ge 0$. One may show that in two 
limiting cases one has
	\begin{align}
		x_n \to & 2 \pi i n \frac y{1+y} + 2\left(\frac{n\pi} y\right)^2, &\lvert n\rvert \ll \frac y{2\pi}, \label{eqn:XNSmallN}\\
			\to & 2\pi i n + \ln \left[-i \frac{2\pi n}y\right], & \lvert n\rvert \gg \frac y{2\pi}. \label{eqn:XNLargeN}
	\end{align}
To proceed further, we note that $\frac \dd{\dd \omega} \left[\omega+i\wc \left(1-e^{i\omega \tau}\right)\right]_{\omega_n} = 1+ y -x_n$. Therefore the residues read
	\begin{equation*}
		\mathrm{Res}_{\omega_n} \left[\frac{i \wc \left(1-e^{i\omega \tau}\right)}{\omega \left[\omega+i\wc\left(1-e^{i\omega\tau}\right)\right]} \left(e^{-i\omega\tau}-1\right) e^{\mp a\omega}\right] = -\frac 1{1+y-x_n} \left(e^{-z x_n}-1\right) e^{\pm i \tilde a x_n}.
	\end{equation*}
	Taking into account that for $J^>$ ($J^<$) only poles $\omega_n$ with positive (negative) real part contribute, $n\ge 1$ ($n\le -1$), we obtain for the pole contribution
	\begin{align*}
		J^\gtrless_0(t) = -\sum_{n=1}^\infty \frac{\mp 2\pi i}{1+y-x_n}
    \left(e^{-z x_{\pm n}}-1\right) e^{\pm i \tilde a x_{\pm n}}.
	\end{align*}
	This expression cannot be evaluated analytically further, but analytical approximations are possible by substituting the poles $x_n$ by their asymptotic behavior, Eqs.~\eqn{XNSmallN}, \eqn{XNLargeN}.

\begin{figure}[ht] 
		\centering{\includegraphics[scale=0.3]{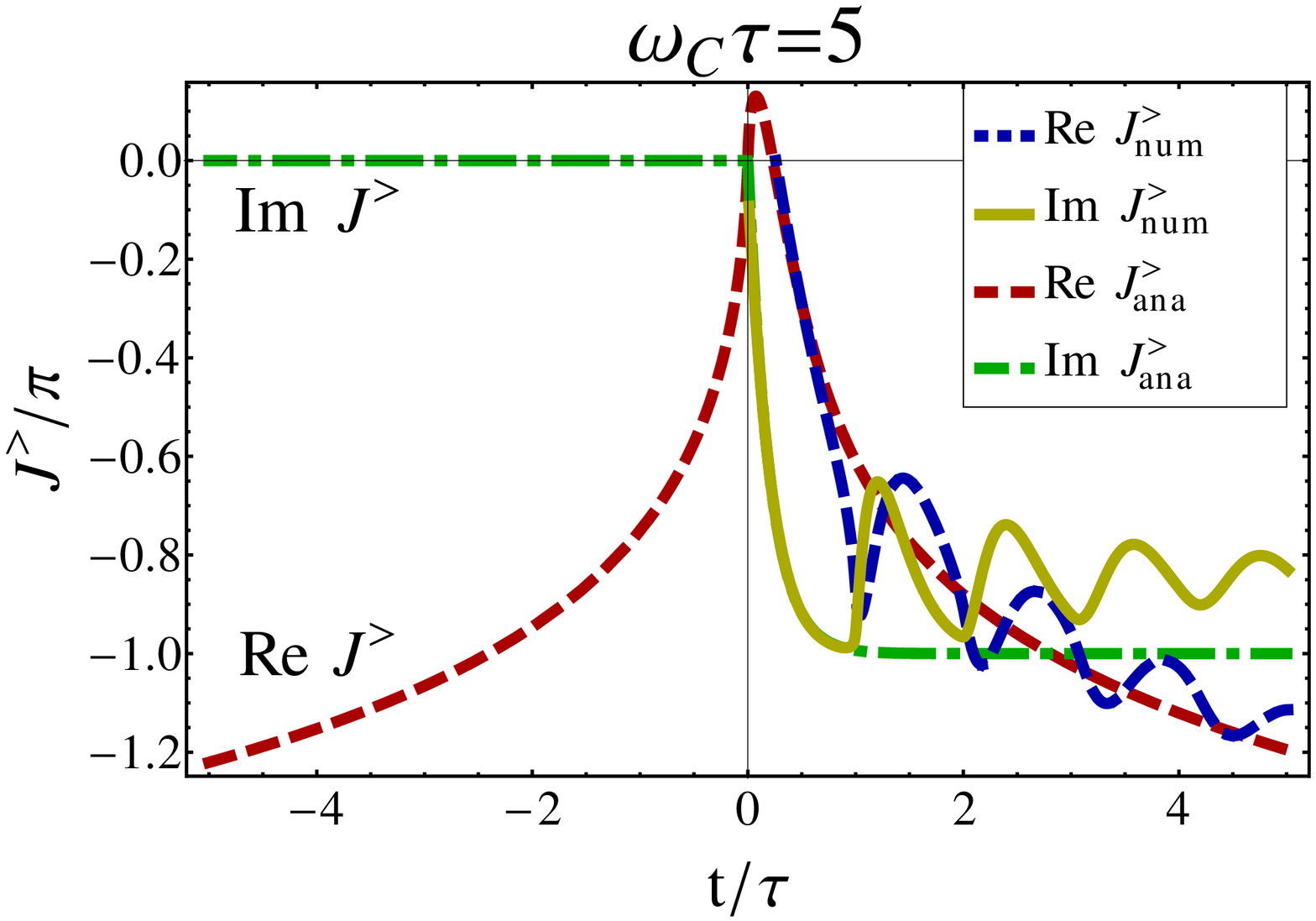}\quad\includegraphics[scale=0.3]{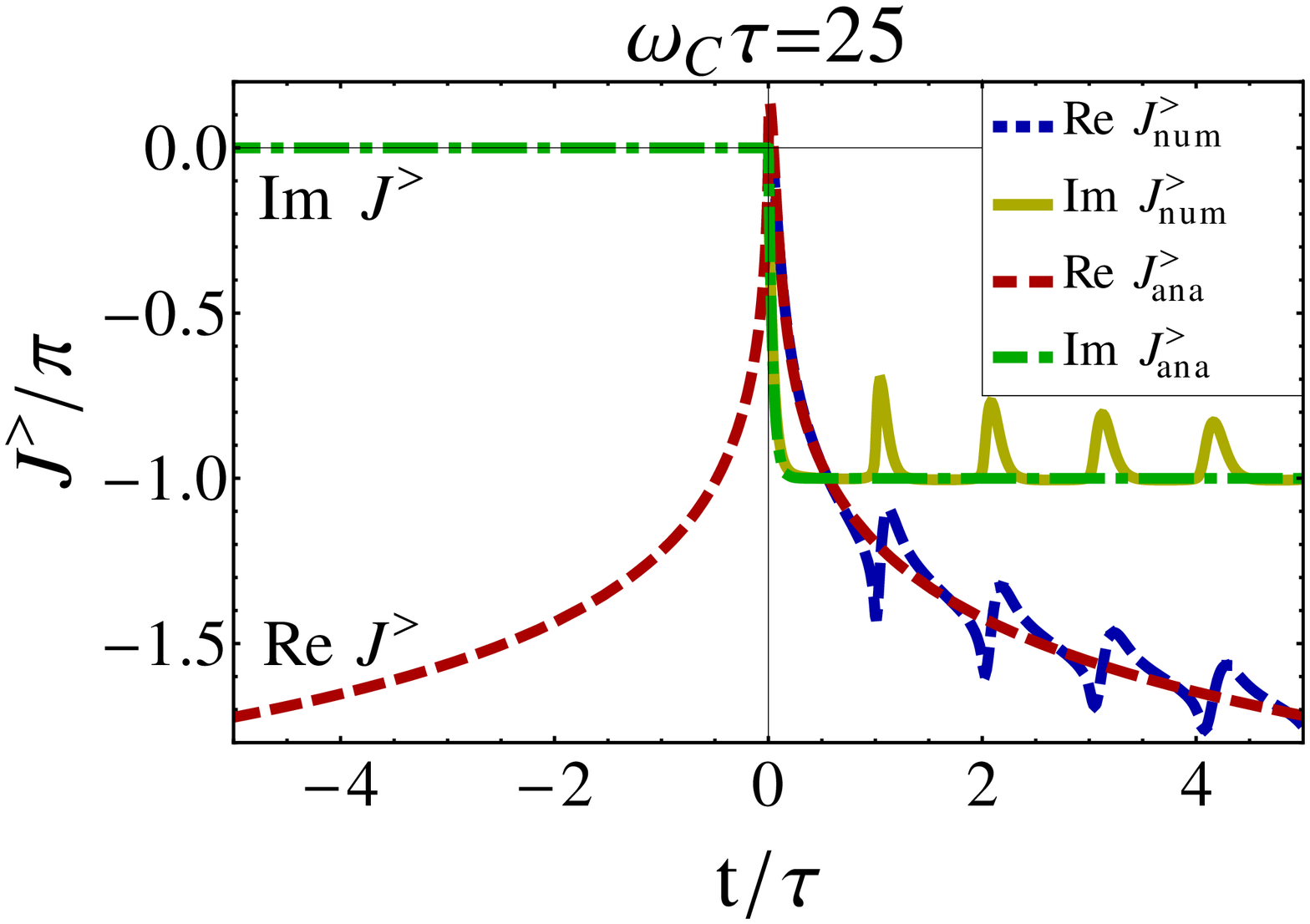}}
		\caption{$J^>$ (real (\emph{blue}) and imaginary (\emph{yellow}) part), numerically evaluated and manually fixed, and analytical continuation $\sim g(-\wc(t-i 0))$ (real (\emph{red}) and imaginary (\emph{green}) part).}
		\label{fig:JPosFix}
	\end{figure}

	We convince ourselves that the short-time divergence, which forced us to introduce the ultraviolet cutoff $\tilde a$, is in fact merely an artifact of our method of calculation, and is cured by taking the sum of $J^\gtrless_1+J^\gtrless_0$. In other words, $J^\gtrless_0$ has to diverge logarithmically for $z\to0$ as well. Of course, any divergence originates from terms with large $\lvert n\rvert \to\infty$, such that for 
our present purpose we may safely use the approximation \eqn{XNLargeN} which yields for $z\to 0$
	\begin{equation*}
		J^\gtrless_0(t) \sim - \sum_{n=1}^\infty \left[\left(\pm i \frac{2\pi n}y\right)^z 
             e^{-2\pi n\left(\tilde a\pm i z\right)} - e^{-2\pi n \tilde a}\right]\frac{1}{n}
			\approx - \ln \frac{1-e^{-2\pi \tilde a}}{1-e^{-2\pi \left(\tilde a \pm iz\right)}}
			\approx  \ln \frac{\tilde a\pm iz}{\tilde a},
	\end{equation*}
	which is exactly what we expected to find. Although the approximation is good enough to estimate the divergency, it is not reliable for obtaining finite offsets. Using $\ln \tilde a = \ln\frac{1-e^{-2\pi \tilde a}}{2\pi}$ we can single out all $\tilde a$-dependencies,
	\begin{multline*}
		J^\gtrless_0(t) + J^\gtrless_1(t) = -\ln\left[\pm 2\pi i z\right] - \sum_{n=1}^\infty \frac{\mp 2\pi i}{1+y-x_{\pm n}} e^{-z x_{\pm n}} \\+ \sum_{n=1}^\infty \left[\frac{\mp 2\pi i }{1+y-x_{\pm n}} \left(e^{\pm i \tilde a\left(x_{\pm n}\mp 2\pi i n\right)}-1\right) + \frac{x_{\pm n} \mp 2\pi i n -1 -y}{\left(1+y-x_{\pm n}\right) n }\right] e^{-2\pi \tilde a n}.
	\end{multline*}
	The $\tilde a$-contribution is just a constant about which we will not care too much presently. For the moment we will fix it manually, by requiring a good agreement between $J_0^\gtrless(t)+J^\gtrless_1(t)$ and the analytical continuation $g(-\wc (t\mp i 0))$ of the result (\ref{eqn:FPIgDef}) obtained for $\Re t<0$. Fig. \ref{fig:JPosFix} shows the corresponding plots for $y=5$ and $y=25$. 

	A numerical study shows that the oscillating contributions decrease in width for large $y$ (while their amplitude remains in the order of unity) and may be therefore neglected in the following. We approximate $J^\gtrless$ by smooth functions $g^\gtrless$, required to be analytical for $\Im t\le 0$ ($\Im t \ge 0$) and
	\begin{align*}
		g^\gtrless(t) =  &-e^{-\wc t} \Gamma(0,-\wc t) - \gamma - \ln\left[-\wc t\right],  &\Re t<0.
	\end{align*}

Since the voltage is assumed to be low $\lvert eV\rvert \ll \wc$ one needs correlation functions for long times $\lvert \wc t\rvert \gg 1$ only and we can use the asymptotic expression (\ref{eqn:FPIgLarge}) for $g$. Therefore 
introducing a short-time cutoff $a\sim \wc^{-1}$ and writing $t_\mp\equiv t\mp i a$ we use the following approximate
relation in our subsequent analysis
	\begin{align*}
		J^\gtrless(t) \approx g^\gtrless(t) = -\gamma - \ln [-\wc t_\mp],
	\end{align*}
	which together with (\ref{eqn:FPIDPhiphiJ}) and $J^\gtrless(t=0)=0$ gives Eqs.~(\ref{eqn:FPIDPhi}) and (\ref{eqn:FPIDPhiZero}). 
\newcommand{\redPot}{{\tilde \mu}} 

\subsubsection{Renormalized Polarization Operators}
	In the real-time instanton approximation, Sect.~\ref{sect:SPA}, virtual fluctuations around the instanton are taken into account by dressing the tunneling polarization operators, Eq.~(\ref{eqn:renPolOp}).
	The phase factor is
	\begin{equation*}
		e^{i \left[D^\gtrless_{\Phi\infty} (t,x_k,x_l) -D_{\Phi}(0)\right]} = e^{\frac\gamma\nuEff}\left(\wc \tau\right)^{\frac 1 \nuEff} \left[\frac{a\pm i t}{a\pm i (t-\tau)}\right]^{\frac{A_{kl}}{2\nuEff}} \left[\frac{a\pm i t}{a\pm i (t+\tau)}\right]^{\frac{A_{kl}}{2\nuEff}}.
	\end{equation*}
	and dressing of the bare polarization operators (\ref{eqn:FPITunPolOp1}) yields ($f_0$ can be found in (\ref{eqn:FermiDistribution}); $\chi$ is put to 0)
	\begin{multline*}
	\tilde \Pi^\gtrless_{kl}(t) = -r_k\bar r_l \frac{(\wc \tau)^{\frac 1\nuEff}}{(2\pi)^2}e^{\frac\gamma\nuEff} e^{-i\epsilon_{kl}\left[2\pi\phi/\phi_0 + (\mu_++\mu_-)\tau\right]} e^{-i eV t}\\
	\times	\left\lbrace\begin{array}{ll}
	             \left[a\pm i t\right]^{\frac 1 \nuEff -2} \left[a\pm i (t-\tau)\right]^{-\frac 1 {2\nuEff}}\left[a\pm i (t+\tau)\right]^{-\frac 1 {2\nuEff}}, & k=l,\\
	             \left[a\pm i t\right]^{-\frac 1 \nuEff} \left[a\pm i (t-\tau)\right]^{\frac 1 {2\nuEff}-1}\left[a\pm i (t+\tau)\right]^{\frac 1 {2\nuEff}-1}, & k\neq l,
	            \end{array}\right.
\end{multline*}
	\begin{figure}
		\centering{\includegraphics[scale=0.3]{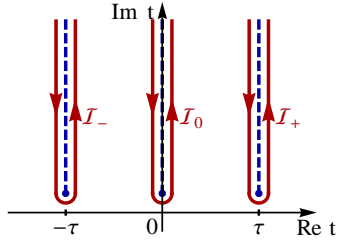}}
		\caption{Analytic structure of $\mathcal I$ in the complex $t$-plane: while the function is analytic in the lower half, it has poles or branchcuts in the upper half, residing at $t^\ast=0, \tau, -\tau$. Integrating $\tilde \Pi^\gtrless_{kl}(t)$ ``around'' these, i.e. along drawn contours, gives $\tilde P^\gtrless_{kl}(t^\ast)$.} \label{fig:FPIBranchcutsI}
	\end{figure}

The dressed polarization operators exhibit non-analytic behavior (poles or branchcuts) around $t\approx 0, \pm \tau$. The double time-integrals (\ref{eqn:FPICurrent}) and (\ref{eqn:FPIinstantonAct}) can be approximately expressed in terms of the integrals $\tilde P^\gtrless_{kl}\equiv \int\!\!\dd t\,\tilde \Pi^\gtrless_{kl}(t)$. Before we demonstrate this statement in the next section, we will devote the remainder of the current section to the evaluation of $\tilde P^\gtrless_{kl}$.

We focus first on $\tilde P^>_{kl}$. To deal with both $k=l$ and $k\neq l$ simultaneously we generically consider the function
\begin{gather*}
	\mathcal I (t,eV) \equiv e^{-ieV t} \left[a+ i t\right]^\eta \left[a + i (t-\tau)\right]^\lambda \left[a+ i (t+\tau)\right]^\lambda,\\
	0> \eta > -2,\quad 0>\lambda >-1,\quad 2\lambda+\eta = -2.
\end{gather*}

Apparently, $\mathcal I$ has branchcuts only in the upper half of the complex $t$-plane (Fig.~\ref{fig:FPIBranchcutsI}), i.e.\ the integral $\int\!\dd t\, \mathcal I(t,eV)$ vanishes whenever the integration contour can be closed in the lower half. Therefore, we assume the nontrivial case $eV<0$. The real-time integrals $\int\!\dd t\, \mathcal I(t,eV)=\mathcal I_-+\mathcal I_0+\mathcal I_+$ consist of three contributions which correspond to integrals along closed contours in the complex $t$-plane.  With $z\equiv \lvert eV \tau\vert\gg 1$ the contour integral around $-\tau$ is
\begin{align*}
	\mathcal I_-
		\approx & 
				-2\pi \frac{2^\lambda}{\Gamma(-\lambda)}   e^{-i z} e^{i\frac \pi 2 \lambda}\ \lvert eV\rvert z^{-2-\lambda}.
\end{align*}
Similarly, one obtains the integral around $+\tau$, $\mathcal I_+ =\mathcal I_-^\ast$.

The situation is slightly less trivial for the integral around $t\approx 0$, since it may be that $\eta=-1$, i.~e. we have a first order pole, or $\eta>-1$, giving rise to a strong divergence. In the first case the integral gives
\begin{equation*}
	\mathcal I_0 = 2\pi \left[a-i \tau\right]^\lambda\left[a+i \tau\right]^\lambda \approx \frac{2\pi}\tau.
\end{equation*}
In the second case we have to go around the singularity with care. We explicitly kept the distance $\delta\ll a$ to the integration contour from the branchcut in the calculations. After weakening the degree of divergence by partial integration we may safely put $\delta\to 0$ and obtain
\begin{align*}
	\mathcal I_0 
	\approx & 2\pi \Gamma(-\eta)^{-1} \lvert eV \rvert z^{2\lambda}.
\end{align*}
Note that in this approximation $\mathcal I_0$ is continuous in $\eta=-1$.

For the direct terms, $k=l$, we have $\eta=\frac 1\nuEff -2$, $\lambda = -\frac 1{2\nuEff}$, i.~e.\ $\eta+\lambda=\frac 1{2\nuEff} -2$, $2\lambda = -\frac 1\nuEff$, hence for large voltages, $z\gg 1$, $\mathcal I_0$ is dominant. For the interference terms, $k\neq l$, we set $\eta=-\frac 1\nuEff$, $\lambda =\frac 1{2\nuEff}-1$, i.~e.\ $\eta+\lambda=-\frac 1{2\nuEff} -1$, $2\lambda = \frac 1 \nuEff -2$, hence the contributions $\mathcal I_\pm$ dominate over $\mathcal I_0$ if and only if $\nuEff>\frac 32$. 

As $\int\!\!\dd t\, \mathcal I(t,eV)$ splits into three contributions, so do $\tilde P^\gtrless_{kl}=\tilde P^\gtrless_{kl}(-\tau)+\tilde P^\gtrless_{kl}(0)+\tilde P^\gtrless_{kl}(+\tau)$. Note that $\tilde \Pi^<(t)=\tilde\Pi^>(-t)\Big\rvert_{eV\mapsto-eV}$ implies $\tilde P_{kl}^<(t^\ast)=\tilde P_{kl}^>(-t^\ast)\Big\rvert_{eV\mapsto-eV}$.

Summarizing, for $z\gg 1$ the dominant integrals of the dressed polarization operators are
\begin{align}
	\tilde P^\gtrless_{kk}(t^{\ast}=0) =& -\theta(\mp eV) \frac{\lvert eV\rvert}{2\pi}\ R_{k\ast}(eV), \label{eqn:FPIDressedPolOpKKAsymp}\\
	\tilde P^\gtrless_{kl}(t^{\ast}=\kappa\tau) =&- \theta(\mp eV) \frac1{2\pi \tau}\ R_{12\ast}(eV)
	\  \frac12 e^{i\kappa\left[\lvert eV\tau\rvert - \frac \pi 2 \left(1+\frac 1{2\nuEff}\right)\right]} e^{-i\epsilon_{kl}\left[2\pi\phi/\phi_0+(\mu_++\mu_-)\tau\right]},\label{eqn:FPIDressedPolOpKL}\\
	\tilde P^\gtrless_{kl}(t^{\ast}=0) =&- \theta(\mp eV) \frac1{2\pi \tau}\ r_k\bar r_l\ \lvert eV\tau\rvert^{\frac 1 \nuEff-1} \frac {(\wc\tau)^{\frac 1 \nuEff} e^{\frac \gamma \nuEff}}{\Gamma(\frac 1\nuEff)}\ e^{-i\epsilon_{kl}\left[2\pi\phi/\phi_0+(\mu_++\mu_-)\tau\right]} \label{eqn:FPIDressedPolOpNu1}
\end{align}
with $k\neq l,\ \kappa=\pm$. In case of $\nuEff>\frac 32$ the contribution (\ref{eqn:FPIDressedPolOpKL}) is dominant, while in the case $\nuEff<\frac 32$ it is (\ref{eqn:FPIDressedPolOpNu1}). We have used definitions (\ref{eqn:FPIrenRefl}) of the renormalized reflection coefficients and assumed for simplicity $r_1 \bar r_2$ to be real.	

\subsubsection{Instanton Action and Current}
We have now everything in place to finalize the calculation of the instanton action (\ref{eqn:FPIinstantonAct}) and the current (\ref{eqn:FPICurrent}). The instanton phases $\delta \Phi_\ast^\gamma(x^k,t)=D^{\gamma\alpha}_\Phi(t-t_1,x^k,x^i)-D^{\gamma\beta}_\Phi(t-t_2,x^k,x^j)$ and thus $i\tilde\act_t[\varphi_\ast]$ are functions of the times $t_1$, $t_2$ over which to integrate in (\ref{eqn:FPICurrent}). A shift of integration variables $t_{3/4}\mapsto t_{3/4}+t_2$ in (\ref{eqn:FPIinstantonAct}) immediately shows that the action $i\tilde\act_t[\varphi_\ast]$ is a function of the difference $t\equiv t_1-t_2$. Hence, the whole integrand of (\ref{eqn:FPICurrent}) is purely a function of $t= t_1-t_2$. Performing a change of integration variables $(t_1,t_2)\mapsto (t=t_1-t_2,T=(t_1+t_2)/2)$, the integral over the center-of-mass time $T$ is seemingly divergent. This simply amounts  to infinite transferred charge $Q=\int_0^{t_0} \dd T\, I$ for a steady current $I$ and an infinite measuring time $t_0\to\infty$. Since our interest lies in the steady current (not on transient effects due to switching of the measuring device) we identify $\int\!\dd t_1\dd t_2=\int\!\dd T\,\dd t=t_0\int\!\dd t$ upon which the current becomes
\begin{align}
	I^{\alpha\beta}_{ij}=e \int\!\!\dd t\, e^{i\tilde\act_t[\varphi_\ast]}\Big\rvert_{t_1-t_2=t}\ e^{2i\epsilon_{ij}\tau\varphi_0}\ \tilde \Pi^{\alpha\beta}_{ij}(t)\Big\rvert_{\chi=0}.
\end{align}

Given that $i\tilde\act_t[\varphi_\ast]$ is non-divergent, large contributions to this current stem from the singularities of $\tilde\Pi^{\alpha\beta}_{ij}(t)$ which we identified in the previous section. The case $\nuEff=1$, i.e.\ $\nuEff<\frac 32$, needs to be treated more carefully and will be considered towards the end of this section. Focussing for now on $\nuEff\ge2$, dominant contributions are then
\begin{align}  \label{eqn:FPICurrentApprox}
	e^{-1}I^{\alpha\beta}_{ij}\approx \left\lbrace
	\begin{array}{ll}
		e^{i\tilde\act_t[\varphi_\ast]}\Big\rvert_{t_1-t_2\approx 0}\ \tilde P^{\alpha\beta}_{ii}(t^\ast=0), & i=j,\\
		e^{i\tilde\act_t[\varphi_\ast]}\Big\rvert_{\lvert t_1-t_2\rvert \approx \tau}\ e^{2i\epsilon_{ij} \tau\varphi_0}\left(\tilde P^{\alpha\beta}_{ij}(t^\ast=+\tau)+\tilde P^{\alpha\beta}_{ij}(t^\ast=-\tau)\right), & i\neq j. 
	\end{array}\right.
\end{align}

We evaluate the $t_3,t_4$-integrals in (\ref{eqn:FPIinstantonAct}) using a similar approximation scheme. Within the given constraints $t_1-t_2\approx 0$ for $i=j$ and $\lvert t_1-t_2\rvert \approx \tau$ for $i\neq j$, the singularity of 
\begin{align*}
	\tilde\Pi_{kk}(t_3-t_4)\sim \frac 1{(t_3-t_4)^{2-1/\nu^\ast}}
\end{align*}
dominates over the ones of
\begin{align*}
	\tilde \Pi_{kl}(t_3-t_4)\sim \frac 1{\left[(t_3-t_4-\tau)(t_3-t_4+\tau)\right]^{1-1/2\nu^\ast}},\quad k\neq l,
\end{align*}
and of the instantons
\begin{align*}
	e^{i\delta\Phi_\ast(x^k,t'+t_2)}&= e^{iD_\Phi(t'-t_1+t_2,x^k,x^i)-iD_\Phi(t',x^k,x^j)}\\
	&\sim \left(\frac{(t'-t_1+t_2)^2}{(t'-t_1+t_2-\tau)(t'-t_1+t_2+\tau)}\right)^{A_{ki}/2\nu^\ast} \left(\frac{(t'-\tau)(t'+\tau)}{t'^2}\right)^{A_{kj}/2\nu^\ast}.
\end{align*}
Hence, again transforming to relative and center-of-mass times, $t=t_3-t_4$, $T=(t_3+t_4)/2$, the dominant contribution to the instanton action stems from the $t\approx 0$-singularity of $\tilde\Pi_{kk}(t)$:
\begin{align*}
	i\tilde\act_t[\varphi_\ast] &\approx \sum_{k=1,2} \sum_{\gamma,\delta=\mp}\,\gamma\delta\ \int\!\!\dd t\,\tilde\Pi_{kk}^{\gamma\delta}(t)\ \int\!\!\dd T\, e^{-i\delta\Phi_\ast^\gamma(x^k,T)+i\delta\Phi_\ast^\delta(x^k,T)}\\
	&=-\sum_{k=1,2} \sum_{\gamma\neq\delta} \int\!\!\dd t\,\tilde\Pi_{kk}^{\gamma\delta}(t)\ \int\!\!\dd T\, \left(e^{iJ^{\gamma\delta}(T)}-1\right) \quad\text{ with } J^{\gamma\delta}(T)=-\delta\Phi_\ast^{\gamma}(x^k,T)+\delta\Phi_\ast^\delta(x^k,T)\\
	& \approx \sum_{k=1,2} \sum_{\gamma\neq\delta}\tilde P^{\gamma\delta}_{kk}(t^\ast=0) \int\!\!\dd T\, \left(e^{iJ^{\gamma\delta}(T)}-1\right).
\end{align*}
The second equality follows from $\tilde \Pi^>_{kl}(t)+\tilde \Pi^<_{kl}(t)=\tilde \Pi^\T_{kl}(t)+\tilde \Pi^\aT_{kl}(t)$. The integrals $\tilde P_{kl}^{\gamma\delta}=\int\!\dd t\,\tilde\Pi^{\gamma\delta}_{kl}(t)$ have been studied in the previous section.

\begin{figure}[t]
	\centering{\includegraphics[scale=0.4]{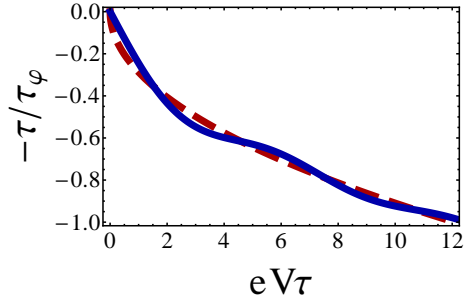}}
	\caption{Dephasing rate as a function of source-drain voltage shown for $\nuEff=2$, $\wc\tau=25$ and
$R_{1\ast}(\epsth)=R_{2\ast}(\epsth)=0.2$. 
The solid line gives the numerical result for (\ref{eqn:FPIdephRateFuller}). The dashed line is the power-law asymptotic given by Eq.~(\ref{eqn:FPIdephRate}).} \label{fig:FPIdephRate}
\end{figure}

Using the definition (\ref{eqn:FPIinstantonPhase}) of the instanton and the relation $(D_\Phi^T-D^>_\Phi)(t,x^k,x^l)=(D_\Phi^<-D^\aT_\Phi)(t,x^k,x^l)=A_{kl}\frac\pi\nuEff \theta(-t) \theta(t+\tau)$ one obtains 
(independent of $\alpha$ and $\beta$!)
\begin{align}
	J^\gtrless(T)=\pm \frac\pi\nuEff \left[A_{ki}\theta(t_1-T)\theta(T-t_1+\tau)-A_{kj} \theta(t_2-T)\theta(T-t_2+\tau)]\right.
\end{align}
With constraints $\lvert t_1-t_2\rvert \ll \tau$ for $i=j$ and $\lvert t_1-t_2\rvert =\tau$ for $i\neq j$ (cf.\ Eq.~(\ref{eqn:FPICurrentApprox})), the instanton action thus reads
\begin{align} \label{eqn:FPIinstantonActResult}
	i\tilde\act_t[\varphi_\ast] = \left\lbrace
	\begin{array}{ll}
		-\lvert t_1-t_2\rvert/\tau_\varphi, & i=j\\
		-\tau/\tau_\varphi -i eV\tau \epsilon_{ij}(R_{1\ast}(eV)-R_{2\ast}(eV)) \pi^{-1} \sin\frac\pi\nuEff, & i\neq j
	\end{array}
	\right.
\end{align}
with dephasing rate
\begin{align} \label{eqn:FPIdephRateFuller}
	\tau_\varphi^{-1} =-4 \sin^2\frac\pi{2\nu^\ast} \sum_{k=1,2}\left(\tilde P^>_{kk}(0)+\tilde P^<_{kk}(0)\right)
\end{align}
which due (\ref{eqn:FPIDressedPolOpKKAsymp}) to becomes the expression (\ref{eqn:FPIdephRate}) in the limit $\lvert eV \tau\rvert\gg 1$, see Fig.~\ref{fig:FPIdephRate}. The purely imaginary contributions to (\ref{eqn:FPIinstantonActResult}) correspond to (perturbatively small) renormalization corrections to bias voltage and will be neglected further on. We will also neglect the instanton action for $i=j$.

Combining these results with (\ref{eqn:FPICurrentApprox}) we obtain for the incoherent and interference corrections to current due to tunneling
\begin{align} 
	\Delta I_\dir &= I_{11}+I_{22} = -\frac{e^2}{2\pi} \left(R_{1\ast}(eV)+R_{2\ast}(eV)\right) V, \label{eqn:FPICurrentFinalResultDir}\\
	I_\osc &= -\frac e{\pi \tau} R_{12\ast}(eV)\ \sgn V\ e^{-\tau/\tau_\varphi}\sin(\lvert eV\tau\rvert-\pi/4\nuEff) \cos\varphi_{\rm AB} \label{eqn:FPICurrentFinalResultAB}
\end{align}
with Aharonov-Bohm phase 
\begin{align}
		\varphi_{\rm AB}=2\pi \phi/\phi_0+(\mu_++\mu_-)\tau-2\tau\varphi_0
\end{align}
which in the limit $\wc\tau\gg 1$ gives (\ref{eqn:FPIABPhase}). For large bias, 
$\wc\gg \lvert eV\rvert \gg \epsth$, this yields the dimensionless conductances given in (\ref{eqn:FPIcond})  and (\ref{eqn:g_osc}).

Concluding this section we turn to the case $\nuEff=1$. According to Eq.~(\ref{eqn:FPIDressedPolOpNu1}) the dominant contribution to current is
\begin{align}
	e^{-1} I_{ij}^{\alpha\beta} \approx e^{i\tilde \act_t[\varphi_\ast]}\Big\rvert_{t_1-t_2\approx 0}\ e^{2i\epsilon_{ij}\tau\varphi_0} \tilde P^{\alpha\beta}_{ij}(t^\ast=0).
\end{align}
The $i\neq j$-contribution of Eq.~(\ref{eqn:FPICurrentApprox}) is also present, but subleading. The instanton action $i\tilde \act_t[\varphi_\ast]$ can be evaluated following the same line of reasoning as for $\nuEff\ge 2$, yielding $i\tilde \act_t[\varphi_\ast]=-\lvert t_1-t_2\rvert/\tau_\varphi$ which can be neglected. The dominant contribution to current is thus
\begin{align*}
	I_{\osc}^{\nuEff=1}=-\frac e{\pi\tau}\ r_1\bar r_2\ \wc\tau e^\gamma\ \cos\varphi_{\rm AB}.
\end{align*}
In the limit $\wc\tau\gg1$ its bias dependence is negligible, in contrast to the contribution $I_\osc$, Eq.~(\ref{eqn:FPICurrentFinalResultAB}). Therefore, while the latter is subleading in \emph{current} for $\nuEff=1$, it yields the leading contribution to \emph{conductance}: $g_\osc=\partial I^{\nuEff=1}_\osc/\partial V+\partial I_\osc/\partial V\approx\partial I_\osc/\partial V$, i.e.\  giving the previous result (\ref{eqn:g_osc}).

\section{Summary and Outlook}
\label{sect:conclusion}

In this article, we have developed a general theoretical framework for
investigation of electronic properties of a broad class of nonequilibrium
nanostructures consisting of one-dimensional channels coupled by tunnel
junctions and/or by impurity scattering. Our formalism is based on
nonequilibrium version of functional bosonization. 

We have derived a nonequilibrium (Keldysh) action for this class of systems,
Eq.~(\ref{eqn:fullTunAction}), that has a form reminiscent of the theory of full
counting statistics. In order to make the further analytical progress
possible, we have developed a method based on a combination of a weak-tunneling
approximation and an instanton (saddle-point) approach. 
 
We have supplemented a detailed exposition of the formalism by
two important applications: (i) tunneling spectroscopy of a biased
Luttinger liquid with an impurity, and (ii) nonequilibrium quantum Hall
Fabry-P\'erot interferometry. One more application, quantum Hall Mach-Zehnder
interferometry, has been  presented recently by
two of us with Schneider in a separate
publication, Ref.~\cite{Schneider11}. 

The developed Keldysh-action formalism has allowed us to explore the rich interaction-induced physics of all the above problems, which includes, in particular, renormalization and dephasing far from
equilibrium as well as an oscillatory voltage dependence of visibility of Aharonov-Bohm oscillations (``lobe structure``).  The theoretical results are in good agreement with experiments on Fabry-P\'erot and Mach-Zehnder interferometers. 

We close the paper by identifying future research perspectives; a work in some of these directions is currently underway.

\begin{itemize}
\item[(i)] 
It is important to see under what conditions and how can one proceed in a controllable way 

\item[(ii)] A further important direction is the analysis of asymptotic behavior of relevant types of Fredholm determinants. Recently, such an analysis has been carried out for Toeplitz determinants arising in the problem of nonequilibrium Luttinger liquids and related models \cite{gutman11}. Required generalizations include, in particular, block Toeplitz determinants which arise naturally in the case of models with several channels coupled by tunneling.

\item[(iii)] One of prospective applications of our formalism is related to edge states of quantum spin Hall topological insulators. Of great interest is a generalization on setups based on fractional quantum Hall (or, more generally, fractional topological insulator) edge states.

\end{itemize}

\section{Acknowledgements}

We thank S.T.~Carr, L.I.~Glazman, I.V.~Gornyi, M.~Heiblum, A.~Kamenev, N.~Ofek, D.G.~Polyakov, and B.~Rosenow  for
useful discussions. This work was supported
by the German-Israeli Foundation and by the Deutsche Forschungsgemeinschaft via CFN and SFB/TR 12.

\begin{appendix}
\section{Derivation of Keldysh Action} \label{app:derivAction}

\begin{figure}[ht]
	\centering{\includegraphics[scale=.4]{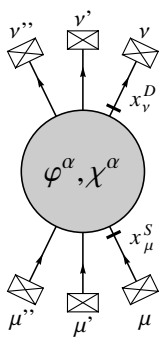}}
	\caption{Sketch of the system (shaded blob) which is connected to reservoirs (rectangles) via source (drain) leads, depicted by incoming (outgoing) lines. The leads $\mu$ ($\nu$) enter (leave) the systems at contact positions $x_\mu^\sr$ ($x_\nu^\dr$).}\label{fig:systemleads}
\end{figure}

The first two Appendices, \ref{app:derivAction} and
\ref{app:nazarov}, are devoted to the proof of (\ref{eqn:fullTunAction}) and
more specifically to the calculation of the Jacobian $J[\varphi,\psi,\bar\psi]$ 
corresponding to the gauge transformation $\psi_\mu^\mp\to e^{i\Theta_\mu^\mp}
\psi^\mp_\mu$ with (\ref{eqn:phaseHSField}). Since this is a linear
transformation in $\psi$, $\bar\psi$, its Jacobian is independent of $\psi$,
$\bar\psi$ and can be therefore computed as
	\begin{equation} \label{eqn:defJac}
		J[\varphi] \equiv e^{i\act_f[\varphi]}\equiv \int \DD(\psi,\bar\psi)\, e^{i\act[\varphi,\psi,\bar\psi]}= \Tr \left\lbrace\hat \varrho_0 \hat U[\varphi^+]^\dagger\hat U[\varphi^-]\right\rbrace
	\end{equation}
with density operator $\hat\varrho_0$ and the (many-particle) time evolution operators $\hat U[\varphi^\mp]$. The latter describes any \emph{single-particle} dynamics the electrons undergo in the system \emph{and the leads} (see Fig.~\ref{fig:systemleads}), say scattering and propagation through time-dependent potentials $\varphi^\alpha(t)$. These potentials may have a non-trivial Keldysh structure, thus the superindex $\alpha=\mp$ which refers to the forward/backward branch of the Keldysh contour, respectively.

\subsection{Integrating out Fermions with Klich's Formula}

There are several ways to evaluate the many-particle trace (\ref{eqn:defJac}).
Here we employ an approach that generalizes a derivation of the
full counting statistics in Ref.~\cite{Klich}.
In Appendix \ref{app:nazarov} we present an alternative derivation
which keeps closely to the spirit of Ref.~\cite{Nazarov08} where an
analogous action was derived for the case of a single compact scatterer. 

A central
mathematical statement proven in Ref.~\cite{Klich} relates traces of
certain \emph{many}-particle operators with determinants of associated
\emph{single}-particle operators. We denote the (many-particle) Fock space
representation of single-particle operators $C$ by $\Gamma(C)\equiv \sum
c^\dagger_i \bra{i} C\ket{j} c_j$. Here, $\lbrace \ket{i}\rbrace_i$ is some
single-particle basis, and $c_i$ ($c^{\dagger}_i$) annihilates (creates) an
electron in state $\ket{i}$. Then, the following identity holds:
	\begin{equation}\label{eqn:mathKlich}
		\Tr \left (e^{\Gamma(A_1)}\cdots e^{\Gamma(A_n)}\right) = \det\left(\xUnit+e^{A_1}\cdots e^{A_n}\right).
	\end{equation}

To proceed we write the density operator in the form
\begin{equation} \label{eqn:densOp}
	\hat \varrho_0 = \frac{e^{\Gamma(F)}}{\det \left(\xUnit+e^F\right)}
\end{equation}
where the single-particle operator $F=\sum \alpha_i N_i$ is a suitable linear combination of (single-particle!) ``number operators'' $N_i=\ket i \bra i $ in the reservoirs. E.g.\ in thermal equilibrium $F=-\beta H_0$ with some appropriate Hamiltonian $H_0=\sum \epsilon_i N_i$.
The many-particle time-evolution operator is canonically discretized as
\begin{equation*}
	\hat U[\varphi^\alpha] = \Texp\left[-i \int_{-\infty}^\infty \!\!\dd t\,\hat H[\varphi^\alpha(t)]\right] = \lim_{\overset{\Delta t\to 0\ }{N\to \infty}} \prod_{i=1}^N\,e^{-i\Delta t\ \Gamma(H[\varphi^\alpha(t_i)])}.
\end{equation*}
Hence, Eq.~(\ref{eqn:defJac}) is a trace over a (infinite) product of operator
exponentials which, according to (\ref{eqn:mathKlich}), is
	\begin{align} \label{eqn:FCS}
		J[\varphi] &= \frac{\Tr \left[e^{\Gamma(F)}\hat U[\varphi^+]^\dagger\hat U[\varphi^-]\right]}{\det(\xUnit+e^F)} = \frac{\det\left(\xUnit+e^F U[\varphi^+]^\dagger U[\varphi^-]\right)}{\det(\xUnit+e^F)} \nonumber \\ &= \det\left(\xUnit+f\left(U[\varphi^+]^\dagger U[\varphi^-]-\xUnit\right)\right)
	\end{align}
	with the single-particle time-evolution operator $U[\varphi^\alpha]$ (not to be confused with $\hat U[\varphi^\alpha]$) and the occupation number operator $f=\left[\xUnit+e^{-F}\right]^{-1}$.

\subsection{Wave packet representation}
\begin{figure}[ht]
		\centering{\includegraphics[scale=0.4]{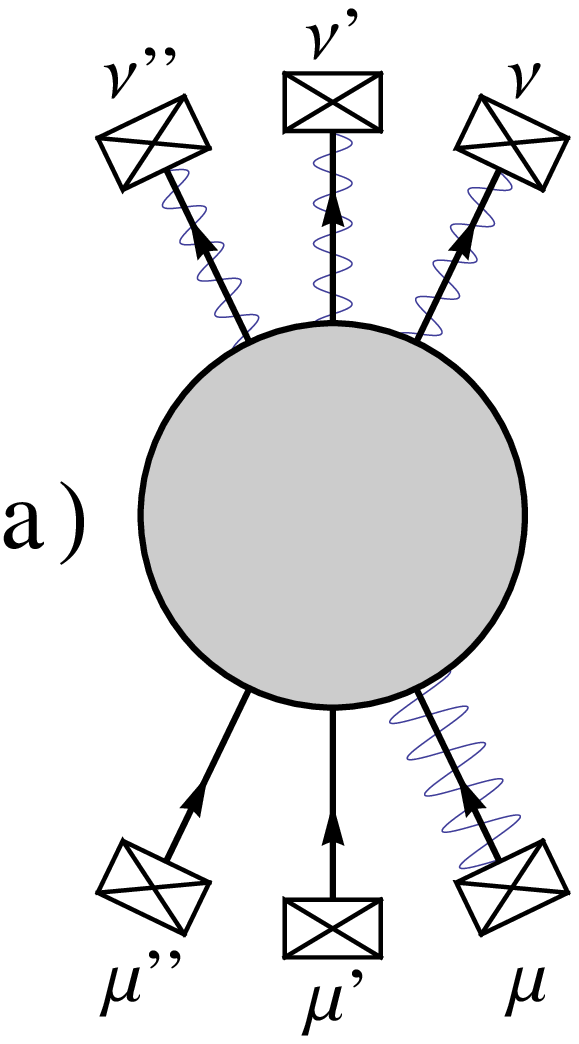}\quad\includegraphics[scale=0.4]{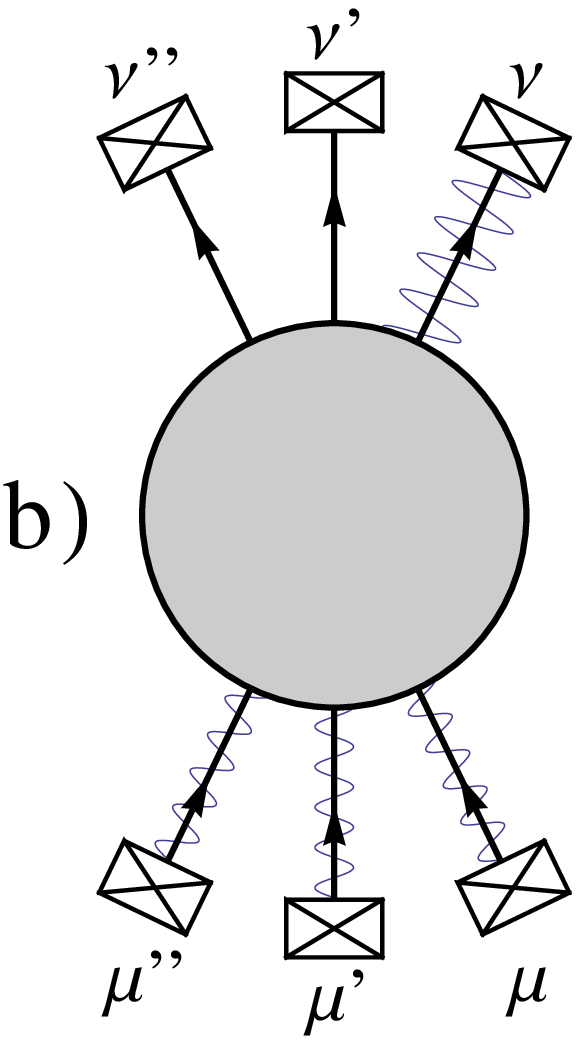}}
		\caption{(a) Spatial distribution of incoming scattering state $\ket {\epsilon \mu}^\In$ (wavy lines). It extends in source lead $\mu$ and arbitrary drain leads $\nu$, but in no other source leads. (b) Spatial distribution of outgoing scattering state $\ket{\epsilon \nu}^\Out$ (wavy lines). It extends in drain lead $\nu$ and arbitrary source leads $\mu$, but in no other drain leads.} \label{fig:scatteringStates}
	\end{figure}

In a next step, we follow Landauer's original idea\cite{Landauer} and represent the time-evolution operators with respect to the wave packet bases, relating them to the single-particle scattering matrices. Using a more compact notation for the single-particle time-evolution operator
	\begin{equation*}
		U^\alpha(t_1,t_0) =\Texp\left[-i \int_{t_0}^{t_1} \!\!\dd t\,H[\varphi^\alpha(t)]\right],\qquad t_0<t_1,
	\end{equation*}
	(hence $U[\varphi^\alpha]=U^\alpha(\infty,-\infty)$),
Eq.~(\ref{eqn:FCS}) can be brought into the form
	\begin{equation} \label{eqn:FCS_timeEvolv}
		J[\varphi] = \lim_{t_{\pm}\to \pm \infty} \det\left\lbrace\xUnit+f\left[U^+(t_+,t_-)^\dagger U^-(t_+,t_-) -\xUnit\right]\right\rbrace.
	\end{equation}
	We fix some time-\emph{in}dependent reference Hamiltonian, say $H_0\equiv H[\varphi\equiv0]$, which contains the lead kinematics as well as scattering, but no interaction or current counting. Then the incoming/outgoing scattering states with respect to $H_0$ form two natural bases of the single-particle Hilbert space, see Fig.~\ref{fig:scatteringStates}. Each state is characterized by its energy $\epsilon$ and the lead $\mu$ through which it enters/leaves the system: $H_0 \ket{\epsilon \mu}^{\In/\Out} =\epsilon \ket{\epsilon \mu}^{\In/\Out}$. The two bases are hence $\left (\ket{\epsilon \mu}^\In\right)_{\epsilon;\mu}$ and $(\ket{\epsilon \mu}^\Out)_{\epsilon;\mu}$.

	For the sake of argument we will assume the lead channels $\mu$ to be one-dimensional (1D), semi-infinite and non-dispersive with constant velocity $v_\mu$. We use the convention $v_\mu>0$, i.e.\ for source channels $-\infty<x<x^\sr_\lambda$ and for drain channels $x^\dr_\lambda<x<\infty$, and choose the normalization such that $\vphantom{\bra{\epsilon' \mu'}}^{\In/\Out}\bracket{\epsilon' \mu'}{\epsilon \mu}^{\In/\Out}=\delta_{\mu\mu'}\delta(\epsilon-\epsilon')$ is satisfied.

	The incoming state $\ket{\epsilon \mu}^\In$ is a \emph{plane wave} in source lead $\mu$ and spreads into drain leads $\nu$. It vanishes in all other source leads $\mu'$: ${}^\In\bracket{x \mu'}{\epsilon \mu}^\In = \delta_{\mu'\mu}\frac 1{\sqrt{2\pi v_\mu}} e^{i\epsilon x /v_\mu}$ (where $\ket{x\mu}$ is the eigenstate of the position operator in channel $\mu$). Analogous statements hold for the outgoing states.

	Let us now construct the incoming (outgoing) wave packet basis at reference time $t_-$ ($t_+$). 
For that we define
	\begin{align*}
		\ket{t\mu}^\In \equiv \int\!\!\frac{\dd \epsilon}{\sqrt{2\pi}}\, e^{i\epsilon (t-t_- -x^{\sr}_\mu/v_\mu)} \ket{\epsilon \mu}, \quad
		\ket{t\lambda}^\Out \equiv \int\!\!\frac{\dd \epsilon}{\sqrt{2\pi}}\, e^{i\epsilon (t-t_+ -x^{\dr}_\lambda/v_\lambda)} \ket{\epsilon \lambda}.
	\end{align*}
	Note that $t$ is not a parameter which describes the time-evolution of a state ``$\ket{\mu}$'' but labels the state $\ket{t\mu}$ similar to $\epsilon$ in $\ket{\epsilon\mu}$. The two new bases are thus $(\ket{t \mu}^\In)_{t;\mu}$ and $(\ket{t \mu}^\Out)_{t;\mu}$.

	To shed light on the meaning of label $t$, we study the time-evolution of the newly constructed states with respect to the reference Hamiltonian $H_0$. For 
	 time $t'$ one has
	\begin{align*}
		\leftexp{\In}{\bra{x\mu'}}e^{-iH_0(t'-t_-)}\ket{t\mu}^\In = \delta_{\mu'\mu}\ \sqrt{v_\mu} \delta(x-x_\mu^S+v_\mu(t-t')).
	\end{align*}
	\begin{figure}[ht]
		\centering{\includegraphics[scale=0.3]{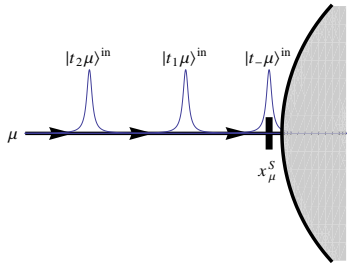}}
		\caption{Zoom into source lead $\mu$: Sketch of spatial distribution of wave packet states $\ket {t_- \mu}^\In$, $\ket {t_1 \mu}^\In$, $\ket {t_2 \mu}^\In$ with $t_2>t_1>t_-$.} \label{fig:arrivingWavePacket}
	\end{figure}

	This is a wave packet in source channel $\mu$ which propagates towards contact $x_\mu^S$, arriving there at time $t$, see Fig.~\ref{fig:arrivingWavePacket}. After entering the system it may split and spread in some complicated way. Similarly, the outgoing state $\ket{t\lambda}^\Out$ may be distributed in some complicated manner inside the system, however tuned such that at time $t$ it arrives at drain contact $x_\lambda^D$ and continues propagation as a single wave packet in drain channel $\lambda$. Summarizing, 
	\begin{align} \label{eqn:wpContactH0}
		e^{-iH_0(t-t_-)}\ket{t\mu}^\In\equiv \sqrt{v_\mu}\ket{x_\mu^S}, \quad e^{-iH_0(t-t_+)}\ket{t\lambda}^\Out\equiv \sqrt{v_\lambda}\ket{x_\lambda^D}
	\end{align}
	are wave packets residing at contacts $x_\mu^\sr$, $x_\lambda^\dr$, and thus being independent of $t_\mp$ and $t$.
	
	Making the assumption that interaction, counting etc.\ is switched on and off adiabatically such that $H(t')=H_0$ for $t'\not\in[t_-,t_+]$, we now argue that the same simple relations hold when taking the full Hamiltonian $H(t)$ into account,
	\begin{align} \label{eqn:wpContactH}
		U^\alpha(t,t_-)\ket{t\mu}^\In= \sqrt{v_\mu}\ket{x_\mu^S}, \quad U^\alpha(t,t_+)\ket{t\lambda}^\Out= \sqrt{v_\lambda}\ket{x_\lambda^D}.
	\end{align}
	These relations are a direct consequence of (\ref{eqn:wpContactH0}) and the fact that potential $\varphi^\alpha$ is restricted spatially and temporally:
	For $t>t'>t_-$ incoming wave packet $U(t',t_-)\ket{t\mu}$ is completely contained in the source lead where we assume $\varphi^\alpha$ to be absent. For $t<t'<t_-$ potential $\varphi^\alpha(t')$ is again ineffective since not switched on yet. Therefore, $U(t,t_-)\ket{t\mu}^\In=e^{-iH_0(t-t_-)}\ket{t\mu}$ for all $t$. The reasoning is analogous for outgoing states.

	We are now able to give the operator $fU^+(t_+,t_-)^\dagger U^-(t_+,t_-)$ in the wave-packet representation. For source channels $\mu,\mu'$ the matrix elements read
	\begin{multline} \label{eqn:matrixElemO}
		\leftexp{\In}{\bra{t\mu}}fU^+(t_+,t_-)^\dagger U^-(t_+,t_-) \ket{t' \mu'}^\In \\= \sum_{\mu''}\,\int\!\!\dd t'' \, f_{\mu\mu''}(t,t'') \sum_{\lambda}\int\!\!\dd t'''\,\leftexp{\In}{\bra{t'' \mu''}}U^+(t_+,t_-)^\dagger \ket{t''' \lambda}^\Out \leftexp{\Out}{\bra{t'''\lambda}}U^-(t_+,t_-)\ket{t'\mu'}^\In.
	\end{multline}
	Since the leads are populated by the reservoirs such that the occupation number of the \emph{incoming} states is fixed, $\leftexp{\In}{\bra{\epsilon \mu}}f \ket{\epsilon' \mu'}^\In= \delta_{\mu \mu'}\delta(\epsilon -\epsilon') f_\mu(\epsilon)$, the distribution function in time domain simplifies to
	\begin{equation*}
		f_{\mu\mu'}(t,t')\equiv \bra{t\mu} f\ket{t'\mu'} = \delta_{\mu\mu'}\ \int \!\frac{\dd \epsilon}{2\pi}\, e^{-i\epsilon (t-t')}\ f_\mu(\epsilon).
	\end{equation*}

	The matrix elements of the time-evolution operator further reduce to
	\begin{equation} \label{eqn:defSM}
		\leftexp{\Out}{\bra{t\lambda}}U(t_+,t_-)\ket{t' \mu}^\In = \leftexp{\Out}{\bra{t\lambda}} U(t_+,t) U(t,t') U(t',t_-) \ket{t'\mu}^\In = \sqrt{v_\lambda v_\mu} \bra{x^\dr_\lambda} U(t,t') \ket{x^\sr_\mu } \equiv S_{\lambda \mu} (t,t')
	\end{equation}
	which defines the scattering matrix $S=S[\varphi]$.

	In the wave-packet representation Eq.~(\ref{eqn:FCS_timeEvolv}) can be
written
	\begin{equation*}
		J[\varphi] = e^{i\act_f[\varphi]}=\det\left[\xUnit- f+ S^{+\dagger} S^- f\right]
	\end{equation*}
where the determinant is to be taken with respect to source lead indices
and arrival times. The log of this determinant appears in our general result stated in
Eq.~(\ref{eqn:fullTunAction}). The retarded and advanced part of polarization operator which are 
present in Eq.~(\ref{eqn:fullTunAction}) are not reproduced within the method of this section since they
represent itself the quantum anomaly. Their structure does not depend on the actual nonequilibrium
state of the system and can be deduced from the analysis of the fermion action in the absence
of tunneling as it was discussed in section~\ref{sect:model}.

\subsection{Construction of Scattering Matrix} \label{sect:scattMatrix}
	According to (\ref{eqn:defSM}) the scattering matrix element $S_{\lambda\mu}(t,t')$ is the transition amplitude for a peak residing in the (incoming) lead $\mu$ at time $t'$ to a peak residing in the (outgoing) lead $\lambda$ at time $t$. One expects that it is the summed amplitude for all possible trajectories which connect $\mu$ with $\lambda$. 
	We will briefly demonstrate this assuming that the scatterer is a network of simple blocks which are connected to each other via ``interface channels'' which may be ``outgoing'' with respect to one block and ``incoming'' with respect to a neighboring one. The electronic state residing in the interface channel $\lambda$ is denoted by $\ket{x_\lambda}$. It corresponds to a wave packet which is leaving one of the blocks and about to enter another one. We further assume that the blocks be simple enough such that each of them can be characterized by a unique dwell time $\tau$ (possibly different for each block), i.e.\ a wave packet which enters the block at time $t$ will definitely leave it at (exactly) $t+\tau$, through whatever channel: $U(t+\tau,t)\ket{x_i} = \sum_f u_{fi}(t)\ket{x_f}$ where $\ket{x_i}$ is an incoming interface state and the sum extends over outgoing interface states $\ket{x_f}$. This defines the functions $u_{fi}(t)$ and the scattering matrix elements
	\begin{equation} \label{eqn:defSmallS}
		s_{fi}(t',t) \equiv \delta(t'-t-\tau) \sqrt{\frac{v_i}{v_f}} u_{fi}(t)
	\end{equation}
where we have assigned a characteristic velocity $v_\lambda$ to each interface channel $\lambda$. 
The full scattering matrix $S_{\lambda\mu}(t',t)$ can be constructed out of elements  $s_{fi}(t',t)$
with the use of following decomposition property.
\begin{figure}
\begin{center}
		\centering{\includegraphics[scale=0.4]{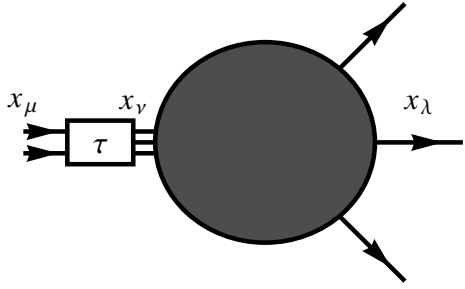}}
		\caption{Sketch of the system: A block characterized by the dwell time $\tau$ is connected to some part of the system (the shaded blob) via interface states $\ket{x_\nu}$.}
		\label{fig:Decomposition}
\end{center}
\end{figure}
Consider the situation sketched at Fig.~\ref{fig:Decomposition}.
Because of the decomposition property $U(t',t)=U(t',t+\tau)U(t+\tau,t)$ of the time-evolution operator, the amplitude for the transition from a peak at $t$ in channel $\mu$ to one at $t'$ in $\lambda$ is 
	\begin{equation} \label{eqn:decomp}
		\sqrt{v_\lambda v_\mu} \bra{x_\lambda} U(t',t)\ket{x_\mu} 
		 = \sum_{\nu} \int\!\!\dd t'' \sqrt{v_\lambda v_\mu} \bra{x_\lambda} U(t',t'') \ket{x_\nu} s_{\nu\mu}(t'',t).
	\end{equation}
	Obviously, the inner transition amplitude can be decomposed further in the same manner, and the full scattering matrix element $S_{\lambda \mu}$ turns out to be the sum of amplitudes $A^{(p)}_{\lambda\mu}$, each of them corresponding to a possible path $p$ connecting the incoming state $\mu$ with the outgoing one $\lambda$. As $p$ passes through a certain number of building blocks, $A^{(p)}_{\lambda\mu}$ is the product of the blocks' scattering matrix elements. For completeness we note that, since each trajectory will end in the outgoing leads, each decomposition will end with
	\begin{equation}
		\sqrt{v_{\lambda'} v_\lambda} \bra{x_{\lambda'}} U(t',t)\ket{x_\lambda} = \delta(t'-t) \delta_{\lambda'\lambda}
	\end{equation}
	for $\lambda,\lambda'\in \Out$.

	\subsubsection*{Simple blocks}	
	Having convinced ourselves of the usefulness of definition (\ref{eqn:defSmallS}) we turn to simple two examples of building blocks: wires with fluctuating potentials and point scatterers. Simple as they are, a broad class of devices, including quantum wire junctions and electronic interferometers, can be modeled as a network of these construction units, and in the following we will restrict ourselves to such systems.
	
	The corresponding scattering matrices are found by considering the time-evolution of wave packets $\psi(t,x)\equiv \bra x U(t,t_0)\ket{x_i}$  which satisfy the Schr\"odinger equation
	\begin{equation} \label{eqn:WPSchroedingerEq}
		i\partial_t \psi(t)=H(t)\psi(t),\quad \text{with initial condition\ }\psi(t_0,x)=\delta(x-x_i).
	\end{equation}
	
	We list the results here, giving all necessary definitions in the subsequent paragraphs:
	\begin{center}
		\begin{tabular}{|lc|l|}
			\hline
			\multicolumn{2}{|l|}{\emph{Construction unit}} & \emph{Scattering matrix}\\
			Chiral wire & \showgraph{scale=0.2}{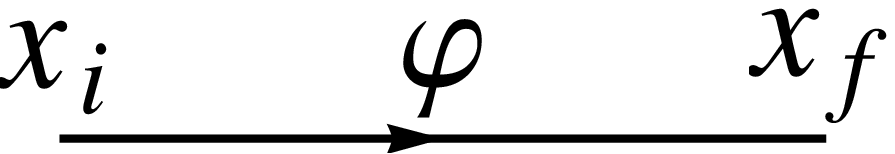}& $ \wire^\alpha(t',t) = e^{i\vartheta_{fi}^\alpha(t')} \Delta(t',t)$\\[.2cm]
			Point scatterer & \showgraph{scale=0.2}{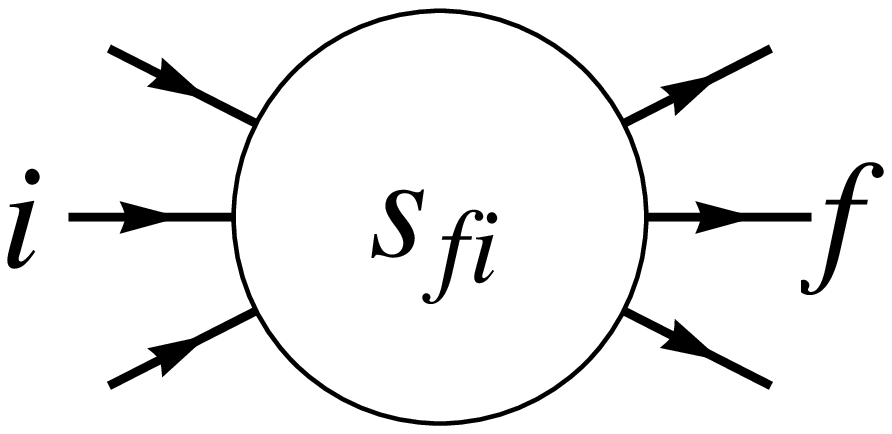} & $s_{fi}(t',t)=s_{fi}\ \delta(t'-t)$\\
			\hline
		\end{tabular}
	\end{center}
	\setcounter{paragraph}{0}
	\paragraph{Chiral wire with fluctuating potential}
	A possible block is a chiral non-dispersive wire where fermions propagate with constant velocity $v$ in a fluctuating potential $\varphi^\alpha(t,x)$. Similar to the leads wires are described by a single coordinate  $x$, extending from $x_i$ to $x_f$, such that the dwell time is $\tau=\frac{x_f-x_i}v$. It is taken into account by the ``delay operator'' $\Delta(t',t)\equiv \delta(t'-t-\tau)$.
	The presence of the potential leads to accumulation of phase 
	%
	\begin{equation} \label{eqn:accPhase}
	 	\vartheta_{fi}(t)=-v^{-1} \int^{x_f}_{x_i}\!\!\dd x'\, \varphi(x',t-(x-x')/v).
	\end{equation}
	The wire connects exactly one incoming to one outgoing channel and the scattering matrix has just one entry $\wire(t',t)$ in channel space. 

	\paragraph{Point scatterer}
	Another possible construction unit is the point scatterer which connects one-dimensional incoming (index $i$) and outgoing (index $f$) channels such that scattering occurs instantaneously (dwell time $\tau=0+$). The scatterer is characterized by the unitary time-independent scattering matrix $s_{fi}$. If all channels $\lambda$ have a linear dispersion with constant velocity $v_\lambda$, then according to (\ref{eqn:WPSchroedingerEq}) the wave packet incident from channel $i$ (at time $t_0$) is 
	\begin{align*}
		\psi(t,x,\lambda) &= \delta_{\lambda i}\ \delta(x_i+v_i(t-t_0)-x))+\sum_f \delta_{\lambda f}\ s_{fi}\ \sqrt{\frac{v_f}{v_i}}\ \delta(x_f+v_f(t-t_0)-x), \qquad t\approx t_0.
	\end{align*}
	Since the state extends over several channels the wave function is a function of both channel index $\lambda$ and channel coordinate $x$, and the sum is to be taken over all outgoing channels $f$.
	\paragraph{Counting Fields}
	Up to now we have not addressed the issue of counting fields, claiming that they can be treated on the same footing as fluctuating potentials, a statement to be proven in this section.

	The number of electrons which flow through a certain point $\tilde x$ in the time interval $\tilde t_0 < t <\tilde t$ is described quantum- mechanically by the operator $N=\int_{\tilde t_0}^{\tilde t}\!\!\dd t\, I(t)$. The current operator $I=v \psi^\dagger(\tilde x) \psi(x)$ becomes time-dependent in Heisenberg representation, $I(t)= e^{i H(t-t_-)} I e^{-i H(t-t_-)}$ ($t_-$ is some reference time at which the initial state of the system is fixed), $v$ is the fermion velocity in the considered channel. According to Levitov and Lesovik~\cite{Levitov93} 
the correct generating functional of charge transfer through $\tilde x$ is
	\begin{align*}
		\mathcal Z(\chi) = \left\langle U_{-\chi}(\tilde t,\tilde t_0)^\dagger U_\chi(\tilde t,\tilde t_0)\right\rangle\quad \text{with } U_\chi(\tilde t,\tilde t_0)=\Texp\left[i\frac \chi2\int_{\tilde t_0}^{\tilde t}\!\!\dd t\, I(t)\right].
	\end{align*}
	From the properties $U_\chi(\tilde t_0,\tilde t_0)=\xUnit$ and $i\partial_{\tilde t}U_\chi(\tilde t,\tilde t_0)=-\frac \chi2 I(\tilde t)U_\chi(\tilde t,\tilde t_0)$ we conclude that
	\begin{align*} 
		U_\chi(\tilde t,\tilde t_0)=e^{iH(\tilde t-t_-)}e^{-i(H-\frac\chi2 I)(\tilde t-\tilde t_0)} e^{-iH(\tilde t_0-t_-)}
	\end{align*}
	is a possible alternative representation.
	Thus,
	\begin{align*}
		\mathcal Z(\chi)=\left\langle \T_\keldC \exp \left[-i \int_\keldC\!\!\dd t'\, H_\chi(t')\right]\right\rangle
	\end{align*}
	is the Keldysh partition sum with respect to the Hamiltonian $H^\alpha_\chi(t)\equiv H+v \int\!\!\dd x\, A_\chi^\alpha(x,t) \psi^\dagger(x)\psi(x)$ where time integration and ordering is to be understood along the Keldysh contour $\keldC$ and we defined the local vector potential $A_\chi^\alpha(t,x) = \alpha\ \frac{\chi(t)} 2 \delta(x-\tilde x)$ with the ``time-dependent'' counting field $\chi(t') \equiv \chi \theta(\tilde t-t)\theta(t-\tilde t_0)$.

	The corresponding scattering matrix reads $s^\alpha (t',t) = \delta(t'-t)\ e^{-i \frac\alpha2 \chi(t)}$ and can be incorporated in the total scattering matrix.

\section{Alternative Derivation of the Action for 1D Systems}
\label{app:nazarov}

\begin{figure}[ht]
	\centering{\includegraphics[scale=.5]{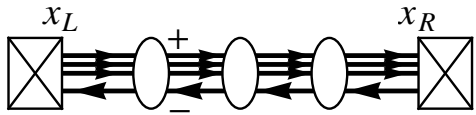}}
	\caption{Exemplary 1D system with 3 right-moving ($+$) and 1 left-moving ($-$) channels and 3 scatterers. The system extends $x_L<x<x_R$.} \label{fig:nonChiralSetup}
\end{figure}

In this appendix we sketch an alternative derivation of the action
${\act_f[\varphi]}$ which holds for one-dimensional (1D)
systems. As shown in Fig.~\ref{fig:nonChiralSetup} they may consist of several
channels. Either direction of propagation is the same in all of them (in case of
which the setup is referred to as ``chiral'') or there are two distinct possible
directions: ``right'' (+) and ``left''(-). The derivation generalizes
that of Ref.~\cite{Nazarov08} and some of the arguments given already
there will be not reiterated here.

We use a single coordinate system $x_L<x<x_R$ to describe all channels with left (right) contact position $x_{L(R)}$, i.e. velocities in right-(left-)moving channels are positive (negative): $\vF$ ($-\vF$). The fermionic action is
\begin{equation*}
	\act_{\mathrm{f}}[\psi,\bar\psi,\scattPot] = \int\!\!\dd x\,\int_\keldC \!\!\dd t\,\Psi^\dagger\left(i\partial_t+i\tau_3 \vF\partial_x -\impS-\scattPot\right)\Psi
\end{equation*}
where $\Psi=(\psi^\alpha_\mu)$ are vectors of Grassmannian fields with Keldysh and channel/direction indices $\alpha$ and $\mu$, resp, $\tau$ are Pauli matrices in direction space, $\vF$ is Fermi velocity, $\impS$ is the self-energy correction due to the coupling to the reservoirs (see below), and $\scattPot$ denotes the (temporally and spatially local) potential under the influence of which the electrons traverse the system. The latter may be the static scattering, Hubbard-Stratonovich or counting potential.

To obtain the integrated action $\act$, first its variation with respect to $\scattPot$ is considered, 
\begin{equation} \label{eqn:varAction}
	i\delta \act = -\int\!\!\dd x \int \dd t\, \Tr\!_\tau \left[\delta\scattPot^-(t,x) G^\T(x,t-0;x,t)-\delta\scattPot^+(t,x) G^\aT(x,t+0;x,t)\right] \equiv  -\int\!\!\dd x\,\Tr \delta \scattPot(x)\sigma_3 G(x,x)
\end{equation}
where $\sigma$ are Pauli matrices in Keldysh space, and the first trace $\Tr_\tau$ is taken over  Keldysh and channel indices, while the ``full'' trace $\Tr$ is additionally taken over real time.
The full fermionic Green's function $G$ is needed at coinciding spatial coordinates, where it is discontinuous because of linear dispersion. The above time shift regularization, which takes into account that fermion fields in the action are normal-ordered, is equivalent to the identification
\begin{equation} \label{eqn:regGF}
	G^{\T(\aT)}(x,t;x',t') \stackrel{x\to x'}{\longrightarrow}\frac 1{2i\vF} \left(g^{\T (\aT)}(x;t,t') - \delta\left(t-t'\right)\right)
\end{equation}
with the quasiclassical Green's function
\begin{align*}
	g^{\alpha\beta}(x,t,t')= i\vF \left(G^{\alpha\beta}(x+0,x;t,t')+G^{\alpha\beta}(x-0,x;t,t')\right) .
\end{align*}

\newcommand{\UU}{\mathcal U}
\newcommand{\GF}{Green's function}
\newcommand{\const}{\mathrm{const.}}
\newcommand{\cst}{\const}
\subsection{Transfer matrices} The Green's function $G(x_1,x_2)$ is related to $G(x_1',x_2')$ via the the single-particle transfer matrices $\tm(x,x')$ for spatial evolution from $x'$ to $x$: $G^{\alpha\beta}(x_1,x_2)=\tm^\alpha(x_1,x_1') G^{\alpha\beta}(x_1',x_2')\tm^\beta(x_2',x_2)^\dagger$ or, for quasiclassical Green's functions, $g^{\alpha\beta}(x)=\tm^\alpha(x,x') g^{\alpha\beta}(x')\tm^\beta(x,x')^\dagger$. 
Defining $\UU^\alpha(x;t,t')=i\vF^{-1}\delta(t-t') \tau_3 (i\partial_{t'}-\scattPot^\alpha(x,t'))$ (or more clearly in energy representation $\UU^\alpha(x)(\epsilon,\epsilon')=i\vF^{-1} \tau_3 (2\pi\delta(\epsilon-\epsilon')\epsilon-\scattPot^\alpha(x,\epsilon-\epsilon'))$), the transfer matrices are given by
\begin{equation*}
	\tm^\alpha(x,x') = \left\lbrace\begin{array}{ll}
	                        	\mathcal O_{x_1} \exp\left[\int_{x'}^x\dd x_1\, \UU^\alpha(x_1)\right], & x\ge 'x,\\
	                        	\tilde {\mathcal O}_{x_1} \exp\left[\int_{x'}^x\dd x_1\, \UU^\alpha(x_1)\right], & x\le 'x,\\
	                        \end{array}\right.
\end{equation*}
where $\mathcal O_x$ ($\tilde{\mathcal O}_x$) orders subsequent operators with respect to their space coordinate $x$, smaller (larger) coordinates ordered to the right. 
Consequently, transfer matrices are diagonal in Keldysh space (with $\tm^\alpha$ being only related to $\scattPot^\alpha$) and satisfy
\begin{equation} \label{eqn:scattPotTM}
	\delta \tm^\alpha(x_R,x_L) = \int_{x_L}^{x_R} \dd x\, \tm^\alpha(x_R,x) \left[-i \vF^{-1} \tau_3 \delta \scattPot^\alpha(x)\right] \tm^\alpha(x,x_L).
\end{equation}
For short, we will also write for the total transfer matrix $\tm\equiv \tm(x_R,x_L)$. Note that in chiral systems considered in Ref.~\cite{Nazarov08} scattering matrices $S$ can be used. While coinciding with the transfer matrices in chiral systems, they differ in non-chiral ones, the two being related via
\begin{equation} \label{eqn:tm_sm}
	S = \tau_3\left(\tau_++\tm\tau_-\right)^{-1} \left(\tm\tau_++\tau_-\right)\tau_3
\end{equation}
with the projectors $\tau_\pm=(\tau_0\pm\tau_3)/2$ in direction space. In chiral, say right-moving, systems $\tau_3=\tau_+=\xUnit$, $\tau_-=0$ and indeed $S=\tm$. Generally, $S$ is unitary, $S S^\dagger=\xUnit$, $\tm$ is  pseudo-unitary, $\tm(x,x')\tau_3\tm(x,x')^\dagger=\tau_3$. By defining $\bar g\equiv \sigma_3g \tau_3$ spatial evolution amounts for the similarity transformation
\begin{align*}
		\bar g(x)=\tm(x,x') \bar g(x')\tm(x,x')^{-1}.
\end{align*}
The factor $\sigma_3$ ensures the normalization property $\bar g(x)^2=\xUnit$.

Using (\ref{eqn:regGF}), (\ref{eqn:scattPotTM}) and $\bar g(x)=\tm(x,x_L)\bar g(x_L) \tm(x_R,x_L)^{-1}\tm(x_R,x)$ the variation of the action (\ref{eqn:varAction}) can be simplified to
\begin{align}
	i\delta \act =& -\int\!\!\dd x\,\Tr \tau_3 \delta \scattPot(x)\ \frac1{2i\vF} \left(\bar g(x)-\sigma_3\tau_3\right)\nonumber\\
		=& -\frac 12 \int\!\!\dd x\,\Tr \left[-i\vF^{-1} \tau_3\delta\scattPot(x)\right] \tm(x,x_L) \bar g(x_L) \tm(x_R,x_L)^{-1} \tm(x_R,x_L) + \const\nonumber\\
		=& -\frac 12 \Tr\bar g(x_L) \tm^{-1} \delta\tm + \const \label{eqn:varAction1}
\end{align}
where we absorb all contributions to the action which are independent of distribution functions in ``$\const$''. We will show later on that they vanish.

\subsection{Reservoir Green's Functions} In a next step, $\bar g(x_L)$ is expressed in terms of the quasiclassical Green's function 
\begin{equation} \label{eqn:reservoirGF}
	g_{L(R)}(t-t') = \begin{pmatrix}
	            	g_{L(R)}^\T &  g_{L(R)}^<\\
			g_{L(R)}^> & g_{L(R)}^\aT
	            \end{pmatrix}_{t-t'}
		= \begin{pmatrix}
			\xUnit -2 f_{L(R)} & - 2 f_{L(R)}\\
			2(\xUnit-f_{L(R)}) & \xUnit -2 f_{L(R)}
		  \end{pmatrix}_{t-t'}
\end{equation}
 of the left (right) reservoirs (with distribution functions $f_{L(R)}$ which may have a non-trivial channel structure $f_{L(R)\mu}$). The Green's functions $\bar g_i=\sigma_3\bar g_i \tau_3$, $i=L,R$, are related to each other via a similarity transformation as follows: First, introducing the Keldysh matrices
\begin{align}
	L\equiv \frac1{\sqrt 2} \begin{pmatrix} \xUnit & -\xUnit \\ \xUnit & \xUnit\end{pmatrix},\quad
	\tilde U_i =\begin{pmatrix} \xUnit & (\xUnit-2f_i)\\ 0 & -\xUnit \end{pmatrix} = \tilde U_i^{-1}
\end{align}
one easily finds
\begin{align}
	L \bar g_i L^{-1} = \begin{pmatrix} \xUnit & 2(\xUnit-2f_i)\\ 0 & -\xUnit\end{pmatrix} \tau_3 = \tilde U_i^{-1} \sigma_3 \tau_3 \tilde U_i
\end{align}
and hence
\begin{gather} \label{eqn:nazGParamU}
	\bar g_i = U_i^{-1} \sigma_3\tau_3 U_i\quad \text{with}\quad U_i = \tilde U_i L = \frac 1{\sqrt 2}\begin{pmatrix} 2f^>_i & - 2 f^<_i\\ -\xUnit & -\xUnit \end{pmatrix}
\end{gather}
with $f_i^<=f_i$, $f_i^>=\xUnit-f_i$. Thus we haven proven
\begin{align}
	\bar g_R = U^{-1} \bar g_L U \quad \text{with}\quad U=U_L^{-1}U_R.
\end{align}

To relate the Green's function $\bar g(x_L)$ inside the system to its counterparts in the reservoirs one assumes that the dynamics inside the leads is governed by some relaxation process, say isotropization of momentum direction due to scattering off static white noise disorder. This is described by the self-energy contribution
\begin{equation*}
	\impS(t_1,x_1;t_2,x_2)=\delta(x_1-x_2) \left(-\frac i{2\tau_{\mathrm{rel}}}\right) \times \left\{\begin{array}{ll}
	                g_{L}(t_1,t_2), & x_1,x_2<x_L\\
									g_{R}(t_1,t_2), & x_1,x_2>x_R
	                                                               \end{array}
 \right..
\end{equation*}
Here $\tau_{\mathrm{rel}}$ denotes the relaxation time. The requirement that $G(x+\Delta x,x)$ vanish for infinite distances $\Delta x$ yields the boundary conditions\cite{Nazarov08},
\begin{equation} \label{eqn:boundCond}
(\xUnit+\bar g_{L})(\xUnit-\bar g(x_-))=0, \qquad (\xUnit-\bar g_{R})(\xUnit-\bar g(x_+))=0,
\end{equation}
again with $\bar g\equiv \sigma_3 g \tau_3$. 
Defining $\bar \tm\equiv U\tm$ the second equation is equivalent to $0=\left(\xUnit-\bar\tm^{-1} \bar g_L\bar \tm\right)(\xUnit+\bar g(x_L))$. Combining it with the first equation gives
\begin{align}
	0=\left(2\xUnit + \bar g_L -\bar\tm^{-1} \bar g_L \bar \tm\right)-\left(\bar g_L+\bar\tm^{-1}\bar g_L\bar\tm\right)\bar g(x_L)
\end{align}
and by inversion
\begin{align}\label{eqn:boundCondSol}
	\bar g(x_L) = \xUnit + 2(\xUnit-\bar g_{L}) \left( \bar g_{L} \bar\tm+\bar\tm \bar g_{L}\right)^{-1} \bar \tm
	\end{align}
where we have made use of $\bar g_L^2=\xUnit$.

	To rewrite this expression we choose a specific basis representation. Since $\bar g_L^2=\xUnit$ there exists one in which $\bar g_L=\diag(\xUnit,-\xUnit)$. In the very same representation we write
	\begin{equation*}
		\bar \tm =\begin{pmatrix}
		         	\bar \tm_{11} & \bar \tm_{12}\\
				\bar \tm_{21} & \bar \tm_{22}
		         \end{pmatrix}.
	\end{equation*}
	Then we have $\bar g_L \bar \tm+\bar \tm \bar g_L = 2\ \diag(\bar\tm_{11},-\bar \tm_{22})$, which is readily inverted, as well as $\xUnit+\bar g_L=2\ \diag(\xUnit,0)$, $\xUnit-\bar g_L=2\ \diag(0,\xUnit)$, and (\ref{eqn:boundCondSol}) gives
	\begin{equation} \label{eqn:basisRepG}
		\bar g(x_L)=\begin{pmatrix}
		            	\xUnit & 0\\
				-2\bar \tm_{22}^{-1} \tm_{21} & -\xUnit
		            \end{pmatrix}.
	\end{equation}
Defining 
\begin{equation} 
\mathcal D\equiv (\xUnit+\bar g_{L})/2 + \bar \tm (\xUnit-\bar g_{L})/2
\label{eq:Denominator}
\end{equation}
one may show that
	\begin{equation} \label{eqn:GDRep}
		\bar g(x_L)=\xUnit-\left(\xUnit-\bar g_{L}\right) \mathcal D^{-1} \bar\tm
	\end{equation}
	is equivalent to (\ref{eqn:basisRepG}). We have thus expressed $\bar g(x_L)$ entirely in terms of $\bar g_{L(R)}$ and the transfer matrix $\tm(x_R,x_L)$.

Substituting this result into (\ref{eqn:varAction1}) and using $\tm(x_R,x_L)^{-1} \delta\tm(x_R,x_L)=\bar \tm^{-1}\delta \bar\tm$ yields
\begin{multline}
	i\delta \act=\frac 12 \Tr (\xUnit-\bar g_L)\mathcal D^{-1} \delta\bar \tm +\const= \Tr \mathcal D^{-1} \delta \mathcal D + \const\\
	\Rightarrow i\act = \Tr \Ln \mathcal D +\const = \Tr\Ln \left[\frac{\xUnit+\bar g_L}2+\bar\tm \frac{\xUnit-\bar g_L}2\right] + \const \label{eqn:nazarovAction}
\end{multline}
\subsection{Role of Drain Distribution Functions} So far we have not been concerned with the channel structure explicitly and merely stated that there are 2 directions of motion, right($+$) and left($-$), each of which is realized by a certain (not necessarily equal) number of channels (possibly even zero in chiral systems). In both left and right reservoirs to each channel $\mu$ was assigned a distribution function $f_{L\mu}$ and $f_{R\mu}$. For a right-(left-)moving channel $f_{R\mu}$ ($f_{L\mu}$) is the \emph{drain} distribution function and one naturally wonders whether it should have an effect on the chiral fermions as long as the latter have not entered the drain reservoirs. We show here that this is not the case.

To this end we introduce the block decomposition with respect to channel indices, e.g.
\begin{align}
	\tm = \begin{pmatrix} \tm_{++} & \tm_{+-}\\ \tm_{-+} & \tm_{--} \end{pmatrix},
	\quad f_i =\begin{pmatrix} f_{i+} & \\ & f_{i-} \end{pmatrix},\quad i=L,R,
\end{align}
where e.g.\ $\tm_{+-}$, $f_{i+}$, $f_{i-}$ still may have channel structure $\left(\tm_{+-}\right)_{\mu\nu}$, $\left(f_{i+}\right)_\mu$, $\left(f_{i-}\right)_\nu$, however, with $\mu$ ($\nu$) extending exclusively over right-(left-)moving channels.

Introducing
\begin{gather}
		Q\equiv  U_R \tm  U^{-1}_L=\begin{pmatrix} Q_{11} & Q_{12} \\ Q_{21} & Q_{22} \end{pmatrix} =
	\begin{pmatrix}
		f_R^> \tm^-+f_R^<\tm^+ & -2 f_R^>\tm^- f_L^< + 2 f_R ^<\tm^+ f_L^>\\
		-\frac 12 \tm^- + \frac 12 \tm^+ & \tm^-f_L^< +\tm^+f_L^>
	\end{pmatrix}
\end{gather}
	and the projectors
\begin{gather}
	P_+ = \frac 12 (1+\sigma_3\tau_3) = \begin{pmatrix} \tau_+ & 0 \\ 0 & \tau_-\end{pmatrix},\quad
		P_- = \frac 12 (1-\sigma_3\tau_3) = \begin{pmatrix} \tau_- & 0 \\ 0 & \tau_+ \end{pmatrix}
\end{gather}
the action (\ref{eqn:nazarovAction}) reads
\begin{align*}
	i\act = \ln \Det \left(P_++QP_-\right)+\const = \ln\Det \left(P_++P_-QP_-\right)+\const
\end{align*}
where
\begin{align*}
	 	P_-QP_- = \begin{pmatrix}
		          	\tau_-Q_{11} \tau_- & \tau_- Q_{12}\tau_+\\
				\tau_+ Q_{21} \tau_- & \tau_+Q_{22}\tau_+
		          \end{pmatrix}.
\end{align*}
Writing out the direction structure explicitly yields
\begin{align*}
		i\act = \ln\Det \begin{pmatrix}
			f_{R-}^> \tm^-_{--}+f^<_{R-}\tm^+_{--} & -2 f^>_{R-}\tm^-_{-+} f^<_{L+}+2 f^<_{R-} \tm^+_{-+} f^>_{L+}\\
			-\frac 12 \tm^-_{+-} + \frac 12 \tm^+_{+-} & \tm^-_{++} f^<_{L+}+\tm^+_{++} f^>_{L+}
		                \end{pmatrix} + \cst
	\end{align*}
	This proves that the action depends only on $f_{L+}$ and $f_{R-}$, i.e.\ the source distribution functions.

\subsection{Tracing out Keldysh Structure}We now return to the expression (\ref{eqn:nazarovAction}). Since we already know that the result does not depend on $f_{L,-}$ and $f_{R,+}$ we may make the choice $f_{L,-}=f_{R,-}$ and $f_{R,+}=f_{L,+}$. In other words, we put $\bar g_L=\bar g_R=\diag(\bar g_{L+},\bar g_{R-})\equiv \bar g_\In\equiv\sigma_3 g_\In\tau_3$. As suggested by the subindex $g_\In$ contains the source (or incoming) distribution functions. It can be parametrized analogously to (\ref{eqn:nazGParamU}),
\begin{align}
	\sigma_3 g_\In = U_\In^{-1} \sigma_3 U_\In \quad \text{with } U_\In=\frac 1{\sqrt 2} \begin{pmatrix} 2 f^> & -2 f^< \\ -\xUnit & -\xUnit \end{pmatrix}
\end{align}
with $f$ being the matrix of source distribution distribution functions. Since $U$ is not needed anymore and hence $\bar \tm=\tm$, we recall the definition of $\mathcal D$ given by Eq.~(\ref{eq:Denominator}) and of scattering matrix S (\ref{eqn:tm_sm}) and obtain
\begin{align}
	\mathcal D =\frac{\xUnit+\sigma_3 g\tau_3}2 +\tm \frac{\xUnit-\sigma_3g\tau_3}2 = \left(\tau_++\tm \tau_-\right) \tau_3 U_\In^{-1} \left[\frac{\xUnit+\sigma_3}2 +U_\In SU_\In^{-1} \frac{\xUnit-\sigma_3}2\right]U_\In.
\end{align}
With
\begin{align*}
	\tilde Q \equiv  U_\In S \ U_\In^{-1} = \begin{pmatrix} Q^{--} & Q^{-+} \\ Q^{+-} & Q^{++} \end{pmatrix}
					=\begin{pmatrix}
             	f^> S^- + f^< S^+ & -2 f^> S^- f^< + 2 f^< S^- f^>\\
							-\frac 12 S^- +\frac 12 S^+ & S^- f^< + S^+ f^>
	                                        \end{pmatrix}
\end{align*}
the action (\ref{eqn:nazarovAction}) is
\begin{align} \label{eqn:actWConst}
	i\act= \ln\Det \mathcal D + \const = \ln \Det Q^{++}+\const = \Tr \Ln \left[\xUnit-f+S^{+\dagger} S^-f\right]+\const
\end{align}
Up to anomalous terms, representing the R- and A-parts of the polarization operator,
and contributions which are independent of distribution functions $f$, this proves Eq.~(\ref{eqn:fullTunAction})

\subsection{Constant contributions}	What are the ``$\const$''-contributions to the action we keep ignoring? All that we know so far about them is their independence of distribution function $f$. According to (\ref{eqn:actWConst}) they can be recovered by substituting $f=0$ in the \emph{full} action (where ``$\const$'' is not neglected). So let us put $f\equiv 0$ in the rest of this section and recalculate the full action. Eq.~(\ref{eqn:boundCondSol}) can be evaluated explicitly now, yielding $g^\T (x_L)=  g_0^-(x_L)$ and $g^\aT (x_L)= g_0^+(x_L)^\dagger$ with 
	\begin{equation*}
		g_0(x_L) = \begin{pmatrix} \xUnit & 0 \\ 2 S_{-+} & \xUnit \end{pmatrix}=\xUnit +2  \begin{pmatrix} 0 & 0 \\ A_{-+}(x_L) & 0 \end{pmatrix}
	\end{equation*}
	where $A_{\nu\mu}(x;t',t)$ is defined as the amplitude for a $\mu$-wave packet at position $x$ and time $t$ to end up as a $\nu$-wave packet at time $t'$ and the same position $x$ (cf. Sect. \ref{sect:scattMatrix}). The above relation between $g^{\T/\aT}$ and $g_0^\mp$ holds for all positions $x$ with $g_0^\alpha(x)\equiv\tm^\alpha(x,x_L) g_0^\alpha(x_L) \tm^\alpha(x,x_L)^\dagger$. To calculate the latter we define (for given $x$) the ``scattering matrix''
	\begin{equation*}
	s \equiv \tau_3\left(\tau_++\tm(x,x_L)\tau_-\right)^{-1} \left(\tm(x,x_L)\tau_++\tau_-\right)\tau_3
	\end{equation*}
	of the region between $x_L$ and $x$ which  takes into account paths extending only within this region.
	\begin{figure}[ht]
		\centering{\includegraphics[scale=0.4]{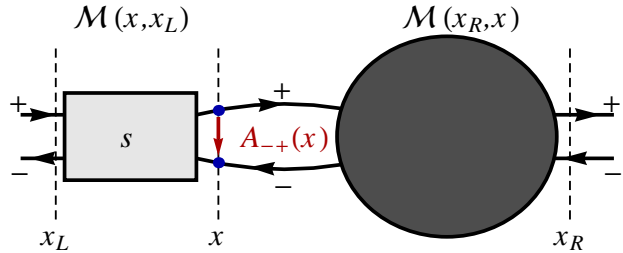}}
		\caption{Spatial evolution of $g$ from $x_L$ to $x$ with $\tm(x,x_L)$; $s$ is the corresponding ``scattering matrix''; the shaded blob represents the rest of the system. In contrast to $s$, paths contributing to $A_{-+}(x)$ may extend throughout the entire system.}
		\label{fig:movingRight}
	\end{figure}
	Using the recursion relations (see Fig.~\ref{fig:movingRight})
	\begin{gather*}
		S_{-+}= s_{-+}+s_{--} A_{-+}(x) s_{++}, \\
		A_{++}(x) = s_{+-} A_{-+}(x), \quad A_{+-}(x) = s_{+-}+s_{+-} A_{-+}(x) s_{+-},\quad A_{--}(x)= A_{-+}(x) s_{+-}
	\end{gather*}
	one obtains
	\begin{equation}
		g_0(x) =\xUnit +2  \begin{pmatrix}  A_{++}(x) & A_{+-}(x) \\ A_{-+}(x) & A_{--}(x) \end{pmatrix}.
	\end{equation}
	For a smooth potential $\scattPot(x)$ all paths contributing to $A_{\nu\mu}(x)$ have a non-vanishing flight time such that $A_{\nu\mu}(x;t,t)=0$. At point scatterers, which are a mere idealization, the Green's function $g(x)$ is not well-defined and we stay sufficiently far away from them. Concluding, for equal times the Green's function (\ref{eqn:regGF}) 
vanishes. Substituting this result into Eq.~(\ref{eqn:varAction}) yields $\delta \act\rvert_{f\equiv 0}=0$, hence $\const=\act\rvert_{f\equiv 0}$ is indeed a physically irrelevant constant.

\section{Weak Tunneling Regularization} \label{app:regularization}
In general, the scattering matrix $S^\alpha=S[\varphi^\alpha]$ depend on time in a 
complicated manner which makes the exact evaluation of the functional determinant (\ref{eqn:fullTunAction}) unfeasible. In this section we present an approximation scheme which applies for systems with weak tunneling.
 A tunneling action $\act_t$ is derived by subtracting a clean action $\act_0$ from the full one $\act$. By construction $\act_b$ is small and may be therefore expanded. 
The crucial step, the separation of actions, can be performed at zero temperature where reservoirs may have differing chemical potentials. As it will turn out, interaction effectively leads to a dressing of the point scatterers by phases $\Phi$.

\subsection{Clean Limit} \label{sect:cleanLimit}
The clean limit was discussed in Sect.~\ref{sect:model}. The action $\act_0$ is obtained from (\ref{eqn:fullTunAction}) by replacing $S$ by the clean scattering matrix $S^\alpha_{\ast} = \diag\left(e^{i\vartheta^\alpha_1} \Delta_1 ,\ldots,e^{i\vartheta^\alpha_N} \Delta_N\right)\equiv e^{i\vartheta^\alpha} \Delta$ with the delay operators $\Delta_\mu$ and the phases $\vartheta_\mu$ accumulated along the complete paths $x^\sr_\mu\to x^\dr_\mu$.

The main step in the quest of the tunneling action
\begin{equation*}
	\act_t = \act -\act_0 =i\Tr\Ln\left[\left(\xUnit-\bar f+S^{+\dagger}  S^-\bar f\right)\left(\xUnit-\bar f+S_\ast^{+\dagger} S_\ast^-\bar f\right)^{-1}\right]
\end{equation*}
is the inversion of the second bracket, a problem which is already dealt with in Ref.~\cite{Braunecker}. Since $S_\ast$ is diagonal in channel space, the operator can be inverted for each channel separately. So we consider
\begin{equation*}
	\left(\xUnit-\bar f+S_\ast^{+\dagger} S_\ast^- \bar f\right)_{\mu\mu}= \xUnit-f_\mu + \Delta_\mu^{-1} e^{i\left(\vartheta_\mu^--\vartheta_\mu^+\right)} \Delta_\mu f_\mu 
\end{equation*}
where $\Delta_\mu(t',t)=\delta(t'-t-\tau_\mu)$ is an appropriate time delay operator. Note that it commutes with the stationary distribution function $f_\mu$. For zero temperature we follow Ref.\cite{Braunecker}, according to which the inversion problem can be reformulated in terms of a certain Riemann-Hilbert problem, which is solved by defining the phases $\vartheta_\mu^\up$ and $\vartheta_\mu^\down$ by
\begin{align} \label{eqn:regPhases}
	\vartheta_\mu^\updown(t) \equiv \frac i{2\pi} \int\!\!\dd t'\, \frac{\vartheta_\mu^-(t')-\vartheta_\mu^+(t')}{t_\pm-t'} \equiv -\int\!\!\dd t'\, \left[ B(t-t')\mp\delta(t-t')\right]\left[\vartheta^-_\mu(t')-\vartheta^+_\mu(t')\right],
\end{align}
with $t_\pm\equiv t\pm i 0$. They satisfy $\vartheta^\up-\vartheta^\down=\vartheta^--\vartheta^+$, hence $\vartheta^--\vartheta^\up = \vartheta^+-\vartheta^\down\equiv \bar \vartheta$, and $f_\mu e^{i\vartheta_\mu^\up} f_\mu = e^{i\vartheta_\mu^\up} f_\mu$, $f_\mu e^{i\vartheta_\mu^\down} f_\mu = f_\mu e^{i\vartheta_\mu^\down}$. Thus, one obtains
\begin{equation*}
	\left(\xUnit-f_\mu + \Delta_\mu^{-1} e^{i\left(\vartheta_\mu^--\vartheta_\mu^+\right)} \Delta_\mu\  f_\mu \right)^{-1} = \Delta_\mu^{-1} \left[e^{-i\vartheta_\mu^\down}\left(\xUnit-f_\mu\right)+e^{-i\vartheta^\up_\mu} f_\mu\right] e^{i\vartheta_\mu^\down} \Delta_\mu,
\end{equation*}
and, taking now the full channel structure into account,
	\begin{equation*}
		\act_t = -i\Tr\Ln\left[\xUnit-\bar f+ \redS^{+\dagger} \redS^- \bar f\right] = i \sum_{n=1}^\infty \frac 1 n \Tr\left[\left(\xUnit- \redS^{+\dagger} \redS^- \right)\bar f\right]^n
	\end{equation*}
with the ``regularized'' scattering matrix
\begin{equation} \label{eqn:defRedS}
	\redS^\mp \equiv e^{-i\bar \vartheta} S^{\mp} \Delta^{-1}e^{-i\vartheta^\updown}\Delta.
\end{equation}

Defining in energy representation $N(\omega)=-\theta(-\omega)$ as the zero temperature limit of the Bose distribution function, we note that the function $B$ introduced in (\ref{eqn:regPhases}) is the ``generalized'' distribution function $B(\omega)=1+2N(\omega)$ ubiquitous in bosonic Keldysh formalism. This fact allows us to endow the phases with a Keldysh structure. We address this issue in the next section before turning to the computation of the regularized scattering matrix in the subsequent one.

\subsection{Phases and Keldysh Structure}
Before we turn to the computation of the regularized scattering matrix let us consider the Keldysh structure of the phases $\vartheta$. To this end we consider the clean path $x^\sr_\mu\to x^\dr_{\mu}$ containing the point $x$. An electron propagating from the source reservoir $x^\sr_\mu$ to $x$ (arrival time $t$) accumulates the phase
\begin{equation} \label{eqn:inPhase}
	\vartheta^\alpha_{\In.\mu}(t,x) \equiv -v^{-1}_\mu \int_{x^\sr_\mu}^{x}\dd x'\, \varphi^\alpha(x',t-(x-x')/v_\mu) =-D^r_{0\mu}\varphi^\alpha_\mu(x,t);
\end{equation}
likewise when traveling from $x$ (departure time $t$) to the drain reservoir $x^\dr_{\mu}$ it accumulates
\begin{equation} \label{eqn:outPhase}
	\vartheta^\alpha_{\Out.\mu}(t,x) \equiv -v^{-1}_\mu\int_{x}^{x^\dr_{\mu}}\dd x\, \varphi^\alpha(x,t+(x'-x)/v_\mu)=D^a_{0\mu}\varphi^\alpha_\mu(x,t)
\end{equation}
where we introduced the retarded/advanced bare electron-hole pair propagators 
\begin{align*}
	D^{r/a}_{0\mu}(t;x',x)=\pm\theta(\pm t)\delta(x'-(x+v_\mu t))
\end{align*}
along the the clean path. Quite obviously, the phase accumulated along the complete clean path satisfies
\begin{equation} \label{eqn:fullPhaseProp}
	\vartheta^\mp_\mu(t)= \vartheta^\mp_{\mu.\In}(x,t-(x^\dr_{\mu}-x)/v_\mu)+\vartheta^\mp_{\mu.\Out}(x,t-(x^\dr_{\mu}-x)/v_\mu)=-\left(D^r_{0\mu}-D^a_{0\mu}\right)\varphi^\mp_\mu(x,t-(x^\dr_{\mu}-x)/v_\mu).
\end{equation}
The propagators $D_{0\mu}^{r/a}$ directly make reference to propagation velocity $v_\mu$ and thus encode spectral (kinematic) properties of the electron-hole pairs. As usual they lack information about the system's state which one expects to be contained in the missing Keldysh component $D^k_{0\mu}$ of the propagator. In this sense the phases $\vartheta$ are related to the potential $\varphi$ via the linear operator $D_0$ which does not have the full Keldysh structure.

To overcome this ``deficiency'' we recall the definition made in the previous section
\begin{equation} \label{eqn:ThetaBar}
	\bar \vartheta_\mu=\vartheta_\mu^\mp-\vartheta_\mu^\updown = \frac 12 (B+\xUnit)\vartheta^-_\mu - \frac 12 
(B-\xUnit)\vartheta^+_\mu.
\end{equation}
Combining now (\ref{eqn:fullPhaseProp}) and (\ref{eqn:ThetaBar}) and using the stationarity of 
$B(t,t')=B_\mu(t-t')$ we obtain
\begin{equation}
	\bar \vartheta_\mu(t+(x^\dr_{\mu}-x)/v_\mu)=-\frac 12 (B+\xUnit)\left(D_{0\mu}^r-D_{0\mu}^a\right)\varphi_\mu^-(x,t)+\frac 12 (B-\xUnit)\left(D_{0\mu}^r-D_{0\mu}^a\right)\varphi_\mu^+(x,t)
\end{equation}
With that the phase that an electron, traveling along a piece of wire between $x_1$ and $x_2$, accumulates is
\begin{align*}
	\vartheta^\mp_{\mu 21}(t)=& -\vartheta^\mp_{\mu.\Out}(x^2,t)+\vartheta^\mp_{\mu.\Out}(x^1,t-(x^2-x^1)/v_\mu)\\
		\equiv& \Theta^\mp_\mu(x^2,t)-\Theta^\mp_\mu(x^1,t-(x^2-x^1)/v_\mu)
\end{align*}
with the phases
\begin{align}
	\Theta^\mp_\mu(x,t)&\equiv-\vartheta^\mp_{\mu.\Out}(x,t)+\bar \vartheta_\mu(t+(x^\dr_{\mu}-x)/v_\mu) \label{eqn:bigThetaDef} \\
		&= -\frac 12 \left[(B+\xUnit)D_{0\mu}^r-(B\mp\xUnit)D_{0\mu}^a\right]\varphi^-_\mu(x,t)+\frac12 \left[(B-\xUnit)D_{0\mu}^r-(B\mp\xUnit)D_{0\mu}^a\right]\varphi^+_\mu(x,t) \nonumber\\
		& = -\left[D_{0\mu}^{T/>}\varphi_\mu^-(x,t)-D_{0\mu}^{>/\aT} \varphi_\mu^+(x,t)\right]. \nonumber
\end{align}
We thus have managed to rewrite the scattering matrix of chiral wires $\wire^\alpha(x_2,x_1)=e^{i\vartheta^\alpha_{21}}\Delta(x_2,x_1)=e^{i\Theta^\alpha(x_2)}\Delta(x_2,x_1)e^{i\Theta^\alpha(x_1)}$ in terms of phases with full Keldysh structure. Formally we started with expressions which did not contain any distribution functions whatsoever and somewhat artificially included them by redefining phases. 
The usefulness of such construction will become apparent in the next section.  

\subsection{Construction of ``Regularized'' Scattering Matrix}
We have previously shown that the full scattering matrix $S$ can be constructed out of simpler units. We prove in this subsection that the same statement holds for the regularized scattering matrix $\redS$.

Since the full scattering matrix elements are amplitude sums $S^\alpha_{\nu\mu}=\sum_p A^{(p)\alpha}_{\nu\mu}$ over all paths $p$ connecting $x^\sr_\mu$ with $x^\dr_\nu$, definition (\ref{eqn:defRedS}) directly implies that $\redS^\mp_{\nu\mu}=\sum_p \tilde A^{(p)\mp}_{\nu\mu}$ is the sum of the \emph{regularized} amplitudes 
\begin{equation} \label{eqn:regAmpl}
\tilde A^{(p)\mp}_{\nu\mu} \equiv  e^{-i\bar \vartheta_\nu} A^{(p)\mp}_{\nu'\mu} \Delta_\mu^{-1}e^{-i\vartheta^\updown_\mu}\Delta_\mu=e^{-i\Theta^\mp_\nu(x^\dr_{\nu})} A^{(p)\mp}_{\nu\mu} e^{i\Theta^\mp_\mu(x^\sr_\mu)}
\end{equation}
over the same paths $p$ with phases $\Theta_{\mu(\nu)}$ along the clean paths $x^\sr_\mu\to x^\dr_{\mu}$ and $x^\sr_\nu\to x^\dr_{\nu}$. Note that $\vartheta^\mp_{\nu.\Out}(x^\dr_{\nu},t)=0$ and $\vartheta^\mp_{\mu.\Out}(x^\sr_\mu,t)=\vartheta^\mp_\mu(t+(x^\dr_{\mu}-x_\mu)/v_\mu)$, and thus definition (\ref{eqn:bigThetaDef}) implies $\Theta^\mp_\nu(x^\dr_{\nu},t)=\bar\vartheta_\nu(t)$ and $\Theta^\mp_\mu(x^\sr_\mu,t)=-\vartheta^\updown_\mu(t+(x^\dr_{\mu}-x^\sr_\mu)/v_\mu)$.

We now consider a generic path $p$ which starts at incoming lead channel $\mu$,  winds through an alternating sequence of wires and point scatterers (numbered $1,2,\ldots, N$ with positions $x_1, x_2,\ldots, x_N$), eventually ends in outgoing channel $\nu'$. The corresponding full amplitude is
\begin{align*}
	A^{(p)\mp}_{\nu\mu}&=\wire^\mp_\mu(x^\dr_{\nu},x^N) s^N_{\nu\lambda} \cdots s^2_{\rho\kappa}\wire^\mp_\kappa(x^2,x^1) s^1_{\kappa\mu} \wire^\mp_\mu(x^1,x^\sr_\mu)\\
		& = e^{i\Theta^\mp_\nu(x^\dr_\nu)}\Delta_\nu(x^\dr_{\nu},x^N) e^{-i\Theta^\mp_\nu(x^N)} s^N_{\nu\lambda} \cdots s^2_{\rho\kappa}e^{i\Theta_\kappa^\mp(x^2)}\Delta_\kappa(x^2,x^1)e^{i\Theta^\mp_\kappa(x^1)} s^1_{\kappa\mu} e^{i\Theta^\mp_\mu(x^1)}\Delta_\mu(x^1,x^\sr_\mu) e^{-i\Theta^\mp_\mu(x^\sr_\mu)}
\end{align*}
and simply becomes
\begin{align*}
	\tilde A^{(p)\mp}_{\nu\mu} &= \Delta_\nu(x^\dr_{\nu},x^N) e^{-i\Theta^\mp_\nu(x^N)} s^N_{\nu\lambda} \cdots e^{i\Theta_\kappa^\mp(x^2)}\Delta_\kappa(x^2,x^1)e^{i\Theta^\mp_\kappa(x^1)} s^1_{\kappa\mu} e^{i\Theta^\mp_\mu(x^1)}\Delta_\mu(x^1,x^\sr_\mu)\\
\end{align*}
upon regularization. This implies that regularization of the amplitudes amounts to regularization of the scattering matrices of the building blocks:
The effect of fluctuating potentials (and counting fields) is removed from all wires and incorporated in the tunneling phases $$\Phi^\mp_{\nu\mu}(x^j,t)\equiv\Theta^\mp_\mu(x^j,t)-\Theta^\mp_\nu(x^j,t)$$ which dress the point scattering amplitudes $s^j_{\nu\mu}$.
This result is summarized in the table below:

	\begin{center}
		\begin{tabular}{|l|l|}
			\hline
			\emph{Construction unit} & \emph{Regularized scattering matrix}\\
			Chiral wire & $ \Delta(t',t)$\\
			Point scatterer & $\tilde s^\alpha_{\nu\mu}(t',t)=s_{\nu\mu}\ e^{i\Phi^\alpha_{\nu\mu}(t,\bar x)}\ \delta(t'-t)$\\
			\hline
		\end{tabular}
	\end{center}

\newcommand{\source}{{\rm source\ }}
\newcommand{\drain}{{\rm drain\ }}
\section{Second Order Expansion} \label{app:secondOrder}
We prove the second order expansion (\ref{eqn:TunActPolOp}) in tunneling strength (``$\tun$'')  starting from (\ref{eqn:TunAct2Exp}). Obviously, $\mathcal A_{\nu\mu}\equiv \sum_\lambda (\tilde S^{+}_{\lambda\nu})^\dagger \tilde S^-_{\lambda\mu}$ is the sum over all paths winding forward and along the forward Keldysh branch (i.e.\ with potentials $\varphi^-$) from source $\mu$ to some drain $\lambda$ and then \emph{backwards} along the backward branch (with potentials $\varphi^+$) to source $\nu$ (for short: ``${\rm source\ } \mu \stackrel - \to {\rm drain\ } \lambda \stackrel + \to {\rm source\ } \nu$''). The backward part $\bar p^+$ is obtained by time-reversal of a physical, forward path $p^+$ (source $\nu\stackrel +\to$ drain $\lambda$) with amplitude $\tilde A^{(p^+)}_{\lambda\mu}$ and has amplitude $\tilde A^{\bar p^+}_{\nu\lambda}\equiv \tilde A^{(p^+)\dagger}_{\lambda\nu}$ (note that hermitian conjugation, $\dagger$, reverses the order of partial amplitudes in the product). Denoting with $p^-$ the forward part of $p=p^-\oplus \bar p^+$, the total amplitude is $\tilde A^{(p)}_{\nu\mu}\equiv\tilde A^{(p^+)\dagger}_{\lambda\nu} \tilde A^{(p^-)}_{\lambda\mu}$.

In the chosen order of accuracy we only take paths with a total of 2 or less tunneling events into account. 
The expanded tunneling action is $\act_t=\act_1+\act_2$ with
\begin{align}
	\act_1\equiv &i\Tr \sum_\mu (\xUnit-\mathcal A)_{\mu\mu}f_\mu=i\sum_\mu\Tr\left[f_\mu-\sum_{p_1}\tilde A^{(p_1)}_{\mu\mu} f_\mu\right], \label{eqn:pathSum1}\\
	\act_2 \equiv &\frac i2 \Tr \sum_{\mu\neq \nu} \mathcal A_{\mu\nu}f_\nu \mathcal A_{\nu\mu} f_\mu = \frac i 2  \sum_{\mu\neq\nu}\sum_{p_2=p'_2\oplus p''_2} \Tr\tilde A^{(p''_2)}_{\mu\nu}f_\nu \tilde A^{(p'_2)}_{\nu\mu}f_\mu. \label{eqn:pathSum2}
\end{align}
The sum $\sum_p$ in (\ref{eqn:pathSum1}) extends over paths $p_1$: ``$\source \mu \stackrel-\to \drain \nu \stackrel+\to \source \mu$'' with 2 or less tunneling events (obviously the number has to be even), while the sum in (\ref{eqn:pathSum2}) extends over paths $p_2=p'_2\oplus p''_2$: 
\begin{align*}
	&\text{``}\underbrace{\source \mu \stackrel-\to \drain \kappa \stackrel+\to \source}_{p'_2} \underbrace{\nu \stackrel-\to \drain \lambda \stackrel+\to\source \mu}_{p''_2}\text{''},\\
	\text{or equivalently, }&\text{``}\underbrace{\vphantom{\mu}\source \nu \stackrel-\to \drain \lambda \stackrel+\to\source}_{p''_2} \underbrace{\mu \stackrel-\to \drain \kappa \stackrel+\to \source \nu}_{p'_2}\text{''}
\end{align*}
where the equivalence is ensured by the cyclic invariance of the trace. Since $\mu\neq\nu$ subpaths $p'_2$ and $p''_2$ each contain at least one tunneling event, i.e.\ in our approximation \emph{exactly one}. Depending on whether tunneling occurs on the forward or backward Keldysh branch $\kappa,\lambda=\mu\ {\rm or\ }\nu$. It is quite obvious that no matter how often a path evolves in time forward and backward (once for $\act_1$, twice for $\act_2$), as long as it starts and ends in the same reservoir and contains exactly 2 tunneling events, it exactly involves 2 different channels $\mu\neq \nu$. We will use this fact for a systematic classification of all paths.

To warm up we consider first the simplest (and least interesting) paths, which contribute to $\act_1$ and are of the form ``${\rm source}\ \mu\stackrel -\to{\rm drain\ } \mu\stackrel +\to{\rm source\ }\mu$'' without any tunneling taking place. Time delay operators are canceled exactly (since forward and backward paths coincide geometrically) and no phases are accumulated (since after regularization they are carried only by the tunneling amplitudes). The total amplitude is thus a product of forward scattering amplitudes $s_{\mu\mu}^j$ of scatterers $j$ along the clean path $x^\sr_\mu\to x^\dr_{\mu}$: 
\begin{align} \label{eqn:amplTunFree}
	\prod_j \lvert s_{\mu\mu}^j\rvert^2 = \prod_j\left(\xUnit - \sum_{\nu\neq\mu} \lvert s_{\nu\mu}^j\rvert^2\right) = \xUnit- \sum_j\sum_{\nu\neq\mu}\lvert s^j_{\nu\mu}\rvert^2 + \mathcal O(\tun^4).
\end{align}

All other relevant paths contain exactly 2 tunneling events. Not surprisingly there is a whole plethora of them and we are well advised to proceed systematically. To this end and according to the observation made before we classify these paths with respect to the pair $(\mu,\nu)$  of different channels $\mu\neq \nu$ involved and the two scatterers $i$ and $j$ (possibly $i=j$) at which tunneling $\mu\to\nu$ and $\nu\to\mu$ respectively occurs. Note that in this classification classes $(ij;\mu\nu)$ and $(ji,\nu\mu)$ are identical. What classes $(ij;\mu\nu)$ are possible, of course, depends on the topology of the considered network (since e.g.\ not all scatterers are even connected to a given channel $\mu$). But, as we will show, once a class is fixed, its contribution to $\act_t$ is essentially independent of topology!

\begin{figure}[ht]
	\centering{\includegraphics[scale=0.3]{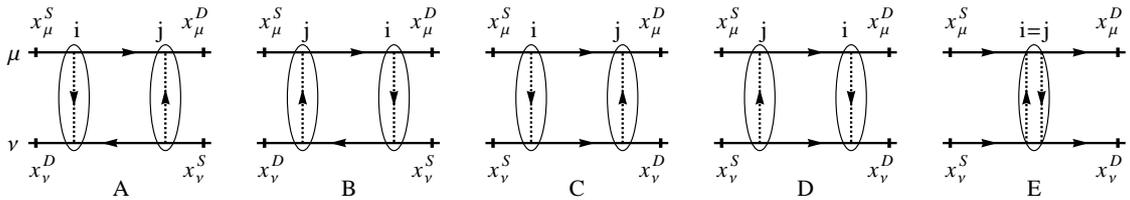}}
	\caption{The 5 topologically distinct configurations for class $(ij;\mu\nu)$. All channels except for $\mu$,$\nu$ and all scatterers except for $i$,$j$ (``distorted white circles'') are dropped. Scatterers $i\neq j$ are different in A-D which allows for 4 different orderings along channels $\mu$ and $\nu$. In E tunneling occurs twice at the same scatterer $i=j$. }\label{fig:networkTop}
\end{figure}
Fig.~\ref{fig:networkTop} shows the 5 topologically distinct configurations of channels $\mu$, $\nu$ and scatterers $i$, $j$ for given class $(ij;\mu\nu)$. Arrows at the scatterers indicate direction of tunneling. All scatterers except for $i$, $j$ are dropped. They are involved only in forward scattering events and thus lead to corrections $\mathcal O(\tun^3)$. In this way propagation from source $x^\sr_\mu$ to scatterer $i$ just leads to an amplitude $\Delta_{i.\mu.\In}$ which accounts for the finite flight time $\tau^i_{\mu.\In}$. Analogously we define $\Delta_{j.\mu.\In}$ and the same for $\nu$, and further
\newcommand{\XX}{{\mathcal X}}
\begin{align*}
	\XX^-_i &\equiv \Delta^{-1}_{i.\nu.\In} \tilde s^{i-}_{\nu\mu} \Delta_{i.\mu.\In}=\Delta^{-1}_{i.\nu.\In} e^{-i\Phi^-_{\mu\nu}(x^i)} \Delta_{i.\mu.\In}\ s^i_{\nu\mu},\\
	\XX^+_i &\equiv \Delta^{-1}_{i.\nu.\In} \left(s^{i+}_{\mu\nu}\right)^\dagger\Delta_{i.\mu.\In}=\Delta^{-1}_{i.\nu.\In} e^{-i\Phi^+_{\mu\nu}(x^i)} \Delta_{i.\mu.\In}\ \bar s^i_{\mu\nu},\\
	\XX^-_j &\equiv \Delta^{-1}_{j.\mu.\In} \tilde s^{j-}_{\mu\nu} \Delta_{j.\nu.\In}=\Delta^{-1}_{i.\nu.\In} e^{i\Phi^-_{\mu\nu}(x^j)} \Delta_{j.\nu.\In}\ s^j_{\mu\nu},\\
	\XX^+_j &\equiv \Delta^{-1}_{j.\mu.\In} \left(s^{j+}_{\nu\mu}\right)^\dagger\Delta_{j.\nu.\In}=\Delta^{-1}_{j.\mu.\In} e^{i\Phi^+_{\mu\nu}(x^j)} \Delta_{j.\nu.\In}\ \bar s^j_{\nu\mu}.
\end{align*}

For paths of given type $p_1$ or $p_2$ fixing the Keldysh branches on which the 2 tunneling events $\mu\to\nu$ (at $i$), $\nu\to\mu$ (at $j$) take place characterizes the corresponding paths uniquely provided they exist. Whether they exist or not depends on topology. Remarkably, they always do  if the 2 tunneling events are required to occur on \emph{different} Keldysh branches.
\begin{figure}[ht]
	\centering{\includegraphics[scale=0.4,angle=-0]{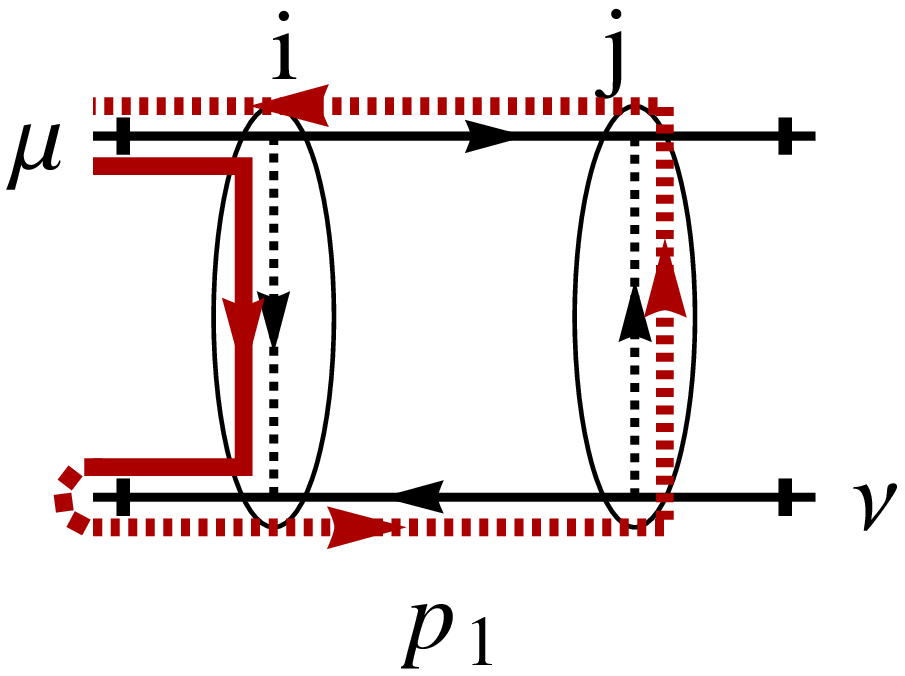}\qquad \includegraphics[scale=0.4,angle=-0]{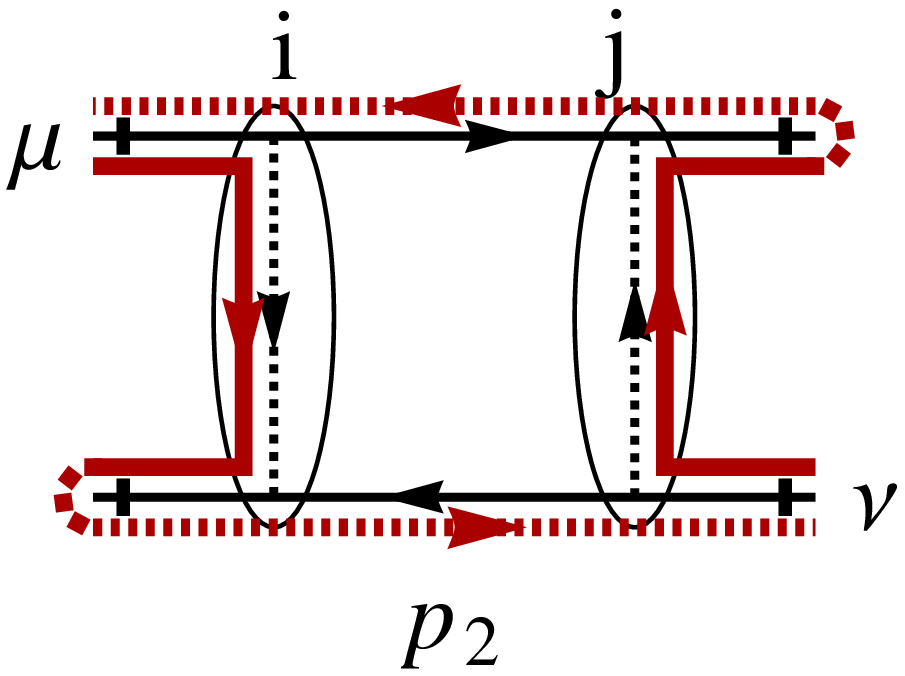}}
	\caption{Paths for topology A with $i:\mu\to\nu$ on the forward, $j:\nu\to\mu$ on the backward branch. The forward parts are represented by  solid thick lines; the backward parts by dashed lines. The first path is of type $p_1$, the second one of type $p_2$.}\label{fig:topALesser}
\end{figure}
Fig.~\ref{fig:topALesser} shows the paths for $i:\mu\to\nu$ occurring on the forward, $j:\nu\to\mu$ on the backward branch for configuration A. Similar paths can be drawn for all other configurations. As an example, we consider the first path in Fig.~\ref{fig:topALesser} which is of type $p_1$. It starts at source $\mu$, tunnels at $i$ on the forward branch, arrives at drain $\nu$, evolves backwards, tunnels at $j$ on the backward branch and returns to source $\mu$. Its amplitude is $\tilde A^{(p_1)}_{\mu\mu}=\XX^+_j\XX^-_i$. The contribution of both paths to $\act_t$ is
\begin{align*}
	\act^<_{ij;\mu\nu} = - i \Tr\left[\XX^+_j \XX^-_if_\mu-\XX^+_jf_\nu \XX^-_if_\mu\right]=-i\Tr\left[e^{-i\Phi^-_{\mu\nu}(x^i)} \Pi_{ij;\mu\nu}^<e^{i\Phi^-_{\mu\nu}(x^j)}\right]
\end{align*}
where tunneling polarization operators $\Pi_{ij;\mu\nu}$ are defined in (\ref{eqn:tunPolOp1}) and we have made use of $s^j_{\mu\nu}=-\bar s^j_{\nu\mu}+\mathcal O(\tun^2)$. Similarly $\act^>_{ij;\mu\nu}$ with $\Pi^>_{ij;\mu\nu}$ is obtained by considering paths with tunneling at $i$ on the backward, at $j$ on the forward branch. As mentioned previously the very same paths can be drawn for all topologies and yield the same action.

The story is less neat if both tunneling events occur on the same Keldysh branch: While $p_2$-type contributions to $\act^\T_{ij;\mu\nu}$ and $\act^{\aT}_{ij;\mu\nu}$ are still independent from topology, $p_1$-type contributions are not as universal. Taking all paths carefully into account results in the table below
\begin{center}
\begin{tabular}{c|c|c}
	& $\Pi^\T_{ij;\mu\nu}(t)/\left[-s^i_{\nu\mu}\bar s^j_{\nu\mu}\right]$ & $\Pi^\aT_{ij;\mu\nu}(t)/\left[-s^i_{\nu\mu}\bar s^j_{\nu\mu}\right]$\\\hline\hline\\[-0.3cm]
	A & $-\tilde f^<_\mu(t) \tilde f^<_\nu(-t)$ & $\tilde f^>_\mu(t) \tilde f^<_\nu(-t) +\tilde f^<_\mu(t) \tilde f^>_\nu(-t)+\tilde f^<_\mu(t) \tilde f^<_\nu(-t)$\\
	B & $\tilde f^>_\mu(t) \tilde f^<_\nu(-t)+\tilde f^<_\mu(t) \tilde f^>_\nu(-t)+\tilde f^<_\mu(t) \tilde f^<_\nu(-t)$ & $-\tilde f^<_\mu(t) \tilde f^<_\nu(-t)$\\
	C & $\tilde f^<_\mu(t) \tilde f^>_\nu(-t)$ & $\tilde f^>_\mu(t) \tilde f^<_\nu(-t)$\\
	D & $\tilde f^>_\mu(t) \tilde f^<_\nu(-t)$ & $\tilde f^<_\mu(t) \tilde f^>_\nu(-t)$\\
	E & $-f_\mu^<(t)f^<_\nu(-t)+\frac 12(f^<_\mu(t)+f^<_\nu(-t))\delta(t)$ & $-f_\mu^>(t)f^>_\nu(-t)+\frac 12(f^<_\mu(t)+f^<_\nu(-t))\delta(t)$
\end{tabular}
\end{center}
with $\tilde f^\gtrless_\mu(t)\equiv f^\gtrless_\mu(t+\tau^i_{\mu.\In}-\tau^j_{\mu.\In})$, $\tilde f^\gtrless_\nu(t)\equiv f^\gtrless_\nu(t-\tau^j_{\nu.\In}+\tau^i_{\nu.\In})$. We have incorporated the tunneling-free contribution (\ref{eqn:amplTunFree}) into the case E which amounts to the $\delta(t)$-terms.

Quite miraculously, using the symmetry relation $f^\gtrless_{\mu/\nu}(t)=-f^\lessgtr_{\mu/\nu}(t)$ for $t\neq 0$, one can show that (\ref{eqn:tunPolOp2}) holds, i.e.\ $\Pi_{ij;\mu}^{\T/\aT}$ can be represented in a form which is independent of topology. We exemplify the proof, which is a straightforward calculation, on configuration A. In this situation, $\tau^j_{\mu.\In}>\tau^i_{\mu.\In}$ and $\tau^i_{\nu.\In}>\tau^j_{\nu.\In}$ and thus $\tilde f_\mu^>(t)=-\tilde f^<_\mu(t)$ for $t>0$ and $\tilde f_\nu^>(-t)=-\tilde f_\nu^<(-t)$ for $t<0$, hence,
\begin{align*}
	-\tilde f^<_\mu(t) \tilde f^<_\nu(-t) &=\theta(t) \tilde f_\mu^>(t)\tilde f^<_\nu(-t)+\theta(-t)\tilde f_\mu^<(t)\tilde f_\nu^>(-t),\\
	\tilde f^>_\mu(t) \tilde f^<_\nu(-t) +\tilde f^<_\mu(t) \tilde f^>_\nu(-t)+\tilde f^<_\mu(t) \tilde f^<_\nu(-t) &= \theta(-t) \tilde f_\mu^>(t)\tilde f^<_\nu(-t)+\theta(t)\tilde f_\mu^<(t)\tilde f_\nu^>(-t),
\end{align*}
proving our statement. In the case E it is necessary to drop a constant (albeit infinite) and thus physically irrelevant contribution to the action to obtain (\ref{eqn:tunPolOp2}) or (\ref{eqn:tunPolOp3}).
(See similar discussion after Eq.~(\ref{eqn:tunPolOp3}) )

\section{Saddle-Point Approximation} \label{app:SPA}
We prove here the formulas of Sect.~\ref{sect:SPA}. To keep things readable we resort on a rather symbolic notation, in which the contributions to the exponent in (\ref{eqn:averageFunctInt}) read
	\begin{align*}
		\act_0[\varphi] &= \frac 12 \varphi V^{-1} \varphi - \varphi \varrho_0, \quad	\act_t[\varphi]=-i e^{-i\Phi(1)}\Pi_{12} e^{i\Phi(2)}, \quad \act_J[\varphi]=-J\varphi
	\end{align*}
	with the tunneling phases $\Phi=-\Phiphi\varphi$ being linear functionals of $\varphi$. The saddle-points of $\act_0+\act_t+\act_J$ and $\act_0+\act_J$ are denoted $\varphi_{\ast\ast}$ and $\varphi_\ast$. We show first that up to the chosen accuracy $\mathcal O(\tun^2)$ the former is not needed and calculation of the latter is sufficient, using that $\varphi_\ast-\varphi_{\ast\ast}=\mathcal O(\tun^2)$. A Gaussian expansion of the full action around $\varphi_{\ast\ast}$ reads
	\begin{align} \label{eqn:gaussExp1}
		\left(\act_0+\act_t+\act_J\right)[\varphi] \approx \left(\act_0+\act_t+\act_J\right)[\varphi_{\ast\ast}]+\frac 12 (\varphi-\varphi_{{\ast\ast}})\ \delta^2\left(\act_0+\act_t+\act_J\right)[\varphi_{\ast\ast}]\ (\varphi-\varphi_{\ast\ast}).
	\end{align}
	Note that the first order term vanishes due to $\varphi_{\ast\ast}$ being the full saddle-point. We now successively replace $\varphi_{\ast\ast}$ by $\varphi_\ast$ in the above expression. Expansion around $\varphi_{\ast\ast}$ and using $\delta^2(\act_0+\act_J)[\varphi_{\ast\ast}]=V^{-1}=\delta^2(\act_0+\act_J)[\varphi_\ast]$ yields
	\begin{align*}
		\left(\act_0+\act_t+\act_J\right)[\varphi_\ast] &= \left(\act_0+\act_t+\act_J\right)[\varphi_{\ast\ast}]+\mathcal O(\tun^4),\\
		\delta^2(\act_0+\act_t+\act_J)[\varphi_{\ast}]&=\delta^2(\act_0+\act_t+\act_J)[\varphi_{\ast\ast}] +\mathcal O(\tun^4).
	\end{align*}
	Writing $\varphi-\varphi_{\ast\ast}=(\varphi-\varphi_\ast)+(\varphi_\ast-\varphi_{\ast\ast})$ and using again $\varphi_\ast-\varphi_{\ast\ast} = \mathcal O(\tun^2)$ Eq.~(\ref{eqn:gaussExp1}) thus becomes
	\begin{multline} \label{eqn:gaussExp2}
		(\act_0+\act_t+\act_J)[\varphi]\approx (\act_0+\act_t+\act_J)[\varphi_\ast]+\frac 12 (\varphi-\varphi_{{\ast}})\ \delta^2\left(\act_0+\act_t+\act_J\right)[\varphi_{\ast}]\ (\varphi-\varphi_{\ast})\\+(\varphi_{\ast}-\varphi_{\ast\ast})\ \delta^2 (\act_0+\act_t+\act_J)[\varphi_\ast]\ (\varphi-\varphi_{\ast\ast}).
	\end{multline}
	Only the last term, which is linear in $\varphi-\varphi_\ast$, contains $\varphi_{\ast\ast}$. To proceed, we define $h\equiv (\varphi_{\ast}-\varphi_{\ast\ast})\ \delta^2 (\act_0+\act_t+\act_J)[\varphi_\ast]=\mathcal O(\tun^2)$ and perform the functional integration (\ref{eqn:averageFunctInt}) in Gaussian approximation (\ref{eqn:gaussExp2}):
	\begin{align}
		\int \DD\varphi\, e^{i(\act_0+\act_t+\act_J)[\varphi]} \approx& e^{i(\act_0+\act_t+\act_J)[\varphi_\ast]} \left(\Det \delta^2(\act_0+\act_t+\act_J)[\varphi_\ast]\right)^{-1/2} \nonumber\\
		 & \times \exp[-\frac i 2 \underbrace{h\ \delta^2(\act_0+\act_t+\act_J)[\varphi_\ast]\ h}_{=\mathcal O(\tun^4)}]\nonumber\\
		= &e^{i(\act_0+\act_J)[\varphi_\ast]} \exp\left[i\act_t[\varphi_\ast]-\frac 12 \Tr\Ln\left(\xUnit+V\delta^2 \act_t[\varphi_\ast]\right)\right] \label{eqn:gaussIntTwoAct}.
	\end{align}
	Hence, $\varphi_\ast$ dropped out completely. The first exponential satisfies $e^{i(\act_0+\act_J)[\varphi_\ast]}=\left\langle e^{i\act_J[\varphi]}\right\rangle_0$ where $\langle \ldots \rangle_0$ denotes the average with respect to the free action $\act_0[\varphi]$. This is shown using $\langle\varphi\rangle_0 =\bar \varphi\equiv V\varrho_0$ and $\langle(\varphi-\bar\varphi)(\varphi-\bar\varphi)\rangle_0=iV$. Since $\act_0[\varphi]$ is Gaussian we have the simple relation
	\begin{align*}
		\left\langle e^{i\act_J[\varphi]}\right\rangle_0 = \left\langle e^{-iJ\varphi}\right\rangle_0 = \exp\left[-iJ\bar \varphi-\frac 12 \left\langle \left[J(\varphi-\bar\varphi)\right]^2)\right\rangle\right]= e^{-iJ\bar \varphi -\frac i 2 JVJ}.
	\end{align*}
	On the other hand it is $i(\act_0+\act_J)[\varphi_\ast]=-\frac i 2 (\varrho_0+J)V(\varrho_0+J) = -\frac i 2 JVJ - i J\bar\varphi -\frac i2 \varrho_0V\varrho_0$. I.e.\ up to the last term $-\frac i2 \varrho_0V\varrho_0$, which is canceled by normalization, we have
	\begin{align*}
		e^{i(\act_0+\act_J)[\varphi_\ast]}=e^{-i J\bar\varphi -\frac i2 JVJ} = \left\langle e^{i\act_J[\varphi]}\right\rangle_0
	\end{align*}
	which proves (\ref{eqn:SPA1stExp}). 

	To deal with the second exponential in (\ref{eqn:gaussIntTwoAct}) we expand the logarithm to leading order in $V$ (again, a Gaussian expansion),
	\begin{align*}
		i\act_t[\varphi_\ast]-\frac 12 \Tr\Ln\left(\xUnit+V\delta^2 \act_t[\varphi_\ast]\right)\approx i\act_t[\varphi_\ast]-\frac 12 \Tr V\delta^2\act_t[\varphi_\ast]=i\act_t[\varphi_\ast]+\underbrace{i\left\langle\delta\act_t[\varphi_\ast](\varphi-\bar \varphi)\right\rangle_0}_{=0}\\+\frac i2\left\langle (\varphi-\bar \varphi)\delta^2\act_t[\varphi_\ast](\varphi-\bar\varphi)\right\rangle_0\approx i\left\langle \act_t[\varphi-\bar\varphi+\varphi_\ast]\right\rangle_0.
	\end{align*}
	Defining the phases $\bar \Phi=-\Phiphi\bar\varphi$, $\Phi_\ast=-\Phiphi\varphi_\ast$ it is
	\begin{align*}
		\left\langle \act_t[\varphi-\bar\varphi+\varphi_\ast]\right\rangle_0 = -i \left\langle e^{-i[\Phi(1)-\bar\Phi(1)+\Phi_\ast(1)]}\  \Pi_{12}\ e^{i[\Phi(2)-\bar\Phi(2)+\Phi_\ast(2)]}\right\rangle_0 = -i e^{-i\Phi_\ast(1)}\ \tilde\Pi_{12}\ e^{i\Phi_\ast(2)},
	\end{align*}
	i.e. Eq. (\ref{eqn:SPA2ndExp}), where $\tilde \Pi_{12} \equiv e^{i \left(D_\Phi(1,2)-D_\Phi(0,0)\right)} \Pi_{12}$ are the renormalized tunneling polarization operators with the phase-phase correlator
	\begin{align*}
		D_\Phi(1,2) = -i\left\langle (\Phi(1)-\bar \Phi(1))(\Phi(2)-\bar \Phi(2))\right\rangle_0= \left(\Phiphi V \Phiphi\right)(1,2).
	\end{align*}

\end{appendix}

\end{document}